\documentclass[
twocolumn,
aps,
prl,
reprint,
superscriptaddress,
amsmath,amssymb,
longbibliography
]{revtex4-2}
\usepackage[pdftex]{graphicx}
\usepackage[pdftex,
            colorlinks=true,
            citecolor=blue,
            linkcolor=blue]{hyperref}

\graphicspath{{./fig},{./fig_SM}}

\usepackage{braket}
\usepackage{times,multirow,amsfonts,bm,xspace,pifont}
\usepackage{soul}
\usepackage[normalem]{ulem}
\usepackage{textcomp}

\begin{document}
\newcommand{\sgn}{\mathrm{sgn}\,}
\newcommand{\tr}{\mathrm{tr}\,}
\newcommand{\Tr}{\mathrm{Tr}\,}

\newcommand{\YY}[1]{\textcolor{magenta}{#1}}
\newcommand{\AD}[1]{\textcolor{blue}{#1}}
\newcommand*{\ADS}[1]{\textcolor{blue}{\sout{#1}}}
\newcommand*\YYS[1]{\textcolor{magenta}{\sout{#1}}}
\newcommand{\reply}[1]{{#1}}
\newcommand{\replyS}[1]{\textcolor{red}{\sout{#1}}}

\renewcommand{\Re}{\mathrm{Re}}
\renewcommand{\Im}{\mathrm{Im}}

\newcommand{\halpha}{{\hat{\alpha}}}
\newcommand{\uint}{{\int_0^\infty}}
\newcommand{\mA}{{\mathcal{A}}}
\newcommand{\mP}{{\mathcal{P}}}
\newcommand{\mB}{{\mathcal{B}}}
\newcommand{\mJ}{{\mathcal{J}}}
\newcommand{\bx}{{\bar{x}}}
\newcommand{\by}{{\bar{y}}}
\newcommand{\bz}{{\bar{z}}}
\newcommand{\bepsilon}{{\bar{\epsilon}}}

\newcommand{\SC}{{\rm{s}}}
\newcommand{\normal}{{\rm{n}}}

\newcommand{\Tc}{{T_{\rm c}}}
\newcommand{\qc}{{q_{\rm c}}}
\newcommand{\Tcn}{{T_{\rm c0}}}
\newcommand{\RK}{{R_{\rm K}}}

\title{Rectification and nonlinear Hall effect by fluctuating finite-momentum Cooper pairs}

\author{Akito Daido}
\affiliation{Department of Physics, Graduate School of Science, Kyoto University, Kyoto 606-8502, Japan}
\email[]{daido@scphys.kyoto-u.ac.jp}
\author{Youichi Yanase}
\affiliation{Department of Physics, Graduate School of Science, Kyoto University, Kyoto 606-8502, Japan}
\date{\today}

\begin{abstract}
Nonreciprocal charge transport is attracting much attention as a novel probe and functionality of noncentrosymmetric superconductors.
In this work, we show that both the longitudinal and the transverse nonlinear paraconductivity are hugely enhanced in helical superconductors under moderate and high magnetic fields, which can be observed 
by second-harmonic resistance measurements.
The discussion is based on the generalized formulation of nonlinear paraconductivity in combination with the microscopically determined Ginzburg-Landau coefficients.
The enhanced nonreciprocal transport would be observable even with the cyclotron motion of fluctuating Cooper pairs, which is elucidated with a Kubo-type formula of the nonlinear paraconductivity. 
Nonreciprocal charge transport in the fluctuation regime is thereby established as a promising probe of helical superconductivity regardless of the sample dimensionality.
Implications for the other finite-momentum superconducting states are briefly discussed.
\end{abstract}

\maketitle

\textit{Introduction}. --- 
Nonreciprocal charge transport (NCT) is attracting much attention as the novel functionality of noncentrosymmetric materials~\cite{Tokura2018-nb,Ideue2021-es,Rikken2001-il,Rikken2005-ew,Ideue2017-vp,Wakatsuki2017-dp,Qin2017-vd,Yasuda2019-jw,Zhang2020-al,Ando2020-om,Lyu2021-sm,Wu2022-ey,Baumgartner2022-lg,Bauriedl2022-nq,Lin2022-cz,Narita2022-od,Mizuno2022-bd,Ideue2020-kg,Yuan2022-pz,Daido2022-ox,He2022-px,Ilic2022-kh,He2023-qo,Du2021-fg,Sodemann2015-vj,Ma2019-om,Kang2019-dr,Kumar2021-wf,Itahashi2022-ja,Zhang2021-hz,Toshio2022-de,Wakatsuki2018-ll,Hoshino2018-sa,Wu2022-kw,Guo2022-ey}.
An example is a diode-like material property known as magnetochiral anisotropy (MCA), which refers to directional resistance, or rectification,
linear in the magnetic field and has been observed in a variety of materials~\cite{Rikken2001-il,Rikken2005-ew,Ideue2017-vp,Wakatsuki2017-dp,Qin2017-vd,Yasuda2019-jw,Zhang2020-al}.
Unidirectional transport even with zero and finite resistance has also been realized, namely
the superconducting diode effect (SDE)~\cite{Ando2020-om,Lyu2021-sm,Wu2022-ey,Baumgartner2022-lg,Bauriedl2022-nq,Lin2022-cz,Narita2022-od,Mizuno2022-bd,Ideue2020-kg,Yuan2022-pz,Daido2022-ox,He2022-px,Ilic2022-kh}.
The nonlinear Hall effect (NHE) is another hot topic~\cite{Du2021-fg,Sodemann2015-vj,Ma2019-om,Kang2019-dr,Kumar2021-wf,Itahashi2022-ja}, by which
a finite transverse resistance can be produced in time-reversal symmetric materials.
These findings pave the way for {next-generation devices}~\cite{Zhang2020-al,Mizuno2022-bd,Ideue2020-kg,Zhang2021-hz,Toshio2022-de}.
Furthermore, NCT would 
{serve} as a versatile electrical probe of inversion-symmetry breaking, applicable even under extreme conditions including high pressure and magnetic fields.
Thus, NCT phenomena are hallmarks of modern condensed matter physics.

The development of NCT techniques may shed light on
the fascinating phenomena of noncentrosymmetric superconductors
that are hardly captured via conventional experiments.
Among other things, helical superconductivity~\cite{Bauer2012-xi,Smidman2017-hb,Agterberg2003-jn,Barzykin2002-eh,Dimitrova2003-mo,Kaur2005-jf,Agterberg2007-vl,Dimitrova2007-hp,Samokhin2008-nv,Yanase2008-yb,Bauer2012-xi,Michaeli2012-gl,Houzet2015-iy} is a 
long-sought 
finite-momentum superconducting state, 
regardless of its predicted ubiquity in magnetic fields.
The pair potential of helical superconductivity has a plane-wave expression known as the Fulde-Ferrell type~\cite{Fulde1964-qq} 
without the modulation of amplitude. This makes its experimental identification more difficult than the Larkin-Ovchinnikov and pair-density-wave states~\cite{Larkin1964-en,Matsuda2007-em,Wosnitza2018-qh,Agterberg2020-gs},
whose detection has been reported
via spatially-resolved techniques in various superconductors~\cite{Kumagai2011-ao,Mayaffre2014-kt,Koutroulakis2016-wg,Kitagawa2018-qh,Kasahara2021-qa,Kinjo2022-bg,Hamidian2016-xz,Ruan2018-fv,Chen2021-qp,Gu2023-ps} including
FeSe~\cite{Kasahara2021-qa}, Sr$_2$RuO$_4$~\cite{Kinjo2022-bg}, Bi$_2$Sr$_2$CaCu$_2$O$_{8+\delta}$~\cite{Hamidian2016-xz,Ruan2018-fv}, CsV$_3$Sb$_5$~\cite{Chen2021-qp}, and UTe${}_2$~\cite{Gu2023-ps}.
Recent theoretical studies~\cite{Daido2022-ox,Ilic2022-kh,Daido2022-gj} have revealed that the characteristic crossover phenomenon of helical superconductivity can be signaled by the sign reversal of the SDE, 
offering a promising probe free of Josephson junctions in contrast to the known methods~\cite{Kaur2005-jf,Kim2016-id}.
Further investigation of NCT would provide us with keys to understanding exotic superconducting states in noncentrosymmetric systems.

The disadvantage of the SDE as a probe of helical superconductivity is to require small-width samples to suppress vortex motion and approach the depairing limit of the critical current~\cite{Tinkham2003-tu}. 
It has also been pointed out that the SDE is sensitive to the conditions around sample edges~\cite{Hou2023-hu,Vodolazov2005-cg}.
Thus, careful microfabrication would be required to study the intrinsic SDE in candidate helical superconductors such as heavy-fermion superlattices~\cite{Naritsuka2017-or,Naritsuka2021-ym} and thin films of Pb~\cite{Sekihara2013-dm} and SrTiO$_3$~\cite{Schumann2020-mw}.
{Toward easier access to helical superconductivity,}
we turn renewed attention to the nonreciprocal paraconductivity, 
i.e., NCT
by fluctuating Cooper pairs{, which is little affected by the edge environments}.
In pioneering works~\cite{Wakatsuki2017-dp,Wakatsuki2018-ll,Hoshino2018-sa}, nonreciprocal paraconductivity was studied focusing on MCA {and was shown to be significantly larger than MCA of normal electrons~\cite{Wakatsuki2017-dp}}.
The theoretical studies not only succeeded in explaining the experiment in MoS$_2$~\cite{Wakatsuki2017-dp}, but also pointed out that spin-singlet and -triplet mixing of Cooper pairs can be detected~\cite{Wakatsuki2018-ll,Hoshino2018-sa}.
However, their formulation is not applicable in the presence of finite-momentum Cooper pairs and/or nonlinear effects of the magnetic field, leaving helical superconductors out of its scope.

In this Letter, 
we generalize the previous formulation of nonreciprocal paraconductivity
and show that the rectification and NHE in the fluctuation regime are hugely enhanced in helical superconductors in moderate and strong magnetic fields.
We also show that the enhanced NCT would still be observable even in the presence of the cyclotron motion of Cooper pairs.
Our formulation is applicable to Fulde-Ferrell-type superconducting states in general, while
thin-film Larkin-Ovchinnikov superconductors may also be explored by applying symmetry-breaking perturbations.
{Our results showcase an interesting example of NCT that originates from the intrinsic nature of exotic Cooper pairs.}

\textit{Notations for NCT.} --- 
We begin by introducing 
the notation for rectification and the NHE,
which is described by the nonlinear conductivity
$j_i=\sigma_1^{ij}E_j+\sigma_2^{ijk}E_jE_k$
or 
the nonlinear resistivity
$E_i=\rho^{ij}j_j=\rho_1^{ij}j_j+\rho_2^{ijk}j_jj_k.$
Here linear and nonlinear resistivities satisfy $\rho_1=\sigma_1^{-1}$ and
\begin{align}
\rho_2^{ijk}=-[\sigma_1^{-1}]_{ia}\sigma_{2}^{abc}[\sigma_1^{-1}]_{bj}[\sigma_1^{-1}]_{ck}.
\label{eq:def_resistivity}
\end{align}
The nonlinear resistivity can be observed via the longitudinal and Hall second-harmonic resistance~\cite{Ideue2017-vp}.
{The nonlinear resistivity} $\rho_2^{xxx}$ gives rise to nonreciprocity in {the longitudinal resistivity} $\rho^{xx}$ {in the electric current $j_x$,}:
\begin{align}
\rho^{xx}=\rho_1^{xx}(1+\eta^{xxx} j_x), \quad
\eta^{xxx}\equiv \rho_2^{xxx}/\rho_1^{xx}.\label{eq:nonreciprocity}
\end{align}
The {longitudinal nonreciprocity} $\eta^{xxx}$ is a natural generalization of the $\gamma$ value for MCA~\cite{Tokura2018-nb,Ideue2021-es} and is used as a quantitative measure of rectification in this paper{: According to Eq.~\eqref{eq:nonreciprocity}, its inverse $1/\eta^{xxx}$ gives a typical current density for nonreciprocity to be visible.}
We also introduce 
\begin{align}
{\eta^{xyy}\equiv \rho_2^{xyy}/\rho_1^{xx}}
\label{eq:Hall_nonreciprocity}
\end{align}
to compare the NHE with rectification {and call it Hall nonreciprocity although the linear Hall effect vanishes in the model studied later.}

Near the transition temperature ${T_{\rm c}}$ of superconductors, {the conductivity} tends to diverge due to the fluctuation of Cooper pairs, which interpolates the finite and vanishing resistance in normal and superconducting states~\cite{Larkin2005-lb}.
The linear and nonlinear conductivities {$\sigma_1$ and $\sigma_2$ can be decomposed into those} in the normal state and the excess contribution by fluctuation {which are specified by the} subscripts ${\rm{n}}$ and ${\rm{s}}$, respectively: $\sigma_1=\sigma_{1{\rm{n}}}+\sigma_{1{\rm{s}}}$ and
$\sigma_2=\sigma_{2{\rm{n}}}+\sigma_{2{\rm{s}}}$.
Our purpose is to obtain the paraconductivity contributions $\sigma_{1{\rm{s}}}$ and $\sigma_{2{\rm{s}}}$ and thereby evaluate {the nonlinear resistivities} $\rho^{xxx}$ and $\rho^{xyy}$ in the fluctuation regime of superconductors.

\textit{Time-dependent GL theory.} --- 
Let us consider a $d$-dimensional superconductor slightly above ${T_{\rm c}}$, with $d=2$ unless otherwise specified.
Following Refs.~\cite{Wakatsuki2017-dp,Wakatsuki2018-ll,Hoshino2018-sa}, we study the fluctuation of Cooper pairs by using the phenomenological time-dependent Ginzburg-Landau (GL) equation in the momentum space~\cite{Supplemental},\begin{subequations}\label{eq:TDGLs}
\begin{align}
\Gamma_0\frac{\partial\psi_{\bm{q}}(t)}{\partial t}&=-\alpha_{\bm{q}}\psi_{\bm{q}}(t)+\zeta_{\bm{q}}(t),
\\
\braket{\zeta^*_{\bm{q}}(t)\zeta_{\bm{q}'}(t')}&=\frac{2\Gamma_0T}{V}\delta(t-t')\delta_{\bm{q},\bm{q}'},
\end{align}\end{subequations}with the GL functional $F[\psi]=V\sum_{\bm{q}}\alpha_{\bm{q}}|\psi_{\bm{q}}|^2$.
The random force $\zeta_{\bm{q}}(t)$ is assumed to be the white noise as in the second line and reproduces $\braket{|\psi_{\bm{q}}|^2}=T/V\alpha_{\bm{q}}$ in equilibrium.
The effect of the electric field $\bm{E}$ is introduced by $\alpha_{\bm{q}}\to \alpha_{\bm{q}-2\bm{A}(t)}$ with $\bm{A}(t)=-\bm{E}t$.
The excess current density by fluctuating Cooper pairs is evaluated with the formula~\cite{Wakatsuki2017-dp,Schmid1969-fw,Supplemental}
\begin{align}
\bm{j}_{\rm{s}}&=\lim_{t\to\infty}-\sum_{\bm{q}}\partial_{\bm{A}}\alpha_{\bm{q}-2\bm{A}(t)}\braket{|\psi_{\bm{q}}(t)|^2}\label{eq:GL_current}\\
&=\frac{4T}{\Gamma_0}\int\frac{d^dq}{(2\pi)^d}\partial_{\bm{q}}\alpha_{\bm{q}}\int^0_{-\infty}dt_1\,e^{-\frac{2}{\Gamma_0}\int^0_{t_1}dt'\,\alpha_{\bm{q}-2\bm{A}(t')}},\notag
\end{align}
{which results from a process where Cooper pairs are formed by fluctuations and then accelerated by the electric field until they vanish after a finite lifetime.}

Within the GL picture, the superconducting transition is triggered by the softening of the mode $\bm{q}=\bm{q}_0$ which minimizes $\alpha_{\bm{q}}$.
This occurs at $\bm{q}_0\neq0$ in helical superconductivity, in contrast to $\bm{q}_0=0$ in conventional superconductors.
Note that the modes around $\bm{q}_0$ dominantly contribute to transport properties in the vicinity of ${T_{\rm c}}$.
Thus, we can expand the GL coefficient in terms of $\delta\bm{q}=\bm{q}-\bm{q}_0$,
\begin{align}
\alpha_{\bm{q}}&=\alpha_{\bm{q}_0}+\alpha_2^{ij}\delta q_i\delta q_j+\alpha_3^{ijk}\delta q_i\delta q_j\delta q_k+O(\delta q^4)\label{eq:alpha_q}\\
&\equiv N_0\left(\epsilon+\sum_i\xi_i^2\delta q_i^2+\bar{\xi}\,{}^3\sum_{ijk}\mathcal{A}^{ijk}\delta q_i\delta q_j\delta q_k\right).\notag
\end{align}
We defined the reduced temperature $\epsilon\equiv (T-{T_{\rm c}})/T$, GL coherence length $\xi_i$, its geometric mean $\bar{\xi}\equiv (\prod_{i=1}^d\xi_i)^{1/d}$, and the dimensionless third-rank tensor ${\mathcal{A}}^{ijk}$, while the overall coefficient $N_0\equiv T\frac{\partial}{\partial T}\alpha_{\bm{q}_0}$ is related to the density of states.
Importantly, cubic anharmonicity ${\mathcal{A}}^{ijk}$ is allowed with $\bm{q}_0\neq0$ and/or without both inversion and time-reversal symmetries.

\textit{Nonlinear Paraconductivity.} ---The GL formula of the fluctuation conductivity can be obtained by plugging Eq.~\eqref{eq:alpha_q} into Eq.~\eqref{eq:GL_current} and expanding it by {the electric field} $\bm{E}$.
We neglect the orbital magnetic field for the time being, while the effect of the Zeeman field can be taken into account.
The linear fluctuation conductivity is {then} given by
$L_z\sigma_{1{\rm{s}}}^{ij}
=\frac{\tau_0T}{2\pi \epsilon}\frac{\xi_i^2}{\xi_x\xi_y}\delta_{ij}$
to the leading order of {the reduced temperature} $\epsilon$, with the sample thickness $L_z$ and {the GL relaxation time} $\tau_0\equiv\Gamma_0/N_0>0$~\cite{Supplemental,Larkin2005-lb,Konschelle2007-da}.
In the absence of anisotropy, this reproduces
$L_z\sigma_{1{\rm{s}}}=1/{16\epsilon}$
for $\tau_0=\pi/8T$~\cite{Larkin2005-lb}.

The nonlinear paraconductivity is similarly obtained~\cite{Supplemental},
\begin{align}
L_z\sigma_{2{\rm{s}}}^{ijk}
&=\frac{\tau_0^2T\sqrt{\xi_x\xi_y}}{4\pi\epsilon^2}{\mathcal{A}}^{ijk},
\label{eq:nonlinear_formula}
\end{align}
to the leading order of {the reduced temperature} $\epsilon$.
The NCT is of $O(\epsilon^{-2})$, as reported previously~\cite{Wakatsuki2017-dp,Wakatsuki2018-ll,Hoshino2018-sa}.
{Notably, it is the anharmonicity parameter ${\mathcal{A}}^{ijk}$ that gives rise to NCT~\cite{Wakatsuki2017-dp,Hoshino2018-sa}, since $\bm{q}_0$ can be traced out from Eq.~\eqref{eq:GL_current} by shifting the momentum.}
Note that $\sigma_{2{\rm{s}}}^{ijk}$ allows not only rectification but also the NHE.
Nonlinear paraconductivities for system dimensions $d=1$ and $3$ are also obtained as
{$L_yL_z\sigma_{2{\rm{s}}}^{xxx}=\frac{3\tau_0^2T\xi_x^2}{8\epsilon^{5/2}}{\mathcal{A}}^{xxx}$ and 
$\sigma_{2{\rm{s}}}^{ijk}=\frac{\tau_0^2T}{16\pi\epsilon^{3/2}}{\mathcal{A}}^{ijk}$,}
respectively, where $L_yL_z$ is the wire cross section.
We emphasize that the obtained formulas allow us to discuss the nonlinear effect of the Zeeman field $\bm{h}$ and, if any, coexisting time-reversal-breaking orders, in contrast to the previous formulas showing $O(h)$ NCT in specific two-dimensional models~\cite{Wakatsuki2017-dp,Wakatsuki2018-ll,Hoshino2018-sa}.
This point is crucial to describe fluctuating finite-momentum Cooper pairs.

To illustrate the formula~\eqref{eq:nonlinear_formula}, we discuss NCT linear in the Zeeman field $\bm{h}$ before studying the nonlinear effects of $\bm{h}$.
In this case, {the anharmonicity parameter}
${\mathcal{A}}^{ijk}$ { is $O(h)$ and} can be rewritten in the form of the cubic spin-orbit coupling (SOC)~\cite{Daido2022-gj}
\begin{align}
{\mathcal{A}}^{ijk}\delta q_i\delta q_j\delta q_k
\equiv\bm{h}\cdot\bm{g}_{\mathcal{A}}(\delta\bm{q}).
\end{align}
The effect of $\bm{h}$ on the other coefficients is $O(h^2)$ and thus is negligible in the low-field region.
For the purpose of symmetry considerations, 
the effective SOC
$\bm{g}_{\mathcal{A}}(\delta\bm{q})$ can be identified with the antisymmetric SOC of the system around the $\Gamma$ point in the Brillouin zone~\cite{Daido2022-gj}.
Typical forms of the effective SOC in Rashba, chiral, Ising, and Dresselhaus systems are illustrated with a unit vector $\hat{n}$ in Table~\ref{tab:SOC}.

\begin{table}[tb]
    \caption{%
    Typical forms of $\bm{g}_{\mathcal{A}}(\hat{n})=\bm{g}_{\mathcal{A}}(\delta\bm{q})/\delta q^3$
    and $\frac{\partial}{\partial \theta}\bm{g}_{\mathcal{A}}(\hat{n})$ for various types of the antisymmetric SOC. 
    Here, we defined $\delta\bm{q}=\delta q\,\hat{n}$ and unit vectors $\hat{n}=(\cos\theta,\sin\theta,0)$ and $\hat{z}=(0,0,1)$.
    }
    \centering      
    \begin{tabular}{lccc}
    \\\hline\hline
        Type of SOC&& $\bm{g}_{\mathcal{A}}(\hat{n})$&$\frac{\partial}{\partial \theta}\bm{g}_{\mathcal{A}}(\hat{n})$ \\\hline
    Rashba     && $\hat{z}\times\hat{n}$&$-\hat{n}$\rule[0mm]{0mm}{4mm}\\
    Chiral && $\hat{n}$&$\hat{z}\times\hat{n}$\rule[0mm]{0mm}{5mm}\\
    Ising&&$\sin 3\theta\,\hat{z}$&$3\cos3\theta\,\hat{z}$\rule[0mm]{0mm}{5mm}\\
    Dresselhaus &&$\sin 2\theta\,\hat{z}\times\hat{n}$& \begin{tabular}{l}
$2\cos2\theta\,\hat{z}\times\hat{n}$\\
    $\quad\,-\sin2\theta\,\hat{n}$    
    \end{tabular}
   \rule[0mm]{0mm}{6.5mm}\\
    \hline\hline
    \end{tabular}
    \label{tab:SOC}
\end{table}
When the electric field with strength $E$ is applied in an in-plane direction $\hat{E}$, the $O(E^2)$ excess current density $\delta^2\bm{j}_{\rm{s}}$ in this direction is ${\hat{E}\cdot\delta^2\bm{j}_{\rm s}}
\propto {\mathcal{A}}^{ijk}\hat{E}_i\hat{E}_j\hat{E}_k$ {from Eq.~\eqref{eq:nonlinear_formula}}, i.e.,
{$\hat{E}\cdot\delta^2\bm{j}_{\rm s}\propto \bm{h}\cdot\bm{g}_{\mathcal{A}}(\hat{E})$.}
The field-angle dependence of rectification
is determined by the effective SOC $\bm{g}_{\mathcal{A}}(\hat{E})$ [Table~\ref{tab:SOC}].
Similarly, the transverse excess current density
is given by
{$[\hat{z}\times\hat{E}]\cdot\delta^2\bm{j}_{\rm s}\propto \bm{h}\cdot \frac{\partial}{\partial\theta}\bm{g}_{\mathcal{A}}(\hat{E})$~\cite{Supplemental}}.
Here the $\theta$ derivative acts on $\hat{E}=(\cos\theta,\sin\theta,0)$ and thus
\begin{align}
[\hat{z}\times\hat{E}]\cdot\delta^2\bm{j}_{\rm{s}}\propto\hat{E}\cdot\bm{h},
\end{align}
e.g., in Rashba systems [Table~\ref{tab:SOC}].
This indicates that NHE occurs for the magnetic field {parallel} to the electric field in contrast to the rectification that occurs in the perpendicular configuration.
The results obtained here give the generalized and convenient description of the known results for the Rashba~\cite{Wakatsuki2018-ll,Hoshino2018-sa,Supplemental} and Ising systems~\cite{Wakatsuki2017-dp,Hoshino2018-sa}.

It should be noted that the nonlinear resistivity $\rho_2$ rather than conductivity $\sigma_{2}$ is directly observed in experiments.
It turns out that not only {the linear resistivity $\rho_1$ but also the nonlinear resistivity $\rho_2$} vanishes as {it approaches the transition temperature} $\epsilon\to0$ in the present framework, due to $\sim\sigma_1^{-3}$ in Eq.~\eqref{eq:def_resistivity}.
Nevertheless, {the nonlinear resistivity} $\rho_2$ can be hugely enhanced in the fluctuation regime before it finally vanishes, reflecting the divergence of the nonlinear conductivity $\sigma_2$.
To estimate {the nonlinear longitudinal and Hall resistivities} $\rho_2^{xxx}$ and $\rho_2^{xyy}$ in the fluctuation regime, we define the reduced temperature $\epsilon_*$ indicating the linear-resistance drop by $25\%$ of the normal-state value{~\cite{Supplemental}.}
{We denote nonlinear resistivities evalated at $\epsilon=\epsilon_*$ by $\rho_{2*}^{xxx}$ and $\rho_{2*}^{xyy}$.}

{In contrast to the nonlinear resistivity, the nonreciprocity of the resistivities $\eta^{xxx}$ and $\eta^{xyy}$ in Eqs.~\eqref{eq:nonreciprocity}~and~\eqref{eq:Hall_nonreciprocity} converges to a finite value as it approaches the transition temperature~\cite{Wakatsuki2017-dp,Wakatsuki2018-ll,Hoshino2018-sa}.
We define this limiting value by
}
\begin{align}
\eta_{{\rm{s}}}^{ijk}\equiv \lim_{\epsilon\to+0}\eta^{ijk}(\epsilon)=-L_z\frac{\pi\sqrt{\xi_x\xi_y}}{T}\left(\frac{\xi_x\xi_y}{\xi_j\xi_k}\right)^2{\mathcal{A}}^{ijk},
\label{eq:def_etaSC}
\end{align}
for two-dimensional superconductors.
{This quantity measures the intrinsic nonreciprocity, which does not depend on the normal-state resistivity.}

\textit{Application to helical superconductivity.} ---By using the GL formula \eqref{eq:nonlinear_formula}, we study rectification and the NHE in atomically thin $s$-wave and $d$-wave Rashba superconductors in the in-plane Zeeman field $\bm{h}$.
The Bloch Hamiltonian is given by
$H_N(\bm{k})=\xi(\bm{k})+[\bm{g}(\bm{k})-\bm{h}]\cdot\bm{\sigma}$,
with the hopping energy $\xi(\bm{k})=-2t(\cos k_x+\cos k_y)-\mu$ and Rashba SOC $\bm{g}(\bm{k})=\alpha_{\rm R}(-\sin k_y,\sin k_x,0)$.
We microscopically determine the GL coefficient $\alpha_{\bm{q}}$~\cite{Supplemental}, 
which gives $\bm{q}_0=q_0\hat{x}$ upon minimization and $\xi_{i}$ and ${\mathcal{A}}^{ijk}$ by taking $\bm{q}$ derivatives.
{The qualitative results do not depend on model parameters $t,\mu,\alpha_{\rm R}$ etc., when $\alpha_{\rm R}\gg T_{\rm c0}$ as is the case in most noncentrosymmetric superconductors. 
{Note that the Rashba energy $\alpha_{\rm R}$ is always dominant over the Zeeman energy $h$ on the entire phase diagram since $h\sim T_{\rm c0}$ is considered.}
Here we denote the transition temperature in the magnetic field $h$ by $T_{\rm c}(h)$ and $T_{\rm c0}\equiv T_{\rm c}(0)$. The parameters adopted for numerical calculations are available in the Supplemental Material~\cite{Supplemental}.}

\begin{figure}
    \centering
    \includegraphics[width=0.5\textwidth]{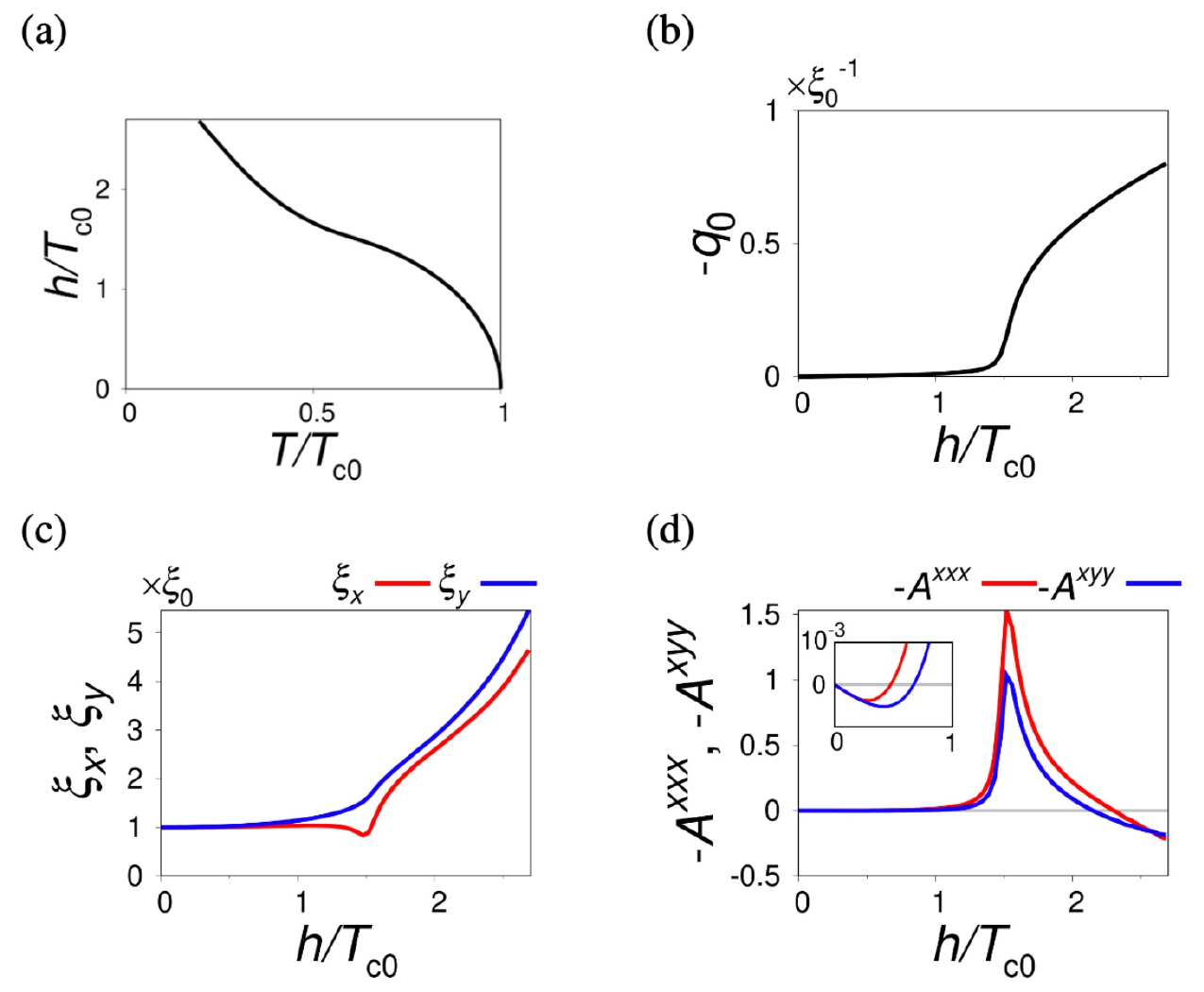}
    \caption{%
    (a) Transition line $({T_{\rm c}}(h),h)$ of the $s$-wave Rashba-Zeeman superconductor, (b) Cooper-pair momentum $-q_0$, (c) GL coherence length $\xi_x$ and $\xi_y$, and (d) anharmonicity parameters $-{\mathcal{A}}^{xxx}$ and $-{\mathcal{A}}^{xyy}$ along the transition line $({T_{\rm c}}(h),h)$.
    Here $\xi_x,\xi_y$ and $q_0^{-1}$ are in units of $\xi_0$, i.e. $\xi_x$ and $\xi_y$ at $h=0$.
    The increasing tendency in $\xi_x$ and $\xi_y$ comes from the decrease of $T_{\rm c}(h)$.
    The inset in (d) shows the region $0\le h/{T_{\rm c0}}\le 1$.    
    }    \label{fig:Tcandq0}
\end{figure}

We show in Fig.~\ref{fig:Tcandq0} the superconducting transition line and GL coefficients of the $s$-wave state.
The Cooper-pair momentum ${q}_0$ of the soft mode along the transition line $({T_{\rm c}}(h),h)$ is shown in Fig.~\ref{fig:Tcandq0}(b), whose finite value indicates the realization of helical superconductivity for $T<{T_{\rm c}}(h)$.
It is shown for $h/{T_{\rm c0}}\sim1.5$ that the system experiences a rapid increase in $|q_0|$ known as the
crossover between weakly and strongly helical states~\cite{Bauer2012-xi,Smidman2017-hb}.
While the coherence lengths $\xi_x$ and $\xi_y$ are always of the same order of magnitude [Fig.~\ref{fig:Tcandq0}(c)],
the anharmonicity parameters ${\mathcal{A}}^{xxx}$ and ${\mathcal{A}}^{xyy}$ are extremely enhanced in the crossover region as shown in Fig.~\ref{fig:Tcandq0}(d):
The rapid change in $\bm{q}_0$ naturally accompanies the anomalous $\bm{q}$ dependence of $\alpha_{\bm q}$ around there.
The inset shows that ${\mathcal{A}}^{xxx}$ and ${\mathcal{A}}^{xyy}$ have tiny linear slopes {corresponding to MCA} in the small magnetic field $h$, as expected.
The $h$-linear behavior is limited to the low-field region and thus the nonlinear effects are essential.
The high-field behavior is discussed below.

\begin{figure}
    \centering
    \includegraphics[width=0.4\textwidth]{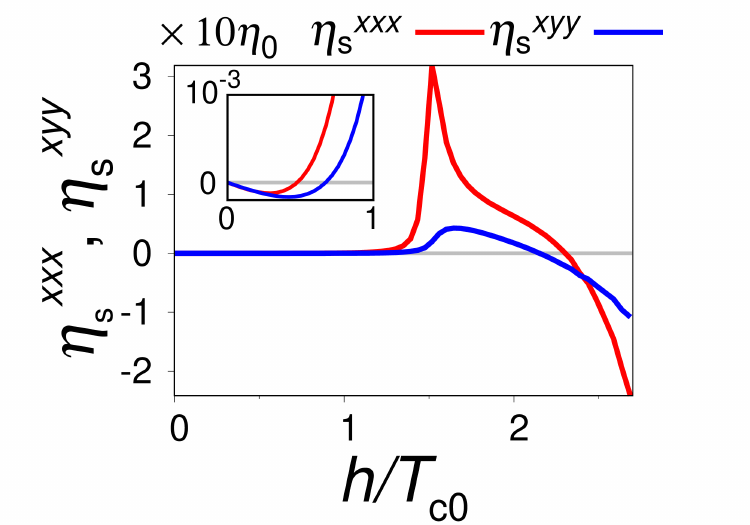}
    \caption{%
    Strength of the rectification $\eta^{xxx}_{\rm{s}}$ and NHE $\eta^{xyy}_{\rm{s}}$ in the $s$-wave Rashba-Zeeman superconductor along the transition line $({T_{\rm c}}(h),h)$.
    The inset shows the region $0\le h/{T_{\rm c0}}\le 1.$
    }
\label{fig:Results}
\end{figure}
The huge increase of the anharmonicity parameters naturally enhances rectification and the NHE as shown in Fig.~\ref{fig:Results}:
Both {the longitudinal and Hall nonreciprocities }$\eta^{xxx}_{\rm{s}}$ and $\eta^{xyy}_{\rm{s}}$ {given in Eq.~\eqref{eq:def_etaSC}} are increased by several orders of magnitude along the transition line.
The enhancement of $\eta^{xxx}_{\rm{s}}$ compared to $\eta^{xyy}_{\rm{s}}$ originates from the increased anisotropy $\xi_y/\xi_x$ in the crossover region [see Fig.~\ref{fig:Tcandq0}(c) and Eq.~\eqref{eq:def_etaSC}].
Similar results are obtained for various parameters and for the $d$-wave states~\cite{Supplemental}.
The values of $\eta_{\rm{s}}^{ijk}$ obtained are 
comparable in units of $\eta_0\equiv L_z\xi_0/{T_{\rm c0}}$, implying large NCT in superconductors with small ${T_{\rm c0}}$ and large $\xi_0$.
For the case of heavy-fermion superlattices~\cite{Naritsuka2021-ym}, 
we obtain $\eta_{\rm{s}}^{xxx}\sim 10\eta_{\rm{s}}^{xyy}\sim 10^{-2}\,\mathrm{\text{\textmu} m^2/\text{\textmu} A}$ while assuming $\xi_0\sim 5\,\mathrm{nm}$, ${T_{\rm c0}}\sim 2\,\mathrm{K}$, and $L_z\sim 10\,\mathrm{nm}$.
This means that $10\%$ rectification is obtained for a current density of approximately $ 10\,\mathrm{\text{\textmu} A/\text{\textmu} m^2}$ at the mean-field transition temperature.
Typical values of the nonlinear resistivity in the fluctuation regime are estimated 
to be $\rho^{xxx}_{2*}\sim 10\rho^{xyy}_{2*}\sim 10^{-4}\,\Omega\mathrm{\text{\textmu} m^3/\text{\textmu} A}$ while assuming $\sigma_{1{\rm{n}}}^{-1}\sim 5\times 10^{-7}\,\Omega\mathrm{m}$.
These values are well within the experimental scope.
Thus, a sharp increase of
rectification and the NHE {in the crossover regime, as opposed to the standard $h$-linear behavior,} serves as a promising probe {of helical superconducivity.}

Interestingly, the anharmonicity parameters take slightly smaller but still sizable values in higher magnetic fields [$h/{T_{\rm c0}}\gtrsim2.5$ in  Fig.~\ref{fig:Tcandq0}(d)].
A large rectification and NHE are obtained there in combination with small ${T_{\rm c}}(h)$ [Fig.~\ref{fig:Results}],
while the sign reversal seen in Fig.~\ref{fig:Results} may be absent or shifted to higher fields, depending on model parameters~\cite{Supplemental}.
It is known that the high-field helical superconductivity resembles in nature the Fulde-Ferrell-Larkin-Ovchinnikov (FFLO) state of centrosymmetric superconductors~\cite{Smidman2017-hb}.
{While these systems do not show nonreciprocal paraconductivity due to the cancellation of fluctuating Cooper pairs with opposite momenta,}
our results imply that the FFLO state, and possibly the pair-density-wave states, might show giant NCT once the symmetry-protected degeneracy of Cooper-pair momenta is externally lifted.
{This} could be achieved by the out-of-plane bias voltage and in-plane magnetic field{, realizing the same symmetry configurations as helical superconductivity.}
Quantitative studies are awaited
for candidate materials such as cuprate thin films~\cite{Agterberg2020-gs,Bollinger2011-ef,Leng2011-bi,Nojima2011-il,Liao2018-ty}.

\textit{Orbital magnetic field.} ---
We have pointed out that the colossal rectification and NHE are promising probes of thin-film helical superconductors.
A natural question is then whether the conclusion still holds in quasi-two-dimensional superconductors where the cyclotron motion of fluctuating Cooper pairs takes place.
To study this problem, we derive a Kubo-type formula of $\sigma_{2{\rm{s}}}^{ijk}$ for the time-dependent GL equation
of the form $\Gamma_0\frac{\partial}{\partial t}\ket{\psi(t)}=\hat{\alpha}\ket{\psi(t)}+\ket{\zeta(t)}$,
\begin{align}
\!\!\!\!\sigma_{2{\rm{s}}}^{ijk}&=\frac{2\Gamma_0^2}{\beta V}\sum_{\mu\nu\lambda}\frac{J_{\mu\nu\lambda}^{ijk}(\alpha_\mu+\alpha_\nu+2 \alpha_\lambda)}{\alpha_\lambda (\alpha_\mu+\alpha_\nu) (\alpha_\mu+\alpha_\lambda)^2 (\alpha_\nu+\alpha_\lambda)^2},\label{eq:formula_general}
\end{align}
with $\hat{\alpha}\ket{\mu}=\alpha_\mu\ket{\mu}$,
$J_{\mu\nu\lambda}^{ijk}=\mathrm{Re}[\braket{\mu|j_i|\nu}\braket{\nu|j_j|\lambda}\braket{\lambda|j_k|\mu}]$, and $j_i=-\partial_{A_i}\hat{\alpha}$~\cite{Supplemental}.
This general formula of the phenomenological nonlinear paraconductivity is applicable to, e.g., systems with orbital magnetic fields as well as multiple pairing channels.

Let us consider bulk noncentrosymmetric superconductors in the magnetic field $B$ in the $y$ direction,
which can be described by
$\hat{\alpha}=\alpha_{\bm{q}}|_{\bm{q}\to\nabla/i-2\bm{A}(\bm{x})}$~\cite{Larkin2005-lb,Abrahams1966-xl,Hoshino2018-sa}.
We focus on the first-order effect of {the anharmonicity parameters} ${\mathcal{A}}^{xxx}$ and ${\mathcal{A}}^{xyy}$ for the purpose of an order estimate of NCT~\cite{Supplemental},
$\sigma_{2{\rm{s}}}^{xxx}=
\frac{\tau_0^2T}{2\pi|{\mathcal{B}}|\sqrt{\bar{\epsilon}}}\,{\mathcal{A}}^{xxx}$ and $\sigma_{2{\rm{s}}}^{xyy}=\frac{3\tau_0^2T}{4\pi\bar{\epsilon}^{3/2}}\,{\mathcal{A}}^{xyy}$,
where ${\mathcal{B}}\equiv B\xi_x\xi_z$ is the magnetic flux threading the area spanned by the coherence length.
The most singular terms regarding the reduced temperature 
in the magnetic field $\bar{\epsilon}\equiv\epsilon+2|{\mathcal{B}}|=(T-{T_{\rm c}}(B))/T$ are kept here, while $\sigma_{2{\rm{s}}}^{yxy}=\sigma_{2{\rm{s}}}^{yyx}=\sigma_{2{\rm{s}}}^{xyy}/2$ is obtained to the leading order of $\bar{\epsilon}$.
{The obtained nonlinear conductivity} indicates that the orbital magnetic field
suppresses the singularity of rectification perpendicular to the field while leaving that of the NHE intact {[see $d=3$ result shown below Eq.~\eqref{eq:nonlinear_formula}].}
See the Supplemental Material for more details of NCT for the orbital magnetic field. 

The obtained expressions of NCT are proportional to {the anharmonicity parameter} ${\mathcal{A}}^{ijk}$, implying that the rapid increase of NCT occurs in bulk samples as well, triggered by the helical-superconductivity crossover.
To estimate the nonreciprocity in the fluctuation regime, we discuss {a typical value of nonreciprocity}
$\eta_{*}^{xxx}\equiv \rho_{2*}^{xxx}/\rho_{1*}^{xx}$ 
since {the intrinsic limiting value} $\eta_{{\rm{s}}}^{xxx}$ vanishes. 
At the reduced temperature defined by $\sigma_{1{\rm{s}}}^{xx}({\bar{\epsilon}}_*)=\sigma_{1{\rm{n}}}^{xx}/3$, we obtain
the nonreciprocity $\eta_*^{xxx}\sim10^{-5}\,\mathrm{\text{\textmu}m^2/\text{\textmu} A}$ for layered helical superconductors in the crossover region, by using  
the coherence lengths $\xi_{x}\sim\xi_y/2\sim 5\,\mathrm{nm}$ and $\xi_z\sim 2\,\mathrm{nm}$, relaxation time $\tau_0\sim \pi/8T$, and $B\sim 2{T_{\rm c0}}\sim{0.4}\,\mathrm{meV}$, as well as ${\mathcal{A}}^{xxx}$ estimated from Fig.~\ref{fig:Tcandq0}(d).
The NHE can be estimated similarly~\cite{Supplemental}.
The obtained rectification and NHE $\rho^{xxx}_{2*}\sim\rho^{xyy}_{2*}\sim 10^{-5}\,\Omega\mathrm{\text{\textmu} m^3/\text{\textmu} A}$ are smaller than those of two-dimensional systems
but are still observable when the fluctuation regime is visible for the experimental resolution of temperature~\cite{Asaba_private}.

\textit{Discussion.} ---
We have demonstrated that rectification and the NHE in the fluctuation regime are promising probes of helical superconductivity regardless of sample dimensions.
The results strongly suggest that the enhanced NCT in moderate and high magnetic fields is observable in realistic thin-film samples with a non-negligible thickness, which would lie between the two-dimensional and three-dimensional limits studied in this work.
In particular, the NHE would serve as a better probe because the linear Hall resistance is absent owing to the $y$-mirror symmetry.
A materials-based study for the candidate helical superconductors~\cite{Naritsuka2017-or,Naritsuka2021-ym,Sekihara2013-dm,Schumann2020-mw} is left as an intriguing future issue, as well as the fully microscopic treatment of NCT including the quantum-mechanical corrections beyond the GL approach.

As a complementary {question}, it is also interesting to consider the effect of helical-superconductivity crossover on NCT caused by {vortices and anti-vortices}.
This occurs below the mean-field transition temperature, and nonreciprocal renormalization of the superfluid density plays an essential role~\cite{Hoshino2018-sa}. 
Since the anharmonicity parameter $\mathcal{A}^{ijk}$ causes such a renormalization, an enhanced NCT is also expected by this mechanism and would smoothly connect with that of paraconductivity above the mean-field transition temperature.
Thus, the enhanced NCT in the crossover regime, both below and above the mean-field transition temperature, will work as the promising probe of helical superconductivity.
Quantitative studies are left as a future issue.


\begin{acknowledgments}
We appreciate inspiring discussions with Yuji Matsuda and Tomoya Asaba.
We also thank helpful discussions with Hikaru Watanabe.
This work was supported by JSPS KAKENHI (Grant Nos. JP18H01178, JP18H05227, JP20H05159, JP21K13880, JP21K18145, JP22H01181, JP22H04476, JP22H04933) and SPIRITS 2020 of Kyoto University.
\end{acknowledgments}

%


\section{Derivation of the fluctuation conductivity}
The derivation of the fluctuation conductivity without the orbital magnetic field can be performed following Ref.~\cite{Wakatsuki2017-dp} with a straightforward generalization.
For clarity, we show the details below.

\subsection{Setup}
We start from the time-dependent GL (TDGL) equation 
\begin{align}
\Gamma_0\frac{\partial\psi(\bm{x},t)}{\partial t}=-\frac{\delta F[\psi,\psi^*]}{\delta \psi^*(\bm{x})}+\zeta(\bm{x},t).
\end{align}
Here, the GL functional is assumed to be the bilinear of the order parameter,
\begin{align}
F[\psi]&=\int d^dx\int d^dx'\psi^*(\bm{x})\alpha(\bm{x}-\bm{x}')\psi(\bm{x}'),
\end{align}
describing the Gaussian fluctuation of Cooper pairs.
The phenomenological parameter $\Gamma_0$ takes into account the relaxation process of Cooper pairs.
In the absence of the electric field, TDGL has the equilibrium solution $\delta F/\delta\psi^*=0$ as its steady state, if $\zeta(\bm{x},t)$ is neglected.
The random force $\zeta(\bm{x},t)$ kicks up the fluctuation of the order parameter $\psi(\bm{x},t)$, which vanishes after a finite lifetime $\propto\Gamma_0/(T-\Tc)$.

It is convenient to switch to the momentum space for our purpose.
We adopt the following convention of the Fourier transform:
\begin{subequations}\begin{align}
\psi_{\bm{q}}(t)&=\frac{1}{V}\int d^dx\, e^{-i\bm{q}\cdot\bm{x}}\psi(\bm{x},t),\\
\zeta_{\bm{q}}(t)&=\frac{1}{V}\int d^dx\,e^{-i\bm{q}\cdot\bm{x}}\zeta(\bm{x},t).
\end{align}\end{subequations}
Here, the momentum is $\bm{q}\in \prod_{i=1}^d(2\pi\mathbb{Z}/L_i)$ with assuming the periodic boundary conditions, with $L_i$ the diameter in the $i$ direction.
The system volume in the $d$ dimensions is defined by $V=\prod_{i=1}^dL_i$.
With this notation, $\psi_{\bm{q}=0}$ coincides with the spatial average of the order parameter.
The GL coefficient in the Fourier space is defined by
\begin{align}
\alpha_{\bm{q}}&=\int d^dx\,e^{-i\bm{q}\cdot\bm{x}}\alpha(\bm{x}),
\end{align}
leading to the GL functional
\begin{align}
F[\psi]&=V\sum_{\bm{q}}\alpha_{\bm{q}}|\psi_{\bm{q}}|^2.
\end{align}
This gives the equilibrium average of the order parameter
\begin{align}\braket{|\psi_{\bm{q}}|^2}&=\frac{\int D[\psi,\psi^*]e^{-\beta F[\psi]}|\psi_{\bm{q}}|^2}{\int D[\psi,\psi^*]e^{-\beta F[\psi]}}\notag\\
&=\frac{1}{\beta V\alpha_{\bm{q}}}\label{eq_SM:psi2_eq},
\end{align}
with the inverse temperature $\beta\equiv1/T$.

With the conventions introduced above, the TDGL equation in the momentum space is given by
\begin{align}
\Gamma_0\frac{\partial\psi_{\bm{q}}(t)}{\partial t}&=-\alpha_{\bm{q}}\psi_{\bm{q}}(t)+\zeta_{\bm{q}}(t).
\end{align}
It should be noted that $\Gamma_0$ can generally depend on $\bm{q}$ and external parameters such as $\bm{h}$ from the microscopic viewpoint.
While the latter dependence is canceled out in the nonreciprocity $\eta_\SC^{ijk}$ in two dimensions, the $\bm{q}$ dependence of $\Gamma_0$ is neglected for simplicity in this paper, which is left as the future issue [fully microscopic treatment would be more suitable to account for such effects].
For the time being, we assume $\Gamma_0\in\mathbb{C}$ with $\Re[\Gamma_0]>0$ for generality, while the imaginary part is expected to be small in the absence of the strong particle-hole asymmetry of the density of states~\cite{Larkin2005-lb}.

The general solution of the TDGL equation is given by
\begin{align}
\psi_{\bm{q}}(t)&=e^{-\frac{1}{\Gamma_0}\int_{t_0}^tdt'\,\alpha_{\bm{q}}(t')}\psi_{\bm{q}}(t_0)\notag\\
&+\frac{1}{\Gamma_0}\int_{t_0}^tdt'\,e^{-\frac{1}{\Gamma_0}\int_{t'}^tdt''\,\alpha_{\bm{q}}(t'')}\zeta_{\bm{q}}(t'),
\end{align}
where $\alpha_{\bm{q}}(t)\equiv\alpha_{\bm{q}-2\bm{A}(t)}$ when the electric field is applied to the system. 
After a sufficiently large period $(t-t_0)\alpha_{\bm{q}}\gg \Re[\Gamma_0]$, the order parameter becomes independent of the initial condition.
We are interested in such a situation, and thus take the limit $t_0\to\infty$ and drop the first term.
Thus, we obtain
\begin{align}
\braket{|\psi_{\bm{q}}(t)|^2}&=\int_{-\infty}^0dt_1\int_{-\infty}^0dt_2\,e^{-\left[\int_{t_1}^0\frac{1}{\Gamma_0^*}+\int_{t_2}^0\frac{1}{\Gamma_0}\right]dt'\,\alpha_{\bm{q}}(t+t')}\notag\\
&\qquad\qquad\cdot\frac{1}{|\Gamma_0|^2}\braket{\xi^*_{\bm{q}}(t+t_1)\zeta_{\bm{q}}(t+t_2)}.
\end{align}
The average over the white noise is defined to reproduce Eq.~\eqref{eq_SM:psi2_eq} by the TDGL equation in equilibrium, i.e. $\bm{E}=0$:
\begin{align}
\braket{\xi^*_{\bm{q}}(t)\xi_{\bm{q}'}(t')}&=\frac{2\Re[\Gamma_0]}{\beta V}\delta(t-t')\delta_{\bm{q},\bm{q}'}.
\end{align}
With this, we obtain the excess electric current density~\cite{Schmid1969-fw}
\begin{align}
\bm{j}_\SC(t)&\equiv-\sum_{\bm{q}}\partial_{\bm{A}}\alpha_{\bm{q}-2\bm{A}(t)}\braket{|\psi_{\bm{q}}(t)|^2}\\
&=\frac{4\Re[\Gamma_0]}{|\Gamma_0|^2\beta}\int\frac{d^dq}{(2\pi)^d}\partial_{\bm{q}}\alpha_{\bm{q}}\int^0_{-\infty}dt_1\,e^{-\frac{2\Re[\Gamma_0]}{|\Gamma_0|^2}\int^0_{t_1}dt'\,\alpha_{\bm{q}-2\bm{A}(t')}}\notag.
\end{align}
To be precise, $\bm{j}_\SC(t)$ in the left hand side should be understood as the current $\bm{j}_\SC(t)L_yL_z$ and the sheet current density $\bm{j}_\SC(t)L_z$ for $d=1$ and $2$, respectively, rather than the current density.
This point is taken into account at the end of the calculation.
Note also that by redefining $|\Gamma_0|^2/\Re[\Gamma_0]\to\Gamma_0$ the expression of $\bm{j}_\SC(t)$ coincides with that for the situation with $\Im[\Gamma_0]=0$.
Thus, we consider the case $\Gamma_0>0$ in the following and in the main text without loss of generality.

\subsection{Expansion by the electric field}
Let us expand $\bm{j}_\SC(t)$ in terms of $\bm{E}$ to obtain linear and nonlinear conductivity.
We start from
\begin{align}
\bm{j}_\SC(t)
 &=\frac{4}{\beta}\int\frac{d^dq}{(2\pi)^d}\partial_{\bm{q}}\alpha_{\bm{q}}\int^0_{-\infty}dt_1\,e^{2\int_0^{t_1}dt'\,\alpha_{\bm{q}-2\bm{A}(t')}},\label{eq_SM:GL_current}
\end{align}
where the domain of the $\bm{q}$ integral is the first Brillouin zone.
Here, we made the replacement $\alpha/\Gamma_0\to\alpha$ for simplicity.
We will recover
the effect of $\Gamma_0$ at the end of the calculation by $\alpha\to\alpha/\Gamma_0$.
The exponent of the exponential is given by
\begin{align}
\int_0^{t_1}dt'\,\alpha_{\bm{q}-2\bm{A}(t')}
&=\alpha_{\bm{q}}t_1+t_1^2E_i\partial_{q_i}\alpha_{\bm{q}}\notag\\
&\quad+\frac{2}{3}t_1^3E_iE_j\partial_{q_i}\partial_{q_j}\alpha_{\bm{q}}+O(E^3),
\end{align}
and thus we can write
\begin{align}
\!\!\!\!\int^0_{-\infty}dt_1\,e^{2\int_0^{t_1}dt'\,\alpha_{\bm{q}-2\bm{A}(t')}}
&\equiv A_0+A_1+A_2+O(E^3),
\end{align}
with
\begin{subequations}\begin{align}
A_0&\equiv\int_0^\infty dt_1\,e^{-2t_1\alpha_{\bm{q}}}=1/2\alpha_{\bm{q}},\\
A_1&\equiv\int_0^\infty dt_1\,2t_1^2E_j\partial_{q_j}\alpha_{\bm{q}}\,e^{-2t_1\alpha_{\bm{q}}}\notag\\
&=\frac{\partial_{q_j}\alpha_{\bm{q}}}{2\alpha_{\bm{q}}^3}E_j,\\
A_2&\equiv\int_0^\infty dt_1\,\Bigl[\frac{1}{2}(2t_1^2E_j\partial_{q_j}\alpha_{\bm{q}})^2\notag\\
&\qquad\qquad-\frac{4}{3}t_1^3E_jE_k\partial_{q_j}\partial_{q_k}\alpha_{\bm{q}}\Bigr]e^{-2t_1\alpha_{\bm{q}}}\notag\\
&=E_jE_k\left[\frac{3}{2}\frac{\partial_{q_j}\alpha_{\bm{q}}\partial_{q_k}\alpha_{\bm{q}}}{\alpha_{\bm{q}}^5}-\frac{\partial_{q_j}\partial_{q_k}\alpha_{\bm{q}}}{2\alpha_{\bm{q}}^4}\right].
\end{align}\end{subequations}
The contribution from $A_0$ is a total derivative of the momentum and therefore vanishes according to the periodicity of the Brillouin zone. This means that the electric current is absent in equilibrium.
In the following, we evaluate the linear and nonlinear fluctuation conductivity tensors determined by $A_1$ and $A_2$.

\subsection{Linear fluctuation conductivity}
We first discuss the linear fluctuation conductivity described by $A_1$,
\begin{align}
\sigma_{1\SC}^{ij}&=\frac{2}{\beta}\int\frac{d^dq}{(2\pi)^d}\frac{\partial_{q_i}\alpha_{\bm{q}}\partial_{q_j}\alpha_{\bm{q}}}{\alpha_{\bm{q}}^3}.\label{eq:LinearFormulaQ}
\end{align}
Since $\alpha_{\bm{q}_0}\equiv\min_{\bm{q}}\alpha_{\bm{q}}\to+0$ as $T\to \Tc+0$, the contribution around $\bm{q}=\bm{q}_0$ diverges and dominates the momentum integral.
Therefore, we can neglect the information on the high-energy modes: We can only consider the domain of the integral $|\bm{q}-\bm{q}_0|<O(\qc)$ with a small cut-off $\qc$ to obtain $\sigma_1^{ij}$ to the leading order of $\alpha_{\bm{q}_0}$.
Near $\bm{q}=\bm{q}_0$, we can write $\alpha_{\bm{q}}$ as
\begin{align}
\alpha_{\bm{q}}&=\alpha_{\bm{q}_0}+\alpha_2^{ij}\delta q_i\delta q_j+\alpha_3^{ijk}\delta q_i\delta q_j\delta q_k+O(\delta q^4),
\end{align}
with $\delta\bm{q}\equiv\bm{q}-\bm{q}_0$.
Note that $\qc$ should be sufficiently small to allow the Taylor expansion of $\alpha_{\bm{q}}$ while sufficiently large to capture the contribution around $\bm{q}_0$.
It turns out that we can choose e.g., $\qc=O(\alpha_{\bm{q}_0}^{1/24})$ for our purpose.
With this choice, the contribution from the outside of the domain is at most of the order $\alpha^{-3}_{\bm{q}}\sim \qc^{-6}\sim \alpha_{\bm{q}_0}^{-1/4}$, which is smaller than the leading term of $\sigma_1^{ij}=O(\alpha_{\bm{q}_0}^{-(2-d/2)})$ as obtained in the following.

For practical calculations, we rewrite the expansion of $\alpha_{\bm{q}}$ as follows:
\begin{align}
\alpha_{\bm{q}}&=\epsilon+\sum_i\xi_i^2\delta q_i^2+\bar{\xi}\,{}^3\sum_{ijk}\mathcal{A}^{ijk}\delta q_i\delta q_j\delta q_k\\
&=\epsilon+\sum_i(\xi_i\delta q_i)^2+\sum_{ijk}a_3^{ijk}(\xi_i\delta q_i)(\xi_j\delta q_j)(\xi_k\delta q_k),\notag
\end{align}
with the reduced temperature $\epsilon$ and the coherence length $\xi_i$.
Note that we are temporarily setting $N_0=T\partial_T\alpha_{\bm{q}_0}\to1$.
This can be recovered by the replacement $\Gamma_0\to \tau_0$ at the end of the calculation.
The coordinate axes are chosen to diagonalize the symmetric tensor $\alpha_2^{ij}=\xi_i^2\delta_{ij}$ with $\xi_i>0$.
For the later convenience, 
we introduce
$a_3^{ijk}=\bar{\xi}^3\mA^{ijk}/(\xi_i\xi_j\xi_k)$ instead of the anharmonicity parameter $\mA^{ijk}$. 
For the linear paraconductivity, however, the $a_3^{ijk}$ term is not necessary to obtain the leading-order contribution,
since $\sigma_{1\SC}^{ij}$ becomes finite without $a_3^{ijk}$ and the contribution from $a_3^{ijk}$ is smaller than the $\delta q^2$ terms by at least $O(\qc)\ll1$.
Higher-order Taylor coefficients of $\alpha_{\bm{q}}$, if considered, are also irrelevant for the same reason.
By using $p_i\equiv\xi_i\delta q_i$ and then $p_i\to\sqrt{\epsilon}\,p_i$, we obtain
\begin{align}
\sigma_{1\SC}^{ij}&=\frac{8T}{\epsilon^{2-d/2}}\frac{\xi_i\xi_j}{\bar{\xi}^d}\int\frac{d^dp}{(2\pi)^d}\frac{p_ip_j}{(1+p^2)^3}\notag\\
&=\begin{cases}
\frac{1}{L_yL_z}\frac{\tau_0T\xi_x}{2\epsilon^{3/2}}&(d=1)\\
\frac{1}{L_z}\frac{\tau_0T}{2\pi\epsilon}\frac{\xi_i^2}{\bar{\xi}^2}\delta_{ij}&(d=2)\\
\frac{\tau_0T}{4\pi\sqrt{\epsilon}}\frac{\xi_i^2}{\bar{\xi}{}^3}\delta_{ij}&(d=3)
\end{cases}
,
\end{align}
reproducing the results in Ref.~\cite{Larkin2005-lb} by $\tau_0\to\pi/8T$.
Here, the domain of $\bm{p}$ integral is set to $\mathbb{R}^d$ since $\xi_i \qc/\sqrt{\epsilon}\gg1$: The contribution from the interval $\xi \qc/\sqrt{\epsilon}<|p|<\infty$ is negligible compared with that from $\mathbb{R}^d$.
We recovered the effect of $\Gamma_0$ in the last line by
multiplying $\tau_0$, considering the expression of Eq.~\eqref{eq:LinearFormulaQ}.
The factors $1/L_yL_z$ and $1/L_z$ are taken into account in the final expression for $d=1,2$ as noted previously. 
For general choice of the coordinate axes, we can use the tensor expression
\begin{align}
\sigma_{1\SC}^{ij}&=\begin{cases}
\frac{1}{L_z}\frac{\Gamma_0T}{2\pi\alpha_{\bm{q}_0}}\frac{\alpha_2^{ij}}{\sqrt{\det[\alpha_2]}}&(d=2)\\
\frac{\Gamma_0T}{4\pi\sqrt{\alpha_{\bm{q}_0}}}\frac{\alpha_2^{ij}}{\sqrt{\det[\alpha_2]}}&(d=3)\end{cases}.
\end{align}

\subsection{Nonlinear fluctuation conductivity}
Next, we evaluate the second-order fluctuation conductivity determined by $A_2$,
\begin{align}
\beta\sigma_{2\SC}^{ijk}E_jE_k&=\int\frac{d^dq}{(2\pi)^d}\Bigl[6\,\frac{\partial_{q_i}\alpha_{\bm{q}}\partial_{q_j}\alpha_{\bm{q}}\partial_{q_k}\alpha_{\bm{q}}}{\alpha_{\bm{q}}^5}\notag\\
&\qquad\qquad\qquad-2\,\frac{\partial_{q_i}\alpha_{\bm{q}}\partial_{q_j}\partial_{q_k}\alpha_{\bm{q}}}{\alpha_{\bm{q}}^4}\Bigr]E_jE_k\notag\\
&\equiv 6B_1-2B_2.
\label{eq_SM:nonlinear1}
\end{align}
Here, $d=1,2,3$ is the system dimensions.
Note that the second term is proportional to the first term.
Actually,
\begin{align}
B_1&=\int\frac{d^dq}{(2\pi)^d}-\frac{1}{4}\partial_{q_j}[\alpha_{\bm{q}}^{-4}]\partial_{q_i}\alpha_{\bm{q}}\partial_{q_k}\alpha_{\bm{q}}E_jE_k\notag\\
&=\frac{1}{4}\int\frac{d^dq}{(2\pi)^d}\alpha_{\bm{q}}^{-4}[\partial_{q_i}\partial_{q_j}\alpha_{\bm{q}}\partial_{q_k}\alpha_{\bm{q}}\notag\\
&\qquad+\partial_{q_i}\alpha_{\bm{q}}\partial_{q_j}\partial_{q_k}\alpha_{\bm{q}}]E_jE_k\notag\\
&\equiv\frac{1}{4}B_2'+\frac{1}{4}B_2.
\end{align}
Here, we defined
\begin{align}
B_2'&\equiv \int\frac{d^dq}{(2\pi)^d}\alpha_{\bm{q}}^{-4}\partial_{q_i}\partial_{q_j}\alpha_{\bm{q}}\partial_{q_k}\alpha_{\bm{q}}E_jE_k\notag\\
&=4\int\frac{d^dq}{(2\pi)^d}\alpha_{\bm{q}}^{-5}\partial_{q_i}\alpha_{\bm{q}}\partial_{q_j}\alpha_{\bm{q}}\partial_{q_k}\alpha_{\bm{q}}E_jE_k\notag\\
&\quad-\int\frac{d^dq}{(2\pi)^d}\alpha_{\bm{q}}^{-4}\partial_{q_j}\alpha_{\bm{q}}\partial_{q_i}\partial_{q_k}\alpha_{\bm{q}}E_jE_k\notag\\
&=4B_1-B_2'.
\end{align}
Thus, we obtain $B_2=2B_1$ and
\begin{align}
\beta\sigma_{2\SC}^{ijk}&=B_2=\int\frac{d^dq}{(2\pi)^d}\frac{\partial_{q_i}\alpha_{\bm{q}}\partial_{q_j}\partial_{q_k}\alpha_{\bm{q}}}{\alpha_{\bm{q}}^4}.
\label{eq:nonlinear_B2}
\end{align}
To evaluate the momentum integral, we again focus on the region $|\delta\bm{q}|<O(\qc)=O(\alpha_{\bm{q}_0}^{1/24})$, since the contribution from the outside is at most $O(\alpha_{\bm{q}_0}^{-8/24})\ll \sigma_2^{ijk}=O(\alpha_{\bm{q}_0}^{-(3-d/2)})$ as clarified below.

Note that $B_2$ vanishes in the absence of $a_3^{ijk}$, since the integrand becomes odd in $\delta\bm{q}$.
Thus, we are interested in the correction by $a_3^{ijk}$.
It is sufficient to keep only the first-order terms in $a_3^{ijk}$ to obtain the leading-order singularity of $\sigma_2^{ijk}$, since the correction is smaller at least by $O(\qc)$.
By using the variable $\xi_i\delta q_i=p_i$, we obtain
\begin{align}
&\partial_{q_i}\alpha_{\bm{q}}\partial_{q_j}\partial_{q_k}\alpha_{\bm{q}}\notag\\
&\simeq\xi_i\xi_j\xi_k[4p_i\delta_{jk}+12a_3^{jkc}p_ip_c+6\delta_{jk}a_3^{ibc}p_bp_c],\notag\\
\alpha_{\bm{q}}^{-4}&\simeq\frac{1}{(\epsilon+p^2)^4}\left(1-4\frac{a_3^{abc}p_ap_bp_c}{\epsilon+p^2}\right).
\end{align}
After plugging these expressions into $B_2$ and $p_i\to\sqrt{\epsilon}\,p_i$, we can now set the domain of integral to $\mathbb{R}^d$.
The term $B_2$ is then given by
\begin{align}
B_2
&=\frac{\xi_i\xi_j\xi_k}{\epsilon^{3-d/2}(2\pi)^d\bar{\xi}\,{}^d}\int d\Omega\,\Bigl(\notag\\
&\quad\left[12\int_0^\infty dp\,\frac{p^{2+d-1}}{(1+p^2)^4}\right]a_3^{jkc}n_in_c\notag\\
&\quad+\left[6\int_0^\infty dp\,\frac{p^{2+d-1}}{(1+p^2)^4}\right]\delta_{jk}a_3^{ibc}n_bn_c\\
&\qquad-\left[16\int_0^\infty dp\,\frac{p^{4+d-1}}{(1+p^2)^5}\right]n_i\delta_{jk}a_3^{abc}n_an_bn_c\Bigr),\notag
\end{align}
where $\bm{n}=(\sin\theta\cos\phi,\sin \theta\sin\phi,\cos\theta)$ and $d\Omega=\sin\theta d\theta d\phi$ for $d=3$; $\bm{n}=(\cos\theta,\sin\theta)$ and $d\Omega=d\theta$ for $d=2$; and $\int d\Omega\to2$ and $n_i\to 1$ for $d=1$.
The momentum integrals are evaluated by
\begin{subequations}\begin{align}
\int_0^\infty dp\,\frac{p^{2+d-1}}{(1+p^2)^4}&=\begin{cases}
\frac{\pi}{32}&(d=1)\\
\frac{1}{12}&(d=2)\\
\frac{\pi}{32}&(d=3),
\end{cases}\\
\int_0^\infty dp\,\frac{p^{4+d-1}}{(1+p^2)^5}&=\begin{cases}
\frac{3\pi}{256}&(d=1)\\
\frac{1}{24}&(d=2)\\
\frac{5\pi}{256}&(d=3).
\end{cases}
\end{align}\end{subequations}
Thus, we obtain
\begin{align}
B_2^{d=1}
&=\frac{3\xi_x^2}{8\epsilon^{5/2}}a_3^{xxx},
\end{align}
for $d=1$,
\begin{align}
B_2^{d=2}&=\frac{\xi_i\xi_j\xi_k}{\epsilon^{2}(2\pi)^2\bar{\xi}^2}\int d\theta\,a_3^{jkc}n_in_c+\frac{1}{2}\delta_{jk}a_3^{ibc}n_bn_c\notag\\
&\qquad-\frac{2}{3}n_i\delta_{jk}a_3^{abc}n_an_bn_c\notag\\
&=\frac{\xi_i\xi_j\xi_k}{\epsilon^{2}(2\pi)^2\bar{\xi}^2}\,\pi a_3^{ijk},
\end{align}
for $d=2$, and
\begin{align}
B_2^{d=3}
&=\frac{\xi_i\xi_j\xi_k}{\epsilon^{3/2}(2\pi)^3\bar{\xi}^3}\frac{\pi}{32}\int d\Omega\,12a_3^{jkc}n_in_c\notag\\
&\quad+6\delta_{jk}a_3^{ibc}n_bn_c-16\cdot\frac{5}{8}n_i\delta_{jk}a_3^{abc}n_an_bn_c\notag\\
&=\frac{\xi_i\xi_j\xi_k}{32\pi\epsilon^{3/2}\bar{\xi}^3}\,2a_3^{ijk},
\end{align}
for $d=3$.
The angular integrals were carried out by employing \textit{Mathematica}, while the case of $d=2$ is also confirmed by hand.
Thus, we obtain
\begin{subequations}
\begin{align}
\sigma_{2\SC}^{xxx}&=\frac{1}{L_yL_z}\frac{3\tau_0^2T\xi_x^2}{8\epsilon^{5/2}}\mA^{xxx}\quad (d=1),\\
\sigma_{2\SC}^{ijk}&=\frac{1}{L_z}\frac{\tau_0^2T\bar{\xi}}{4\pi\epsilon^2}\mA^{ijk}\quad (d=2),\\
\sigma_{2\SC}^{ijk}&=\frac{\tau_0^2T}{16\pi\epsilon^{3/2}}\mA^{ijk}\quad (d=3),
\end{align}
\end{subequations}
with reproducing the effect of $\Gamma_0$ by multiplying $\tau_0^2$ [see Eq.~\eqref{eq:nonlinear_B2}] as well as $1/L_yL_z$ and $1/L_z$ for $d=1$ and $2$, respectively.
These expressions are valid for an arbitrary choice of the coordinate axes since $\bar{\xi}$ and $\mA^{ijk}$ are the scalar and tensor, respectively, while we can also write
\begin{subequations}
\begin{align}
\sigma_{2\SC}^{ijk}&=\frac{1}{L_z}\frac{\Gamma_0^2T\alpha_3^{ijk}}{4\pi\alpha_{\bm{q}_0}^2\sqrt{\det[\alpha_2]}}\quad (d=2),\\
\sigma_{2\SC}^{ijk}&=\frac{\Gamma_0^2T\alpha_3^{ijk}}{16\pi\alpha_{\bm{q}_0}^{3/2}\sqrt{\det[\alpha_2]}}\quad (d=3),
\end{align}
\end{subequations}
by using $\alpha_2^{ij}$ and $\alpha_3^{ijk}$.
In particular, 
the intrinsic nonreciprocity in two-dimensional systems is given by
\begin{align}
\eta_\SC^{ijk}
&=-L_z\frac{\pi\bar{\xi}}{T}\frac{\bar{\xi}{}^4}{\xi_j^2\xi_k^2}\mA^{ijk}\quad (d=2).
\end{align}
The expression for the arbitrary choice of the coordinate axes is given by
\begin{align}
\eta_{2\SC}^{ijk}&=-L_z\frac{\pi\sqrt{\det[\alpha_2]}}{T}\alpha_3^{ibc}[\alpha_2^{-1}]_{bj}[\alpha_2^{-1}]_{ck}\quad (d=2).
\end{align}

\section{Angle dependence of NCT under small magnetic fields}
As noted in the main text, MCA by the fluctuation contribution is determined by the effective cubic spin-orbit coupling $\bm{g}_3(\bm{k})$.
For example in Rashba and Ising systems, this is given by $\sim (k_x^2+k_y^2)(-k_y,k_x,0)$ and $\sim (0,0,k_y^3-3k_yk_x^2)$.
It follows that $\bm{g}_3(\hat{E})=(\hat{E}_x^2+\hat{E}_y^2 )\hat{z}\times\hat{E}=\hat{z}\times\hat{E}$ and $\bm{g}_3(\hat{E})=\hat{z}(\sin^3\theta-3\sin\theta\cos^2\theta)=-\sin3\theta\hat{z}$ for $\hat{E}=(\cos\theta,\sin\theta,0)$, respectively.
$\bm{g}_3(\hat{E})$ for a given antisymmetric spin-orbit coupling can be obtained in the same way.

The $\bm{h}$-linear NHE is given as follows.
\begin{align}
&\frac{4\pi\epsilon^2}{\tau_0^2T\bar{\xi}}[\hat{z}\times\hat{E}]\cdot \frac{L_z\delta^2\bm{j}_\SC}{E^2}\notag\\
&=[\hat{z}\times\hat{E}]_i\hat{E}_j\hat{E}_k\mA^{ijk}\notag\\
&=\frac{1}{3}[\hat{z}\times\hat{E}]\cdot\lim_{\bm{q}\to\hat{E}}\partial_{\bm{q}}[\mA^{abc}q_aq_bq_c]\notag\\
&=\frac{1}{3}[\hat{z}\times\hat{E}]_i\lim_{\bm{q}\to\hat{E}}\partial_{q_i}\bm{g}_3(\bm{q})\cdot\bm{h}\notag\\
&=\frac{\bm{h}}{3}\cdot\lim_{\delta\to0}\frac{\bm{g}_3(\hat{E}+\delta[\hat{z}\times\hat{E}])-\bm{g}_3(\hat{E})}{\delta}.
\end{align}
By introducing the unit vector 
\begin{align}
\hat{E}_{\delta\theta}&\equiv\frac{(\hat{E}+\delta\theta[\hat{z}\times\hat{E}])}{\sqrt{1+\delta\theta^2}}=\hat{E}(\theta+\delta\theta)+o(\delta\theta),
\end{align}
we obtain
\begin{align}
&\frac{4\pi\epsilon^2}{\tau_0^2T\bar{\xi}}[\hat{z}\times\hat{E}]\cdot \frac{L_z\delta^2\bm{j}_\SC}{E^2}\notag\\
&=\frac{\bm{h}}{3}\cdot\lim_{\delta\theta\to0}\frac{(1+\delta\theta^2)^{3/2}\bm{g}_3(\hat{E}_{\delta\theta})-\bm{g}_3(\hat{E})}{\delta\theta}\notag\\
&=\frac{\bm{h}}{3}\cdot\partial_\theta\bm{g}_3(\hat{E}(\theta)).
\end{align}
To obtain the second line, note that $\bm{g}_3(\bm{k})$ contains only $O(k^3)$ terms.
Thus, the formula in the main text is obtained.

Note that the obtained angle dependence is consistent with Ref.~\cite{Wakatsuki2018-ll}, which studies Rashba systems.
To see this, let us consider the electric current in the $x$ direction for the Rashba system.
In the presence of $C_\infty$ rotational symmetry assumed in Ref.~\cite{Wakatsuki2018-ll}, we obtain $\bm{g}_3(\bm{q})\propto (q_x^2+q_y^2)(-q_y,q_x,0)$ and thus
$\bm{g}_3(\hat{E})\propto\hat{z}\times\hat{E}$.
We can write
\begin{align}
\delta^2j_{\SC }^x&=\hat{x}\cdot\left(\hat{E}[\hat{E}\cdot\delta^2\bm{j}_\SC]+[\hat{z}\times\hat{E}][\hat{z}\times\hat{E}\cdot\delta^2\bm{j}_\SC]\right)\notag\\
&=C\Bigl(\cos\theta_E\cos(\theta_E+\pi/2-\theta_h)\notag\\
&\qquad-\frac{1}{3}\cos(\theta_E+\pi/2)\cos(\theta_E-\theta_h)\Bigr),
\end{align}
with a prefactor $C$.
Here, the angles $\theta_E$ and $\theta_h$ for the electric and magnetic fields are measured from the $x$ axis.
This is equivalent to 
\begin{align}
\delta^2j_{\SC}^x=\frac{C}{3}\left[2\sin\theta_h+\sin(\theta_h-2\theta_E)\right],
\label{eq:Cinfty_delta2jx}
\end{align}
which agrees with Ref.~\cite{Wakatsuki2018-ll}.

The Rashba model studied in this paper is classified into the point group $C_{4v}$ in the absence of $\bm{h}$, and does not have the $C_\infty$ symmetry.
In the presence of tetragonal anisotropy, we generally have another component in $\bm{g}_3(\bm{q})$ proportional to $(q_x^2-q_y^2)(q_y,q_x,0)$, which breaks $C_\infty$ symmetry but belongs to the identity representation of $C_{4v}$.
This term additionally contributes to $\bm{g}_3(\hat{E})$ by $C'\cos 2\theta(\sin\theta,\cos\theta,0)$ with another prefactor $C'$ independent of $C$.
Accordingly, the field-angle dependence of $\delta^2j_\SC^x$ may deviate from Eq.~\eqref{eq:Cinfty_delta2jx} in realistic materials.
Such tetragonal anisotropy is also seen in our numerical results, because
$3\mA^{xyy}=\mA^{xyy}+\mA^{yxy}+\mA^{yyx}$ should coincide with $\mA^{xxx}$ up to $O(h)$ if the anisotropy were absent.

\section{Details of the numerical calculations for NCT in Rashba-Zeeman model}
Here we show the details of the numerical calculations of the GL coefficients in the Rashba-Zeeman model.
The GL coefficients can be evaluated with the formula
\begin{gather}
  \alpha_{\bm{q}}=\frac{\sum_m 1}{4U_\varphi}+\frac{1}{2V}\sum_{\bm{k},m,n}F_{mn}(\bm{k},\bm{q})Q_{mn}(\bm{k},\bm{q}),\label{eq:alpha_q_micro}\\
  F_{mn}(\bm{k},\bm{q})=\frac{f\bigl(\epsilon_m(\bm{k}+\bm{q}/2)\bigr)-f\bigl(-\epsilon_n(-\bm{k}+\bm{q}/2)\bigr)}{\epsilon_m(\bm{k}+\bm{q}/2)+\epsilon_n(-\bm{k}+\bm{q}/2)},\notag\\
  {Q_{mn}(\bm{k},\bm{q})=|\braket{ u_m(\bm{k}+\bm{q}/2)|\varphi(\bm{k})|u_n^*(-\bm{k}+\bm{q}/2)}|^2.}\notag
\end{gather}
Here, $U_\varphi>0$ is the attractive interaction in the pairing channel with the form factor $\varphi(\bm{k})$, while $\epsilon_m(\bm{k})$ and $\ket{u_m(\bm{k})}$ are the $m$-th energy dispersion and eigenstate of the normal-state Bloch Hamiltonian $H_N(\bm{k})$.
We consider the $s$-wave and $d$-wave states whose form factors are $\varphi(\bm{k})=i\sigma_y$ and $\varphi(\bm{k})=(\cos k_x-\cos k_y)i\sigma_y$, respectively. 
We adopt the system parameters
\begin{align}
(t,\alpha_{\rm R},\mu,U_s,U_d)&=(1,0.1,-1,0.58,0.475),
\label{eq_SM:parameteres}
\end{align}
where $U_s$ and $U_d$ are the attractive interaction in the $s$-wave and $d$-wave channels chosen to give $\Tc\sim0.01$.

We are interested in the NCT along the transition line $(h,\Tc(h))$. 
In the following, we explain the calculation procedure taking the $s$-wave case as an example. 
The transition temperature $\Tc(h)$ is determined by the bisection method with the threshold $|\alpha_{\bm{q}_0}|<10^{-4}$, by adopting $L_x=12000$ as well as $L_y=500$ and $1500$ for $0\le h\le0.02$ and $0.02\le h\le 0.027$, respectively.
Here, $\bm{q}_0=q_0\hat{x}$ and $\alpha_{\bm{q}_0}$ are evaluated by first minimizing $\alpha_{\bm{q}}$ among discrete points $q_x\in(2\pi/L_x)\mathbb{Z}$ and next using Lagrange interpolation of the three data points on the mesh $q_x\in(2\pi/L_x)\mathbb{Z}$ around the minimum. 
The bottom of the obtained square-fitting function gives 
$q_0$.
We then adopt $L_x=12000$ and $L_y=12000$ to evaluate the other GL coefficients.
$\alpha_2^{xx}$ and $\alpha_3^{xxx}$ are evaluated by using the Lagrange interpolation of the five data points on the mesh around the minimum (i.e. fitting by quartic polynomials), and then evaluating the derivative of the interpolation function at the value of $q=q_0$ obtained above.
To calculate $\alpha_2^{yy}$ and $\alpha_3^{xyy}$, we introduce
\begin{align}
\!\!\!\!\!\!\alpha_2^{yy}(q_x)
&=\frac{\alpha_{(q_x,\delta q_y)}-2\alpha_{(q_x,0)}+\alpha_{(q_x,-\delta q_y)}}{2\delta q_y^2}+O(\delta q_y^2),
\end{align}
with $\delta q_y=2\pi/L_y$.
We evaluate $\alpha_2^{yy}(q_x)$ at the three points on the mesh around $q_0$. After the square-function fitting by the Lagrange interpolation, $\alpha_2^{yy}$ and $\alpha_3^{xyy}$ are obtained by substituting $q_0$ for $q_x$ in the interpolation function and in its $q_x$ derivative, respectively.

\begin{figure}
    \centering
    \begin{tabular}{ll}
    (a)&(b) \\
    \includegraphics[width=0.225\textwidth]{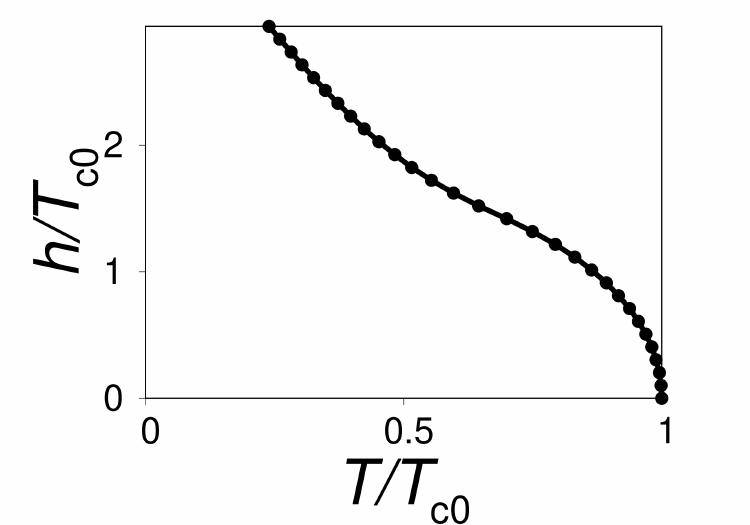}&
    \includegraphics[width=0.25\textwidth]{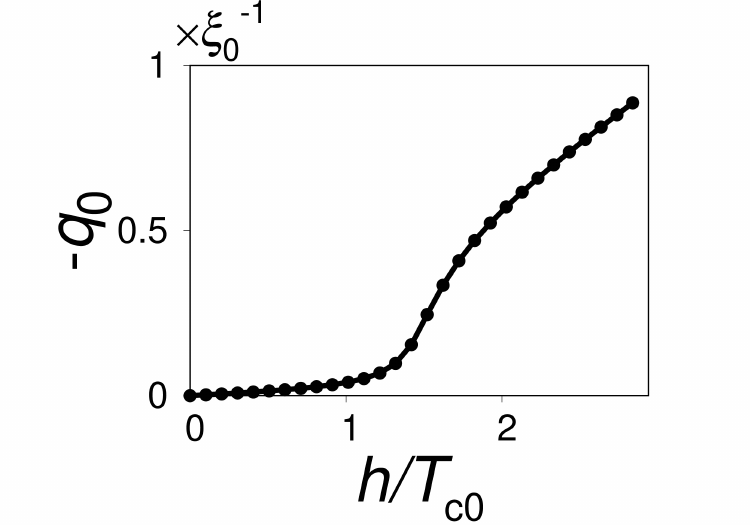}\\
    (c) &(d)\\
    \includegraphics[width=0.25\textwidth]{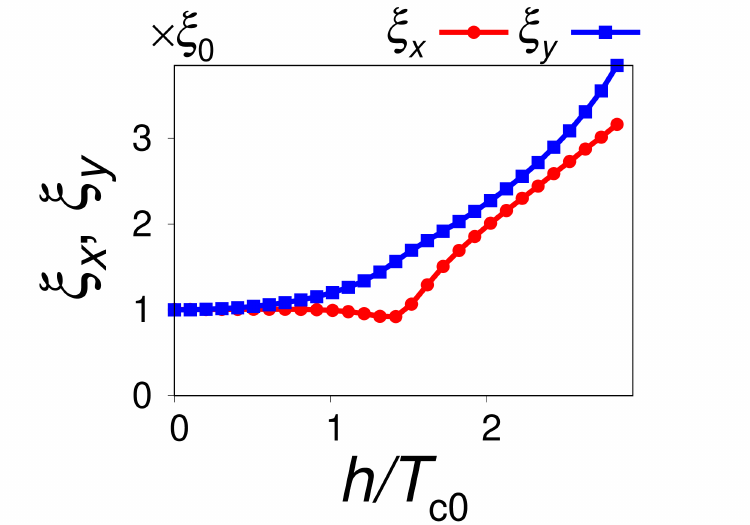}&
    \includegraphics[width=0.25\textwidth]{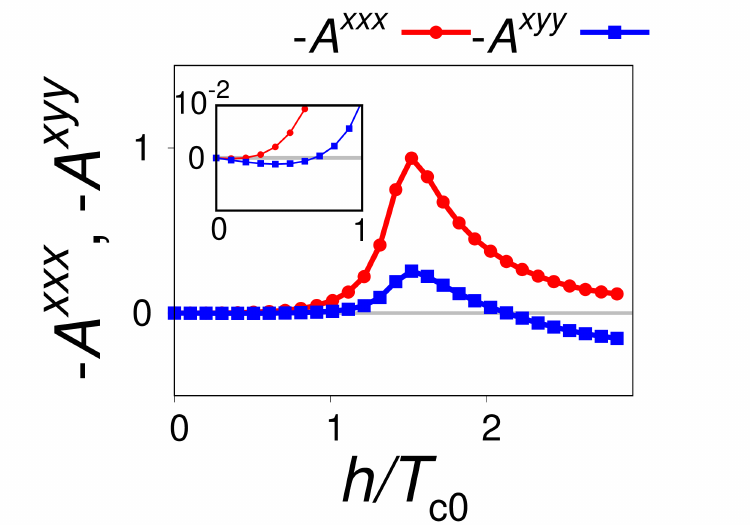}
    \end{tabular}
    \leftline{(e)}\\
    \includegraphics[width=0.4\textwidth]{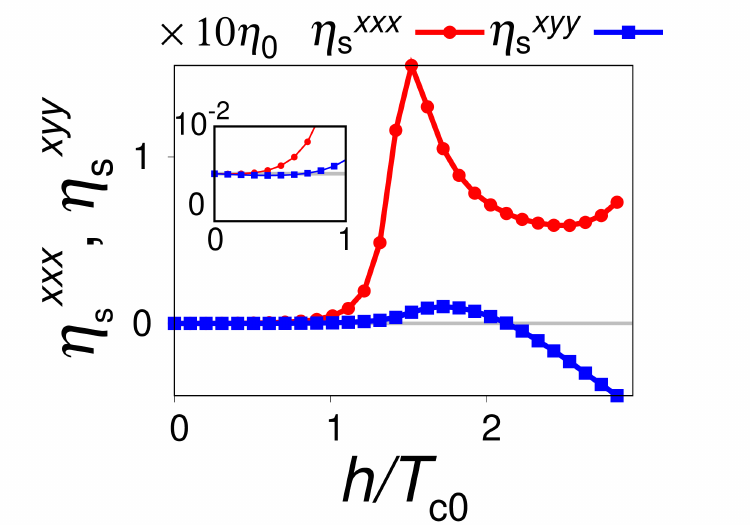}
    \caption{(a) The transition line $\Tc(h)$ of a $d$-wave Rashba-Zeeman superconductor, and (b) Cooper-pair momentum $-q_0$, (c) GL coherence length $\xi_x,\xi_y$, (d) the asymmetry parameter $-\mA^{xxx},-\mA^{xyy}$, and (e) strength of rectification $\eta^{xxx}_\SC$ and NHE $\eta^{xyy}_\SC$ 
    along the transition line $(\Tc(h),h)$.
   Here, $T_c(h)$ and $q_0(h)$ are indicated by black disks.
    $\xi_x$, $\mA^{xxx}$, $\eta_\SC^{xxx}$ are indicated by red disks while $\xi_y$, $\mA^{xyy}$, $\eta_\SC^{xyy}$ by blue squares.
    The black, red, and blue lines are the guide for the eye.}
    \label{fig_SM:Tcandq0_dwave}
\end{figure}

In addition to the results for the $s$-wave state shown in the main text, we here show the results for the $d$-wave state in Fig.~\ref{fig_SM:Tcandq0_dwave} with the parameters in Eq.~\eqref{eq_SM:parameteres}.
The transition temperature and $q_0$ are determined by $L_x=12000$ and $L_y=500$, while $L_x=12000$ and $L_y=4000$ are used for GL coefficients.
Overall, the obtained anharmonicity parameters and NCT are of the same order in magnitude as those of the $s$-wave states in units of $\eta_0$.
Thus, the enhanced NCT under moderate and strong magnetic fields is a general feature regardless of the pairing symmetry.
The difference from the $s$-wave state 
is the behavior of $\eta_\SC^{xxx}$ at high fields:
The sign reversal of $\eta_\SC^{xxx}$ seen in the $s$-wave state does not occur in the $d$-wave state for the range of $h/\Tcn$ shown here.
We find a sign reversal for $h/\Tcn\lesssim4$ [data not shown], but larger $L_x$ and $L_y$ are necessary to conclude its presence due to the large coherence lengths at low temperatures.
It should also be noted that the quantum-fluctuation corrections may be important for such low temperatures.

\begin{figure}
    \centering
    \begin{tabular}{ll}
    (a) &(b) \\
    \includegraphics[width=0.225\textwidth]{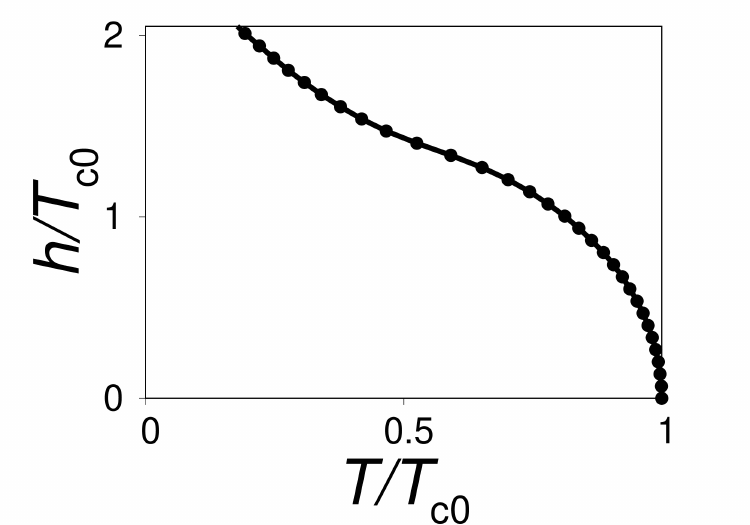}&
    \includegraphics[width=0.25\textwidth]{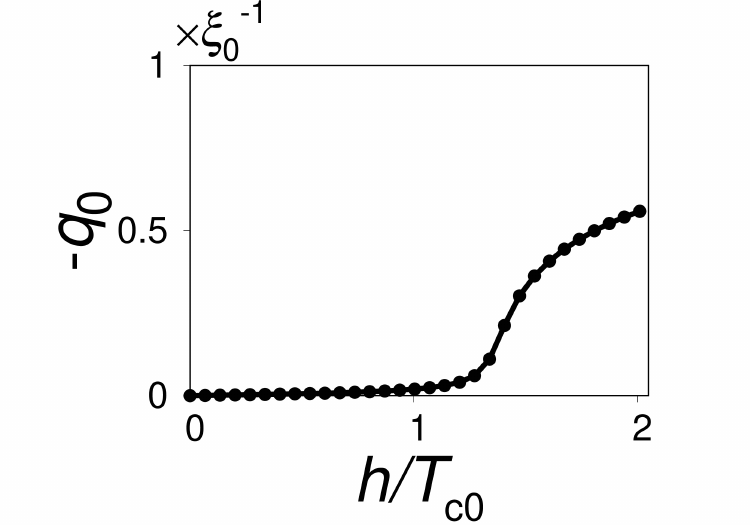}\\
    (c) &(d)\\
    \includegraphics[width=0.25\textwidth]{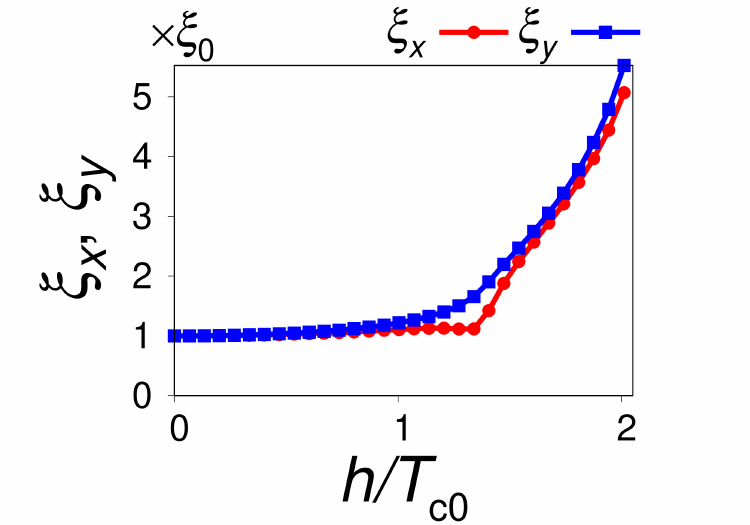}&
    \includegraphics[width=0.25\textwidth]{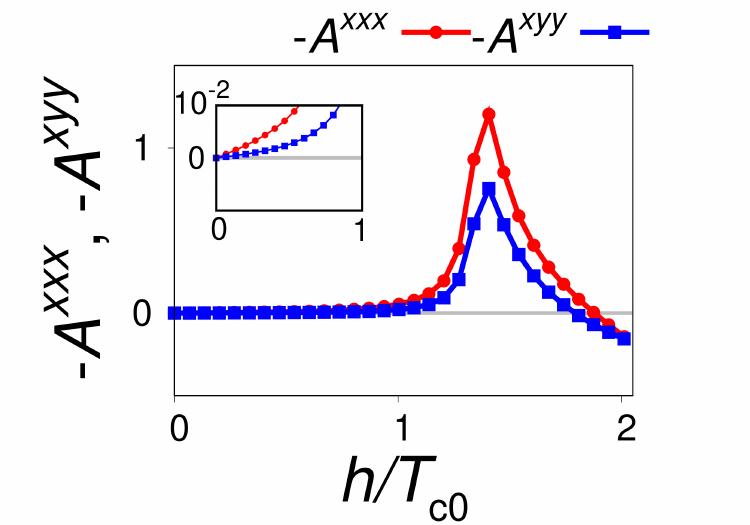}
    \end{tabular}
    \leftline{(e)}\\
    \includegraphics[width=0.4\textwidth]{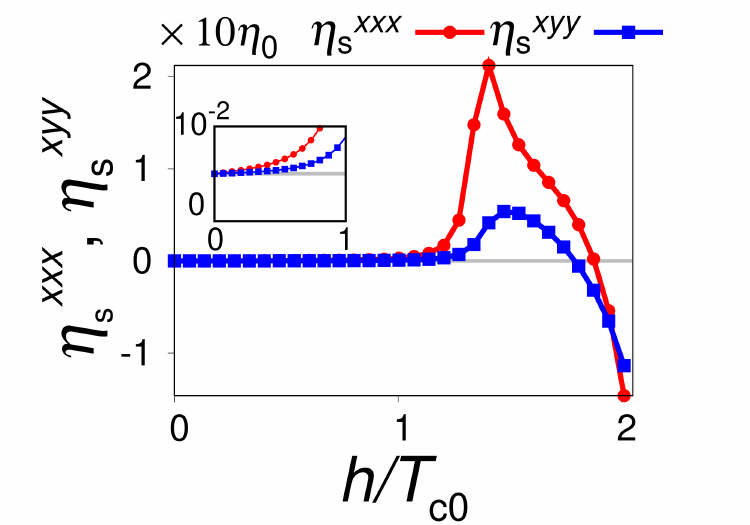}
    \caption{(a) The transition line $\Tc(h)$ of the $s$-wave Rashba-Zeeman superconductor with a high $\Tcn$, and (b) Cooper-pair momentum $-q_0$, (c) GL coherence length $\xi_x,\xi_y$, (d) the asymmetry parameter $-\mA^{xxx},-\mA^{xyy}$, and (e) strength of rectification $\eta^{xxx}_\SC$ and NHE $\eta^{xyy}_\SC$ along the transition line $(\Tc(h),h)$.
    Here, $T_c(h)$ and $q_0(h)$ are indicated by black disks.
    $\xi_x$, $\mA^{xxx}$, $\eta_\SC^{xxx}$ are indicated by red disks while $\xi_y$, $\mA^{xyy}$, $\eta_\SC^{xyy}$ by blue squares.
    The black, red, and blue lines are the guide for the eye.}
    \label{fig_SM:results_swave2}
\end{figure}
\begin{figure}
    \centering
    \begin{tabular}{ll}
    (a) &(b) \\
    \includegraphics[width=0.225\textwidth]{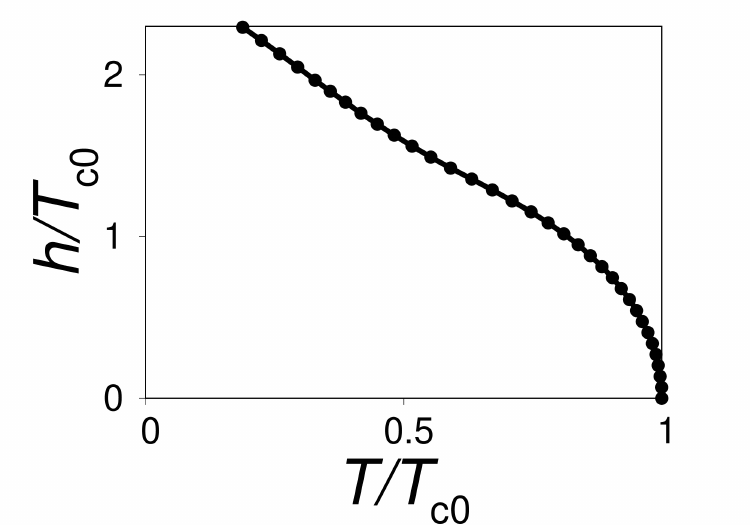}&
    \includegraphics[width=0.25\textwidth]{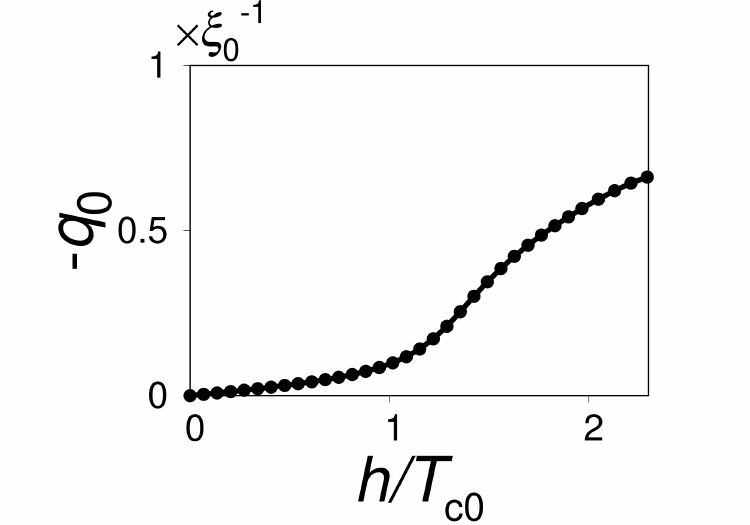}\\
    (c) &(d)\\
    \includegraphics[width=0.25\textwidth]{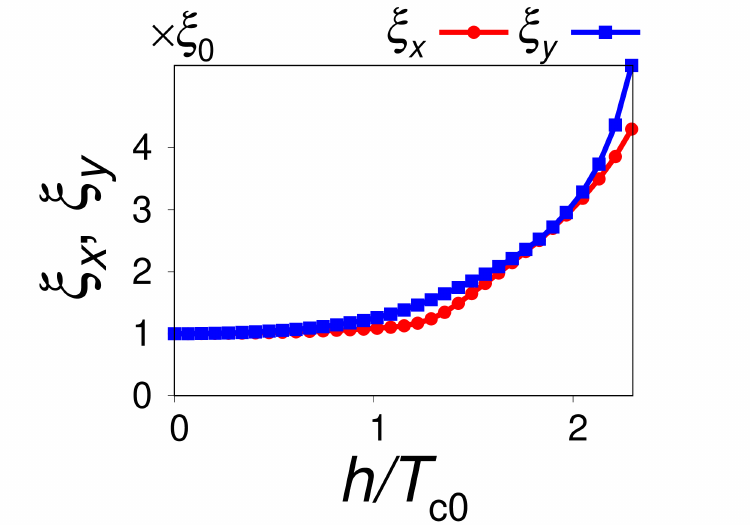}&
    \includegraphics[width=0.25\textwidth]{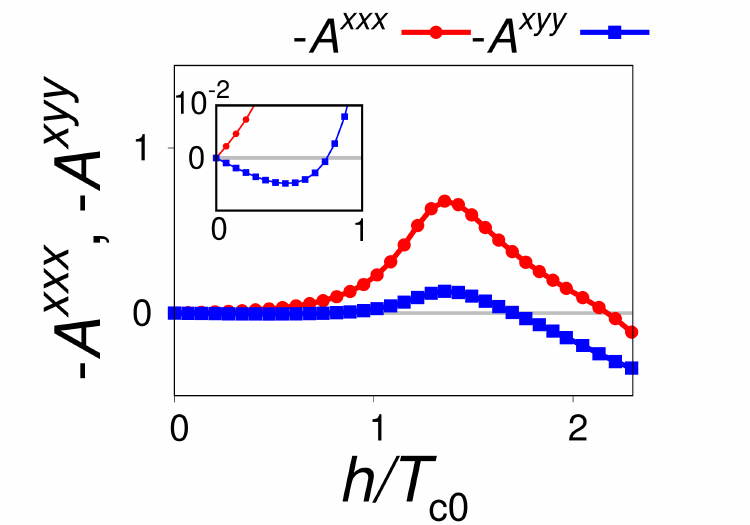}
    \end{tabular}
    \leftline{(e)}\\
    \includegraphics[width=0.4\textwidth]{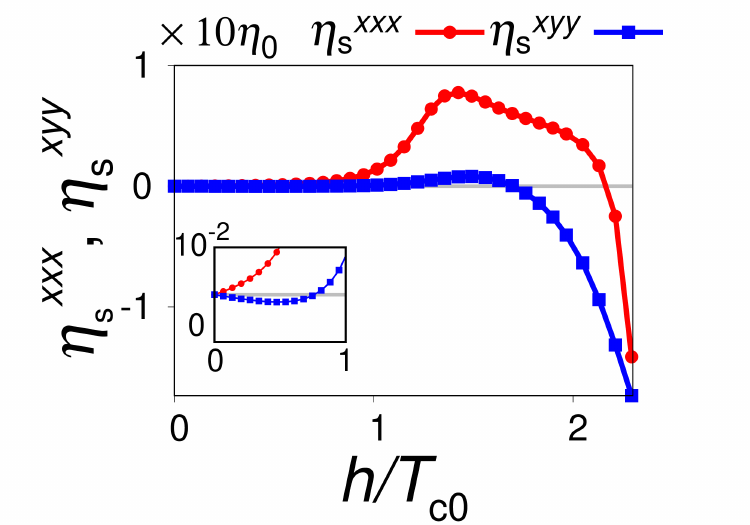}
    \caption{(a) The transition line $\Tc(h)$ of the $d$-wave Rashba-Zeeman superconductor with a high $\Tcn$, and (b) Cooper-pair momentum $-q_0$, (c) GL coherence length $\xi_x,\xi_y$, (d) the asymmetry parameter $-\mA^{xxx},-\mA^{xyy}$, and (e) strength of rectification $\eta^{xxx}_\SC$ and NHE $\eta^{xyy}_\SC$ along the transition line $(\Tc(h),h)$.
    Here, $T_c(h)$ and $q_0(h)$ are indicated by black disks.
    $\xi_x$, $\mA^{xxx}$, $\eta_\SC^{xxx}$ are indicated by red disks while $\xi_y$, $\mA^{xyy}$, $\eta_\SC^{xyy}$ by blue squares.
    The black, red, and blue lines are the guide for the eye.}
    \label{fig_SM:results_dwave2}
\end{figure}

To further study the quantitative aspects of NCT, we show the results for another parameter set
\begin{align}
(t,\alpha_{\rm R},\mu,U_s,U_d)&=(1,0.3,-1,0.945,0.75),
\label{eq_SM:parameteres}
\end{align}
in Figs.~\ref{fig_SM:results_swave2} and~\ref{fig_SM:results_dwave2}.
The strength of the interaction $U_s$ and $U_d$ are chosen to give a high transition temperature $\Tcn\sim 0.1$ for both $s$- and $d$-wave states.
We also choose a larger value of $\alpha_{\rm R}=0.3$ to ensure $\alpha_{\rm R}>\Tcn$.
We used $L_x=6000$ and $L_y=500$ to determine $\Tc(h)$ and $q_0(h)$, while $L_x=6000$ and $L_y=4000$ are used to evaluate GL coefficients.

Qualitatively the same results are obtained for the strong-coupling superconductors, except for the behavior of $h/\Tcn\gtrsim1$ in the $d$-wave states.
The important point is that the enhanced NCT is obtained in the crossover region of strong-coupling superconductors as well.
Furthermore, NCT is comparable to that of weak-coupling superconductors in units of $\eta_0$.
Thus, it is established that the enhanced NCT is a universal property of the helical superconductivity regardless of $\Tcn$ when scaled with $\eta_0\sim \xi_0/\Tcn$.

\subsection{Estimate of rectification and NHE}
Let us make an order estimate of NCT obtained by the microscopic calculations.
We start from the formula
\begin{align}
-L_z^{-1}\eta^{ijk}_{\SC}&=\frac{\pi\sqrt{\xi_x\xi_y}}{T}\left(\frac{\xi_x\xi_y}{\xi_j\xi_k}\right)^2\mA^{ijk}.\label{eq:jinvtemp}
\end{align}
The numerical results are comparable
in units of  $\eta_0=L_z\xi_0/\Tcn$.
Note that we have set the elementary charge $e$ and the Dirac constant $\hbar$ to unity in the derivation of the formula.
Recovering $1=e\RK/2\pi$ with
\begin{align}
\RK\equiv\frac{2\pi\hbar}{e^2}\sim 2.6\times 10^4\ \Omega,
\end{align}
which is abbreviated in the formula, we obtain 
\begin{align}
\eta_0=L_z\frac{\xi_0}{\Tcn}\frac{e\RK}{2\pi}
\sim 5\times 10^{-5} \, \mathrm{\text{\textmu} m^2/\text{\textmu} A}\frac{L_z\xi_0/[\mathrm{nm}^2]}{\Tcn/[\mathrm{K}]}.
\end{align}
In the following, we estimate typical values of $\eta_0$ and thereby estimate NCT.

Let us assume a strongly-correlated superconductor $\xi_0\sim 5\, \mathrm{nm}$ and $\Tcn\sim 2\, \mathrm{K}$, which corresponds to CeCoIn$_5$ superlattices~\cite{Naritsuka2021-ym}.
We also assume a sample thickness $L_z\sim {10}\ \mathrm{nm}$.
These values lead to
\begin{align}
\eta_0
&\sim {10^{-3}}\, \mathrm{\text{\textmu} m^2/\text{\textmu} A}.
\end{align}
Thus, we obtain $\eta_\SC^{xxx}\sim 10\eta_0\sim {10^{-2}}\,\mathrm{\text{\textmu} m^2/\text{\textmu} A}$ and $\eta_\SC^{xyy}\sim \eta_0\sim {10^{-3}}\, \mathrm{\text{\textmu} m^2/\text{\textmu} A}$.

To compare the obtained rectification with that expected from the parity-mixing mechanism~\cite{Wakatsuki2018-ll}, we focus on the MCA (i.e. $h$-linear rectification) by the anharmonicity parameter.
In our calculation, MCA (multiplied by $h$) is obtained as $\eta^{xxx}_\SC\sim 10^{-4}$ in units of $10\eta_0$, which is $10^{-4}$ times smaller than the rectification in the crossover region.
According to Ref.~\cite{Wakatsuki2018-ll}, the ratio of MCA caused by the anharmonicity parameters to that by the parity-mixing mechanism is $\sim r_{t,s}(E_F/\Tc)^2$, which is $O(10^3)$ by assuming $E_F/\Tc\sim t/\Tc\sim 10^2$ and the ratio of spin-triplet to -singlet pairing glues $r_{t,s}\sim 0.1$ adopted in Ref.~\cite{Wakatsuki2018-ll}.
Thus, the presumable MCA which would be obtained when the parity mixing was taken into account in our model is smaller by one order in magnitude than the rectification by the helical-superconductivity crossover.
Thus, the enhancement by the helical-superconductivity crossover is always visible and dominant even when the parity mixing is considered.
Note also that the parity-mixing mechanism requires the odd-parity pairing interaction comparable to the even-parity one, which would not be satisfied in all the noncentrosymmetric superconductors.

We also estimate typical values for nonlinear resistivity by taking the linear resistivity of a heavy-fermion superlattice
$\rho_{1\normal}\sim 5\times 10^{-7}\ \Omega\mathrm{m}$~\cite{Naritsuka2021-ym} as an example.
For simplicity, we assume 
\begin{align}
\sigma^{xx}_{1\normal}=\sigma^{yy}_{1\normal}=\sigma_{1\normal}=\rho_{1\normal}^{-1}.
\end{align}
Neglecting the contribution of the normal-state nonlinear conductivity $\sigma_{2\normal}^{ijk}$,
the nonlinear resistivity at the reduced temperature $\epsilon$ is given by
\begin{align}
\rho_{2}^{xii}(\epsilon)&=-[\sigma^{-1}_1(\epsilon)]^{xa}\sigma_{2\SC}^{abc}[\sigma^{-1}_1(\epsilon)]^{bi}[\sigma^{-1}_1(\epsilon)]^{ci}\\
&=-\rho_{1\normal}\left[\frac{\sigma_{1\normal}}{\sigma_{1\normal}+\sigma_{1\SC}^{xx}(\epsilon)}\right]\left[\frac{\sigma_{1\SC}^{ii}(\epsilon)}{\sigma_{1\normal}+\sigma_{1\SC}^{ii}(\epsilon)}\right]^{2}\eta^{xii}_{\SC}\notag\\
&=-\left[1+\frac{\sigma^{xx}_{1\SC}(\epsilon)}{\sigma_{1\normal}}\right]^{-1}\left[{1+\frac{\sigma_{1\normal}}{\sigma_{1\SC}^{ii}(\epsilon)}}\right]^{-2}\rho_{1\normal}\eta^{xii}_{\SC}.\notag
\end{align}
To obtain the second line, we used $\sigma^{xy}_1=0$ ensured due to the $y$-mirror plane and
\begin{align}
\sigma_{2\SC}^{xbc}(\epsilon)[\sigma^{-1}_{1\SC}(\epsilon)]^{bi}[\sigma^{-1}_{1\SC}(\epsilon)]^{ci}=\eta_{\SC}^{xii}.
\end{align}
In the following, we evaluate rectification and NHE separately.

\subsubsection{Rectification}
To estimate the rectification in the fluctuation regime, let us consider the reduced temperature $\epsilon_*$ defined by
\begin{align}
\frac{1}{3}\sigma_{1\normal}=\sigma_{1\SC}^{xx}(\epsilon_*)=\frac{1}{L_z}\frac{\xi_x^2}{16\epsilon_*\xi_x\xi_y}.\label{eq_SM:def_es}
\end{align}
The longitudinal resistivity is estimated to be
\begin{align}
\rho^{xxx}_{2*}&=-\frac{3}{4}\left(\frac{1}{4}\right)^2\rho_{1\normal}\eta_\SC^{xxx}\sim 10^{-4}\, {\rm \Omega\text{\textmu}m^3/\text{\textmu} A}.
\end{align}
Note that nonreciprocity at $\epsilon=\epsilon_*$,
\begin{align}
\eta_*^{xxx}=\left[{1+\frac{\sigma_{1\normal}}{\sigma_{1\SC}^{xx}(\epsilon_*)}}\right]^{-2}\eta^{xxx}_{\SC}=\frac{1}{16}\eta_{\SC}^{xxx},
\end{align}
is smaller than $\eta_{\SC}^{xxx}$ since $\epsilon_*$ is larger than $\epsilon=0$.
This is estimated to be
\begin{align}
\eta_*^{xxx}\sim 10^{-3}\, {\rm \text{\textmu} m^2/\text{\textmu} A}.
\end{align}

\subsubsection{NHE}
To estimate NHE in the fluctuation regime, let us consider $\epsilon_*$ defined by 
\begin{align}
\frac{1}{3}\sigma_{1\normal}=\sigma_{1\SC}^{yy}(\epsilon_*),
\end{align}
instead of Eq.~\eqref{eq_SM:def_es}.
This corresponds to the $25\%$ drop of the longitudinal resistance under concern, i.e. the applied electric current in the $y$ direction.
The transverse resistivity $\rho_2^{xyy}$ is then given by
\begin{align}
\rho^{xyy}_{2*}&=-\left[1+\frac{1}{3}\frac{\sigma_{1\SC}^{xx}(\epsilon_*)}{\sigma_{1\SC}^{yy}(\epsilon_*)}\right]^{-1}\left(\frac{1}{4}\right)^2\rho_{1\normal}\eta_\SC^{xyy}\notag\\
&\sim 10^{-5}\, {\rm \Omega\text{\textmu} m^3/\text{\textmu} A},
\end{align}
by using
\begin{align}
\frac{\sigma^{xx}_{1\SC}(\epsilon)}{\sigma^{yy}_{1\SC}(\epsilon)}&=\frac{\xi_x^2}{\xi_y^2}\sim \frac{1}{4},
\end{align}
for the crossover region.

\section{Derivation of the general formula for the nonlinear fluctuation conductivity}
In the following, we derive the formula of the nonlinear fluctuation conductivity represented by the eigenstates and eigenvalues of the GL coefficient $\halpha$.
We consider the GL free-energy functional of the form
\begin{align}
F[\ket{\psi}]&=V\braket{\psi|\halpha|\psi}.
\end{align}
Here, the GL coefficient operator $\halpha$ is arbitrary: For example, it can include both $\bm{x}$ and $\nabla/i$ in the presence of the orbital magnetic field, while is a matrix when the system has several pairing channels.
Accordingly, the phenomenological TDGL equation is given by
\begin{align}
\Gamma_0\partial_t\ket{\psi(t)}&=-\halpha(t)\ket{\psi(t)}+\ket{\zeta(t)},
\end{align}
which recasts into the standard expression by $\psi_{\bm{q}}(t)=\braket{\bm{q}|\psi(t)}$ and so on.
The electric field is incorporated into $\halpha(t)$ by the vector potential $A_i(t)=-E_it$.
We assume $\Gamma_0>0$ for simplicity and set $\Gamma_0\to1$ in the following.
$\Gamma_0$ will be recovered by $\hat{\alpha}\to\hat{\alpha}/\Gamma_0$ at the end of the calculation.
The random force $\ket{\zeta(t)}$ satisfies
\begin{align}
\ket{\zeta(t)}\bra{\zeta(t')}\to\frac{2}{\beta V}\delta(t-t')\hat{1},
\end{align}
after taking the noise average.
The identity operator $\hat{1}$ is abbreviated in the following.
We also abbreviate the subscript ``s" representing the paraconductivity contribution in the following for simplicity.

We are interested in the electric current carried by the steady-state solution of the TDGL equation,
\begin{align}
\ket{\psi(t)}&=\int_0^{\infty}dt'\,U(t,t-t')\ket{\zeta(t-t')},
\end{align}
where the time-evolution operator is given by
\begin{align}
U(t,t-t')\equiv T\exp\left(-\int_{t-t'}^t d\tau\,\halpha(\tau)\right).
\end{align}
Note that we can write
\begin{align}
\halpha(t)=U_\chi(t)\halpha U_\chi(t)^\dagger,
\end{align}
with $\halpha\equiv\halpha(0)$ and the unitary operator $U_\chi(t)$ representing the gauge transform.
Thus, we obtain $U(t,t-t')=U_\chi(t)U(0,-t')U_\chi(t)^\dagger$ and
\begin{align}
\ket{\psi(t)}=U_\chi(t)\int_0^{\infty}dt'\,U(0,-t')U_\chi(t)^\dagger\ket{\zeta(t-t')}.
\end{align}
The electric-current operator at the time $t$ is given by
\begin{align}
j_i(t)\equiv-\partial_{A_i}\halpha(t)=U_\chi(t)j_iU_\chi(t)^\dagger.
\end{align}
Thus, the electric current is given by
\begin{align}
\braket{j_i(t)}&\equiv\braket{\psi(t)|j_i(t)|\psi(t)}\notag\\
&=\uint dt_1\uint dt_2\,\Tr[U_\chi(t)^\dagger\ket{\zeta(t-t_2)}\notag\\
&\quad\cdot\bra{\zeta(t-t_1)}U_\chi(t)U(0,-t_1)^\dagger j_iU(0,-t_2)]\notag\\
&\to\frac{2}{\beta V}\uint ds\,\Re\Tr[j_iU(s)e^{-2\halpha s}U(s)^\dagger],
\end{align}
which is independent of $t$ owing to the balance of the applied electric field and the relaxation process.
Here we took the noise average to obtain the last line, defining
\begin{align}
U(s)\equiv\,T\exp\left(-\int^0_{-s}d\tau\,\halpha(\tau)\right)e^{\halpha s},
\end{align}
which represents the deviation of the time evolution under the electric field from the one in the absence of the field.

The nonlinear fluctuation conductivity is obtained by expanding $U(s)$ in terms of the electric field.
For this purpose, let us define
\begin{align}
\delta\halpha(t)\equiv \halpha(t)-\halpha,\quad \mathcal{E}(t)\equiv-e^{-\halpha t}\delta\halpha(-t)e^{\halpha t}.
\end{align}
The operator $\mathcal{E}(t)$ can be expanded by the electric field as $\mathcal{E}(t)=\sum_{n=1}^\infty\mathcal{E}_n(t)$ with $\mathcal{E}_n(t)=O(E^n)$.
The first two terms are given by
\begin{subequations}
\begin{align}
\mathcal{E}_1(t)&=-A_j(-t)e^{-\halpha t}\partial_{A_j}\delta\halpha(-t)e^{\halpha t}\notag\\
&=t\,E_je^{-\halpha t}j_je^{\halpha t},\\
\mathcal{E}_2(t)&=\frac{t^2}{2}E_jE_k\,e^{-\halpha t}\partial_{A_j}j_ke^{\halpha t}.
\end{align}
\end{subequations}
Since $U(s)$ satisfies
\begin{align}
\partial_sU(s)&=U(s)\mathcal{E}(s),
\end{align}
we obtain the integral equation
\begin{align}
U(s)&=1+\int_0^sdt_1\,U(t_1)\mathcal{E}(t_1)\notag\\
&=1+\int_0^sdt_1\,\mathcal{E}(t_1)\notag\\
&\quad+\int_0^sdt_1\int^{t_1}_0dt_2\,\mathcal{E}(t_2)\mathcal{E}(t_1)+O(E^3).
\end{align}
Thus, the electric current of $O(E^n)$ ($n=0,1,2$) is obtained by $\delta^nj_i$ as follows.
The zero-th order term vanishes,
\begin{align}
\delta^0j_i&=\frac{2}{\beta V}\int_0^\infty ds\,\Re\Tr[j_ie^{-2\halpha s}]\notag\\
&=-\frac{2}{\beta V}\int_0^\infty ds\,\frac{1}{2s}\partial_{A_i}\Re\Tr[e^{-2\halpha s}]\notag\\
&=0,
\end{align}
according to the gauge invariance.
The first-order electric current is given by
\begin{align}
\delta^1j_i&=\frac{4}{\beta V}\int_0^\infty ds\int_{0}^sdt_1\,\Re\Tr[j_i\mathcal{E}_1(t_1)e^{-2\halpha s}].
\end{align}
It is easy to see that $\delta^1j_i=\sigma_1^{ij}E_j$ gives
\begin{align}
\sigma_1^{ij}=\frac{\Gamma_0}{\beta V}\sum_{\mu\nu}\frac{\Re[\braket{\mu| j_i|\nu}\braket{\nu|j_j|\mu}]}{\alpha_\mu\alpha_\nu(\alpha_\mu+\alpha_\nu)},\label{eq_SM:linear_formula82}
\end{align}
which reproduces the formulas in Refs.~\cite{Abrahams1966-xl,Larkin2005-lb} as well as Eq.~\eqref{eq:LinearFormulaQ} by $j_i\to 2\partial_{q_i}\alpha_{\bm{q}}$.
Here and hereafter, we use the notation $\halpha\ket{\mu}=\alpha_\mu\ket{\mu}.$

The second-order electric current consists of three terms,
\begin{align}
\delta^2j_i&=\delta^2_aj_i+\delta^2_bj_i+\delta^2_cj_i,
\end{align}
where $\delta^2_aj_i$ is given by
\begin{align}
\delta^2_aj_i&=\frac{4}{\beta V}\int_0^\infty ds\int_{0}^sdt_1\,\Re\Tr[j_i\mathcal{E}_2(t_1)e^{-2\halpha s}]\notag\\
&=\frac{4}{\beta V}\uint dt_1\int_{t_1}^\infty ds\,\Re\Tr[j_i\mathcal{E}_2(t_1)e^{-2\halpha s}],
\end{align}
for example.
After similar procedures, we obtain
\begin{widetext}
\begin{subequations}
\begin{align}
\delta^2_aj_i
&=\frac{4}{\beta V}\uint dt_1\int_0^\infty ds\,\Re\Tr[j_i\mathcal{E}_2(t_1)e^{-2\halpha(s+t_1)}],\\
\delta^2_bj_i
&=\frac{4}{\beta V}\int_0^\infty dt_2\int_{0}^\infty dt_1\int_{0}^\infty ds\,\Re\Tr[j_i\mathcal{E}_1(t_1+t_2)e^{-2\halpha (s+t_1+t_2)}\mathcal{E}_1(t_2)^\dagger],\\
\delta^2_cj_i
&=\frac{4}{\beta V}\uint dt_2\int_0^\infty dt_1\int_0^{\infty}ds\,\Re\Tr[j_i\mathcal{E}_1(t_2)\mathcal{E}_1(t_1+t_2)e^{-2\halpha (s+t_1+t_2)}].
\end{align}
\end{subequations}
\end{widetext}

Before proceeding, we erase $\partial_{A_j}j_k$ from the expression of $\delta_a^2j_i$.
Let us consider an auxiliary quantity
\begin{align}
T_i&\equiv\Re\Tr[\partial_{A_i}j_je^{-\halpha t_1}j_k e^{-\halpha (2s+t_1)}]E_jE_k\notag\\
&=2\partial_{A_i}T_0-\Re\Tr[j_j\partial_{A_i}(e^{-\halpha t_1})j_k e^{-\halpha (2s+t_1)}]E_jE_k\notag\\
&\qquad-\Re\Tr[j_je^{-\halpha t_1}j_k \partial_{A_i}(e^{-\halpha (2s+t_1)})]E_jE_k\notag\\
&\qquad-\Re\Tr[j_je^{-\halpha t_1}\partial_{A_i}j_k e^{-\halpha (2s+t_1)}]E_jE_k,
\end{align}
with $2T_0\equiv \Re\Tr[j_je^{-\halpha t_1}j_ke^{-\halpha(2s+t_1)}]E_jE_k$.
The last line is equivalent to $-T_i$ by interchanging the dummy variables $j$ and $k$.
We also obtain
\begin{subequations}
\begin{align}
\partial_{A_i}e^{-\halpha t_1}&=\int_0^1d\theta\,e^{-\theta\halpha t}\partial_{A_i}(-\halpha t_1)e^{\halpha t_1(1-\theta)}\notag\\
&=\int_0^{t_1}dt_2\,e^{-\halpha t_2}j_ie^{-\halpha(t_1-t_2)},\\
\partial_{A_i}e^{-\halpha(2s+t_1)}&=\partial_{A_i}(e^{-\halpha 2s})e^{-\halpha t_1}+e^{-2s\halpha}\partial_{A_i}(e^{-\halpha t_1})\notag\\
&=2\int_0^sdt_2\,e^{-\halpha 2t_2}j_ie^{-\halpha[2(s-t_2)+t_1]}\notag\\
&\quad+\int_0^{t_1}dt_2\,e^{-\halpha (2s+t_2)}j_ie^{-\halpha(t_1-t_2)}.
\end{align}
\end{subequations}
Thus, we obtain
\begin{align}
T_i
&=\partial_{A_i}T_0-\frac{1}{2}\int_0^{t_1}dt_2\,[f_i(t_2,t_1-t_2,2s+t_1)\notag\\
&\qquad+f_i(2s+t_2,t_1-t_2,t_1)]\notag\\
&\qquad-\int_0^sdt_2\,f_i(2t_2,2(s-t_2)+t_1,t_1),
\end{align}
with 
\begin{align}
f_i(t_a,t_b,\Tc)&\equiv \Re\Tr[e^{-\halpha t_a}j_ie^{-\halpha t_b}j_je^{-\halpha \Tc}j_k]E_jE_k\notag\\
&=f_i(t_b,t_a,\Tc).
\end{align}
The integrand of $\delta^2_aj_i$ is, by using $\partial_{A_j}j_i=\partial_{A_i}j_j$,
\begin{align}
&\Re\Tr[j_i\mathcal{E}_2(t_1)e^{-\halpha 2(s+t_1)}]\notag\\
&=\frac{t_1^2}{2}\Re\Tr[j_ie^{-\halpha t_1}\partial_{A_j}j_ke^{-\halpha(2s+t_1)}]E_jE_k\notag\\
&=E_j\partial_{A_j}S_i-\frac{t_1^2}{2}T_i\\
&\qquad-\frac{t_1^2}{2}\Re\Tr[j_i\partial_{A_j}(e^{-\halpha t_1})j_ke^{-\halpha(2s-t_1)}]E_jE_k\notag\\
&\qquad-\frac{t_1^2}{2}\Re\Tr[j_ie^{-\halpha t_1}j_k\partial_{A_j}(e^{-\halpha(2s-t_1)})]E_jE_k\notag,
\end{align}
with $S_i\equiv(t_1^2/2)\Re\Tr[j_ie^{-\halpha t_1}j_ke^{-\halpha(2s+t_1)}]E_k$.
The total derivatives of the vector potential $E_j\partial_{A_j}S_i$ and $\partial_{A_i}T_0$ vanish according to the gauge invariance.
Thus, we obtain after changing the integral variables, 
\begin{align}
\delta^2_aj_i&=\frac{2}{\beta V}\iiint_{t_1,t_2,s}\left[(t_1+t_2)^2a_1^i(t_1+t_2,t_2,s)\right.\notag\\
&\qquad\left.+t_1^2a_2^i(t_1,t_2,s+t_2)\right],
\end{align}
with
\begin{align}
\iiint_{t_1,t_2,s}\equiv\uint dt_1\uint dt_2\uint ds, 
\end{align}
and
\begin{widetext}
\begin{subequations}
\begin{align}
a_1^i(t_1+t_2,t_2,s)&=\frac{1}{2}[f_i(t_2,t_1,2s+t_1+t_2)+f_i(2s+t_2,t_1,t_1+t_2)]\notag\\
&\quad-f_i(2s+t_1+t_2,t_2,t_1)-f_i(t_1,t_1+t_2,2s+t_2),\\
a_2^i(t_1,t_2,s+t_2)&=f_i(2t_2,2s+t_1,t_1)-2f_i(2s+t_1,t_1,2t_2).
\end{align}
\end{subequations}
\end{widetext}
We also obtain
\begin{subequations}
\begin{align}
\delta^2_bj_i&=\frac{2}{\beta V}\iiint_{t_1,t_2,s}\,2t_1(t_1+t_2)f_i(t_1,t_1+t_2,2s+t_2),\\
\delta^2_cj_i&=\frac{2}{\beta V}\iiint_{t_1,t_2,s}2t_2(t_1+t_2)f_i(2s+t_1+t_2,t_2,t_1).
\end{align}
\end{subequations}
Combining these terms, the second-order electric current $\delta^2j_i$ is obtained by
\begin{align}
\delta^2j_i&=u_i+v_i+w_i,
\end{align}
with
\begin{subequations}
\begin{align}
u_i&\equiv\frac{2}{\beta V}\iiint_{t_1,t_2,s}\frac{1}{2}(t_1+t_2)^2[f_i(t_2,t_1,2s+t_1+t_2)\notag\\
&\qquad\quad+f_i(2s+t_2,t_1,t_1+t_2)],\\
v_i&\equiv \frac{2}{\beta V}\iiint_{t_1,t_2,t_3}t_1^2[f_i(2s+t_1,2t_2,t_1)\notag\\
&\qquad\quad-2f_i(2s+t_1,t_1,2t_2)],\\
w_i&\equiv\frac{2}{\beta V}\iiint_{t_1,t_2,s}(t_1^2-t_2^2)[f_i(t_1,t_1+t_2,2s+t_2)\notag\\
&\qquad-f_i(2s+t_1+t_2,t_2,t_1)].
\end{align}
\end{subequations}
Now, the integrals can be straightforwardly performed
as the products of gamma functions, by introducing the current matrix element by
\begin{align}
J_{\mu\nu\lambda}^i&\equiv\Re[\braket{\mu|j_i|\nu}\braket{\nu|j_j|\lambda}\braket{\lambda|j_k|\mu}]E_jE_k\notag\\
&=J^i_{\nu\mu\lambda},
\end{align}
as well as expanding $f_i$ by
\begin{align}
f_i(t_a,t_b,\Tc)=\sum_{\mu\nu\lambda}J^i_{\mu\nu\lambda}e^{-(\alpha_\mu t_a+\alpha_\nu t_b+\alpha_\lambda \Tc)}.
\end{align}
The results are
\begin{widetext}
\begin{subequations}
\begin{align}
u_i&=\frac{2}{\beta V}\sum_{\mu\nu\lambda}\frac{J^i_{\mu\nu\lambda}}{2\alpha_\mu\alpha_\lambda(\alpha_\nu+\alpha_\lambda)}\left(\frac{1}{(\alpha_\nu+\alpha_\lambda)^2}+\frac{1}{(\alpha_\nu+\alpha_\lambda)(\alpha_\mu+\alpha_\lambda)}+\frac{1}{(\alpha_\mu+\alpha_\lambda)^2}\right),\\
v_i&=\frac{2}{\beta V}\sum_{\mu\nu\lambda}\frac{J^i_{\mu\nu\lambda}}{2\alpha_\mu}\left(\frac{1}{\alpha_\nu(\alpha_\mu+\alpha_\lambda)^3}-\frac{2}{\alpha_\lambda(\alpha_\mu+\alpha_\nu)^3}\right),\\
w_i&=\frac{2}{\beta V}\sum_{\mu\nu\lambda}\frac{J^i_{\mu\nu\lambda}}{\alpha_\nu\alpha_\lambda(\alpha_\mu+\alpha_\nu)}\left(\frac{1}{(\alpha_\mu+\alpha_\nu)^2}-\frac{1}{(\alpha_\nu+\alpha_\lambda)^2}\right).
\end{align}
\end{subequations}
\end{widetext}
Summing up these terms, we finally obtain the formula
\begin{align}
\!\!\!\!\delta^2j_i&=\frac{2}{\beta V}\sum_{\mu\nu\lambda}\frac{J^i_{\mu\nu\lambda}(\alpha_\mu+\alpha_\nu+2 \alpha_\lambda)}{\alpha_\lambda (\alpha_\mu+\alpha_\nu) (\alpha_\mu+\alpha_\lambda)^2 (\alpha_\nu+\alpha_\lambda)^2},\label{eq_SM:formula2}
\end{align}
after symmetrizing the summand with respect to $\mu$ and $\nu$ by employing \textit{Mathematica}.
The formula in the main text is obtained by $\alpha,j_i\to \alpha/\Gamma_0,j_i/\Gamma_0$ and so on.
Note that Eq.~\eqref{eq_SM:formula2} reproduces Eq.~\eqref{eq_SM:nonlinear1},
since
\begin{align}
\delta^2j_i\to \frac{2}{\beta V}\sum_q\frac{\partial_{q_i}\alpha_{\bm{q}}\partial_{q_j}\alpha_{\bm{q}}\partial_{q_j}\alpha_{\bm{q}}}{\alpha^5_{\bm{q}}}E_jE_k,
\end{align}
by $J^i_{\mu\nu\lambda}\to 8\partial_{q_i}\alpha_{\bm{q}}\partial_{q_j}\alpha_{\bm{q}}\partial_{q_j}\alpha_{\bm{q}}$.

\section{Effect of the orbital magnetic field on nonreciprocal transport}
Here we study the effect of the orbital magnetic field.
We consider the GL coefficient
\begin{align}
\halpha&=[\alpha_{\bm{q}_0}+\alpha_2^{ij}\delta q_i\delta q_j+\alpha_3^{ijk}\delta q_i\delta q_j\delta q_k]_{\bm{q}\to\frac{\nabla}{i}-2\bm{A}(\bm{x})}\\
&=N_0\left[\epsilon+\sum_{i}\xi_i^2P_i^2+\sum_{ijk}a_3^{ijk}(\xi_iP_i)(\xi_jP_j)(\xi_kP_k)\right].\notag
\end{align}
Here, we defined the operator
\begin{align}
\bm{P}&\equiv \frac{\nabla}{i}-\bm{q}_0-2\bm{A}(\bm{x}).
\end{align}
The vector potential $\bm{A}(\bm{x})$ represents the magnetic field in the $y$ direction.
We rescale the coordinates by $x_i\to \bx_i=x_i/\xi_i$, and thus 
\begin{align}
\mP_i&\equiv \xi_i P_i=\frac{1}{i}\partial_{\bx_i}-2\mA_i(\bm{\bx}),
\end{align}
with
\begin{align}
\mA_i(\bm{\bx})\equiv\xi_iq_{0i}/2+\delta_{ix}\mB \bz,\quad
\mB\equiv B\xi_x\xi_z.
\end{align}
Thus, we focus on
\begin{align}
\halpha&=N_0\left[\epsilon+\bm{\mP}^2+a_3^{ijk}\mP_i\mP_j\mP_k\right].
\end{align}
$\halpha$ commutes with $\mP_y$ and $\mP_x$, and 
\begin{align}
[\mP_z,\mP_x]=2i\mB.
\end{align}
The system volume in the $\bm{\bx}$ coordinates is given by $\bar{V}\equiv\bar{L}_x\bar{L}_y\bar{L}_z$ with $\bar{L}_i\equiv L_i/\xi_i$.
Since we are interested in the first-order effect of $a_3^{ijk}$, we can consider each component of $a_3^{ijk}$ separately.
In the following, we set $N_0\to1$, which will be recovered at the end of the calculation.


\subsection{Effect of $a_3^{xyy}$}
We first consider the situation where only $a_{3}^{xyy}=a_3^{yxy}=a_3^{yyx}$ is finite.
We have 
\begin{align}
\halpha&=\epsilon+(\mP_x+3a_3^{xyy}\mP_y^2/2)^2+\mP_y^2+\mP_z^2,
\end{align}
up to $O(a_3)$.
Thus, $\halpha$ is diagonalized by the Landau levels as well as the plane wave in the $\by$ direction.
The annihilation operator is given by
\begin{align}
a\equiv\frac{1}{\sqrt{4|\mB|}}[\mP_z+i\,\sgn[\mB](\mP_x+3a_3^{xyy}\mP_y^2/2)],\label{eq_SM:a_OP}
\end{align}
and we obtain
\begin{align}
\halpha&=\epsilon+\mP_y^2+4|\mB|(a^\dagger a+1/2)\notag\\
&=\bar{\epsilon}+\mP_y^2+4|\mB|a^\dagger a,
\end{align}
with $\bar{\epsilon}\equiv \epsilon+2|\mB|$.
Here, $m$-th Landau level has the eigenvalue
\begin{align}
    \alpha_m(p_y)=\bepsilon+p_y^2+4|\mB|m,
\end{align}
where $p_y$ and $m$ are the eigenvalues of $\mP_y$ and $a^\dagger a$, respectively,
and has the degeneracy
\begin{align}
g=\frac{|\mB| \bar{L}_x\bar{L}_z}{\pi}.
\end{align}
The electric current operator is given by
\begin{subequations}
\begin{align}
j_x&=2\xi_x\partial_{\mP_x}\halpha\notag\\
 &=4i\xi_x\sgn[\mB]\sqrt{|\mB|}(a^\dagger-a),\\
j_{y}+\delta j_y&\equiv 2\xi_y\partial_{\mP_y}\halpha\notag\\
&=4\xi_y\mP_y\left(1+\frac{3a_3^{xyy}}{4\xi_x}j_x\right),
\end{align}\label{eq_SM:currentOP}\end{subequations}
and thus their matrix elements between Landau levels $a^\dagger a\ket{n}=n\ket{n}$ are obtained by
\begin{subequations}
\begin{align}
|\braket{m|j_x|n}|^2&=16\xi_x^2|\mB|(m\delta_{m,n+1}+n\delta_{m+1,n}),\\
\braket{m|\delta j_y|n}&=4\xi_y\mP_y\frac{3a_3^{xyy}}{4\xi_x}\braket{m|j_x|n}.
\end{align}\end{subequations}

{In this case, only $\sigma_2^{xyy}$ and $\sigma_2^{yxy}=\sigma_2^{yyx}$can be finite:
Indeed, $\sigma_2^{ijk}$ is not affected by $\alpha_3^{xyy}$ when neither of $i,j,k$ is $y$, while vanishes when either one or three of $i,j,k$ is $y$ according to the $y$ mirror plane.
$\sigma^{ijk}_2$ with $\set{i,j,k}=\set{z,y,y}$ also vanishes by considering the matrix element (e.g., $\Re[\braket{m|j_x|n}\braket{n|j_z|m}]=0$).}
We obtain  with $p_y$ the eigenvalue of $\mP_y$,
\begin{align}
\delta M_{\mu\nu\lambda}^{xyy}&\equiv\Re[\braket{\mu|j_x|\nu}\braket{\nu|\delta j_y|\lambda}\braket{\lambda|j_y|\mu}]]\notag\\
&=16\xi_y^2\frac{3a_3^{xyy}}{4\xi_x}p_y^2\delta_{lm}|\braket{m|j_x|n}|^2,
\end{align}
for $\mu=(m,p_y)$, $\nu=(n,p_y)$ and $\lambda=(l,p_y)$ [Here and hereafter, the indices distinguishing degenerate Landau levels are abbreviated for simplicity].
The matrix element is thus given by
\begin{subequations}
\begin{align}
J^{xyy}_{\mu\nu\lambda}&\equiv\Re[\braket{\mu|j_x|\nu}\braket{\nu|(j_y+\delta j_y)|\lambda}\braket{\lambda|(j_y+\delta j_y)|\mu}]\notag\\
&=J^{xyy}_{\mu\nu\lambda}|_{a_3\to0}+\delta M_{\mu\nu\lambda}^{xyy}+\delta M_{\nu\mu\lambda}^{xyy}+O(a_3^2),\\
J^{yyx}_{\mu\nu\lambda}&=J^{xyy}_{\nu\lambda\mu}.\label{eq_SM:Jyyx}
\end{align}\end{subequations}
Thus, we obtain up to $O(a_3)$,
\begin{align}
\sigma_2^{xyy}&=\frac{2g}{\beta V}\sum_{\mu\nu\lambda}\frac{[\delta M_{\mu\nu\lambda}^{xyy}+\delta M_{\nu\mu\lambda}^{xyy}](\alpha_\mu+\alpha_\nu+2\alpha_\lambda)}{\alpha_\lambda(\alpha_\mu+\alpha_\nu)(\alpha_\mu+\alpha_\lambda)^2(\alpha_\nu+\alpha_\lambda)^2}\notag\\
&=\frac{4g}{\beta V}\sum_{p_ynml}\frac{\delta M_{\mu\nu\lambda}^{xyy}[3\alpha_m(p_y)+\alpha_n(p_y)]}{4\alpha_m^3(p_y)[\alpha_m(p_y)+\alpha_n(p_y)]^3}.
\end{align}
The leading order term in 
$\bepsilon\propto \alpha_0(p_y=0)$ comes from $m=0$.
Thus, we obtain
\begin{align}
\sigma_2^{xyy}&\sim \frac{4g}{\beta V}\sum_{p_y}\frac{16\xi_y^2\frac{3a_3^{xyy}}{4\xi_x}p_y^2\,16\xi_x^2|\mB|}{4\alpha_0^3(p_y)[\alpha_0(p_y)+\alpha_1(p_y)]^2}\notag\\
&=C^{xyy}\int\frac{dp_y}{2\pi}\frac{p_y^2}{(\bar{\epsilon}+p_y^2)^3[\bar{\epsilon}+p_y^2+2|\mB|]^2}\notag\\
&\sim C^{xyy}{\frac{1}{64\mB^2\bar{\epsilon}^{3/2}}.}
\end{align}
The coefficient is 
\begin{align}
C^{xyy}
&=\frac{g}{\beta V}{\bar{L}_y}16\xi_y^2\frac{3a_3^{xyy}\,16\xi_x^2|\mB|}{4\xi_x}\frac{1}{4}\notag\\
&=\frac{48\mB^2}{\pi \beta\xi_x\xi_y\xi_z}[\xi_x\xi_y^2a_3^{xyy}].
\end{align}
Thus, we obtain
\begin{align}
\sigma_2^{xyy}&\sim \tau_0^2\frac{48\mB^2}{\pi\beta \xi_x\xi_y\xi_z}[\xi_x\xi_y^2a_3^{xyy}]{\frac{1}{64\mB^2\bar{\epsilon}^{3/2}}}\notag\\
&={\frac{3\tau_0^2T}{4\pi\bar{\epsilon}^{3/2}}\mA^{xyy}.}
\end{align}
Here we recovered $\Gamma_0/N_0=\tau_0$.

Similarly, we obtain from Eq.~\eqref{eq_SM:Jyyx},
\begin{align}
\sigma_2^{yyx}&=\frac{2g}{\beta V}\sum_{\mu\nu\lambda}\frac{[\delta M_{\nu\lambda\mu}^{xyy}+\delta M_{\lambda\nu\mu}^{xyy}](\alpha_\mu+\alpha_\nu+2\alpha_\lambda)}{\alpha_\lambda(\alpha_\mu+\alpha_\nu)(\alpha_\mu+\alpha_\lambda)^2(\alpha_\nu+\alpha_\lambda)^2}\notag\\
&=\frac{2g}{\beta V}\sum_{\mu\nu\lambda}\frac{\delta M_{\mu\nu\lambda}^{xyy}(2\alpha_\mu+2\alpha_\nu)}{\alpha_\nu(2\alpha_\mu)(\alpha_\mu+\alpha_\nu)^4}\notag\\
&\qquad+\frac{2}{\beta V}\sum_{\mu\nu\lambda}\frac{\delta M_{\mu\nu\lambda}^{xyy}(\alpha_\nu+3\alpha_\mu)}{4\alpha_\mu^3(\alpha_\nu+\alpha_\mu)^3}\notag\\
&=\frac{1}{2}\sigma_2^{xyy}\\
&+\frac{2g}{\beta V}\sum_{p_ymn}\frac{\delta M_{\mu\nu\lambda}^{xyy}}{\alpha_m(p_y)\alpha_n(p_y)(\alpha_m(p_y)+\alpha_n(p_y))^3}\notag.
\end{align}
The second term is less singular in terms of $\bepsilon$ than the first term and thus is negligible.
We obtain
\begin{align}
\sigma_2^{yyx}=\sigma_2^{yxy}\sim\frac{1}{2}\sigma^{xyy}_2
&\sim{\frac{3\tau_0^2T}{8\pi\bar{\epsilon}^{3/2}}\mA^{xyy}.}
\end{align}

\subsection{Effect of $a_3^{yyy}$}
We here study $\sigma^{yyy}_{2\SC}$ from theoretical interest, while it vanishes in the presence of the $y$ mirror plane and therefore in the Rashba-Zeeman model studied in the main text.
This component is different from $\sigma^{xyy}_{2\SC}$ and $\sigma^{xxx}_{2\SC}$ in that it refers to the purely one-dimensional transport along the magnetic-field direction.
$\sigma^{yyy}_{2\SC}$ can be finite in the presence of e.g., $a_3^{yyy}$, and for this reason we consider $\sigma^{yyy}_{2\SC}$ induced by $a_3^{yyy}$, although the other components may also be induced by $a_3^{yyy}$.

The effect of $a_3^{yyy}$ is the shift of the eigenvalues by $\alpha_\mu\to \alpha_\mu +a_3^{yyy}p_y^3$, as well as the change in the current operator $j_y$,
\begin{align}
j_y+\delta j_y
&=4\xi_y\mP_y+6a_3^{yyy}\xi_y\mP_y^2.
\end{align}
The leading-order contribution is obtained by the lowest Landau level,
and thus we obtain
\begin{align}
\sigma_2^{yyy}&\sim\frac{2g}{\beta V}\sum_{p_y}\frac{(4\xi_y p_y+6a_3^{yyy}\xi_yp_y^2)^3\,4\alpha_0(p_y)}{32\alpha^6_0(p_y)}\notag\\
&=\frac{16|\mB|\xi_y^3}{\pi\beta \xi_x\xi_y\xi_z\bar{\epsilon}^3}\int\frac{dp_y}{2\pi}\frac{p_y^3(1+3\sqrt{\bar{\epsilon}}a_3^{yyy}p_y/2)^3}{(1+p_y^2+\sqrt{\bar{\epsilon}}a_3^{yyy}p_y^3)^5}.
\end{align}
The leading-order contribution is given by
\begin{align}
&\int\frac{dp_y}{2\pi}\frac{p_y^3(1+3\sqrt{\bar{\epsilon}}a_3^{yyy}p_y/2)^3}{(1+p_y^2+\sqrt{\bar{\epsilon}}a_3^{yyy}p_y^3)^5}
\sim\frac{3\sqrt{\bar{\epsilon}}a_3^{yyy}}{128}.
\end{align}
Thus, we obtain
\begin{align}
\sigma_2^{yyy}&\sim \tau_0^2\frac{16|\mB|[\xi_y^3a_3^{yyy}]}{\pi\beta\xi_x\xi_y\xi_z\bar{\epsilon}^{5/2}}\frac{3}{128}\notag\\
&=\frac{3\tau_0^2T|\mB|}{8\pi\bar{\epsilon}^{5/2}}\mA^{yyy}.
\end{align}
We recovered $\Gamma_0/N_0=\tau_0$. 
The obtained result $\sigma_{2\SC}^{yyy}\propto 1/\bepsilon^{5/2}$ indicates the effective one-dimensional transport under the orbital magnetic field, as is the case for the linear paraconductivity.

\subsection{Effect of $a_3^{xxx}$}
In this case, we start from
\begin{align}
\halpha&=\bar{\epsilon}+\mP_y^2+4|\mB|a^\dagger a+a_3^{xxx}\mP_x^3.
\end{align}
Here, $a$ and $a^\dagger$ are obtained by $a_3^{xyy}\to0$ in Eq.~\eqref{eq_SM:a_OP}.
Since $a_3^{xxx}\mP_x^3\propto (a-a^\dagger)^3$, it does not change the eigenvalues of $\halpha$ up to the first order in $a_3^{xxx}$, while changes the current operator as well as the eigenstates of $\halpha$.
The modified current operator is given by
\begin{align}
j_x+\delta j_x
&=j_x+\frac{3a_3^{xxx}}{8\xi_x}j_x^2,
\end{align}
with
\begin{align}
j_x=ij_0(a^\dagger-a),\quad j_0=4\xi_x\sgn[\mB]\sqrt{|\mB|}.
\end{align}
According to the $y$ mirror plane, $\sigma_2^{xxx},\sigma_2^{xzz}$, $\sigma_2^{zxx},\sigma_2^{zzz}$, $\sigma_2^{xyy}$, $\sigma_2^{zyy}$ (and their permutations) can be finite, while the latter two vanish 
since they are proportional to e.g., $J_{\mu\nu\lambda}^{xyy}\propto \braket{\mu(a_3^{xxx})|j_x(a_3^{xxx})|\mu(a_3^{xxx})}=-\partial_{A_x}\alpha_\mu(a_3^{xxx})=0$ (the algebra of $a$ and $a^\dagger$ does not change by $\mP_x\to \mP_x-2A_x$).
In the following, we focus on $\sigma_2^{xxx}$ to study the inplane transport.

Let us focus on the matrix element
\begin{align}
    J^{xxx}_{\mu\nu\lambda}+\delta     J^{xxx}_{\mu\nu\lambda}&=\Re[\braket{\mu+\delta\mu|j_x+\delta j_x|\nu+\delta\nu}\\
    &\quad\qquad\cdot\braket{\nu+\delta\nu|j_x+\delta j_x|\lambda+\delta\lambda}\notag\\
    &\qquad\qquad\cdot\braket{\lambda+\delta\lambda|j_x+\delta j_x|\mu+\delta\mu}].\notag
\end{align}
In the absence of $a_3^{xxx}$, the matrix element is given by
\begin{align}
    J^{xxx}_{\mu\nu\lambda}&=\Re\Tr[P_\mu j_xP_\nu j_xP_\lambda j_x],
\end{align}
with $P_\mu\equiv\ket{\mu}\bra{\mu}$ and so on.
We are interested in the change of $\sigma_2^{xxx}$ caused by $\delta J_{\mu\nu\lambda}^{xxx}$ up to $O(a_3^{xxx})$.
Since $J_{\mu\nu\lambda}^{xxx}+\delta J_{\mu\nu\lambda}^{xxx}$ is totally symmetric with respect to $\mu,\nu$ and $\lambda$, we can symmetrize the summand of $\sigma_{2\SC}^{xxx}$.
Then, we obtain
\begin{align}
\sigma_2^{xxx}
&=\frac{2g}{3\beta V}\sum_{\mu\nu\lambda}\frac{[J_{\mu\nu\lambda}^{xxx}+\delta J_{\mu\nu\lambda}^{xxx}](\alpha_\mu+\alpha_\nu+\alpha_\lambda)}{\alpha_\mu\alpha_\nu\alpha_\lambda(\alpha_\mu+\alpha_\nu)(\alpha_\nu+\alpha_\lambda)(\alpha_\lambda+\alpha_\mu)}\notag\\
&=\frac{2g}{\beta V}\sum_{\mu\nu\lambda}\frac{[A_{\mu\nu\lambda}+\tilde{B}_{\mu\nu\lambda}](\alpha_\mu+\alpha_\nu+\alpha_\lambda)}{\alpha_\mu\alpha_\nu\alpha_\lambda(\alpha_\mu+\alpha_\nu)(\alpha_\nu+\alpha_\lambda)(\alpha_\lambda+\alpha_\mu)}.\label{eq_SM:summand_xxx}
\end{align}
Here, we defined
\begin{subequations}
\begin{align}
A_{\mu\nu\lambda}
&=\Re\Tr[P_\mu\delta j_x\,P_\nu j_{x} P_\lambda j_{x}],\\
\tilde{B}_{\mu\nu\lambda}
&\equiv \frac{1}{2}\Re\Tr[\delta P_\mu j_x\,P_\nu j_x P_\lambda j_x]+(\nu\leftrightarrow\lambda).
\end{align}
\end{subequations}
The change of the projection operator $P_\mu$ is given by
\begin{align}
\delta P_\mu&=\sum_{\eta\neq\mu}\frac{P_\eta\delta\alpha P_\mu+P_\mu\delta\alpha P_\eta}{\alpha_\mu-\alpha_\eta}\notag\\
&=\frac{a_3^{xxx}}{16\xi_x j_0^2}\sum_{\eta\neq\mu}\frac{P_\eta j_x^3P_\mu+P_\mu j_x^3P_\eta}{\mu-\eta}.
\end{align}
Here, $\mu-\eta$ means the difference between the Landau-level indices of $\mu$ and $\eta$.
We thus obtain
\begin{subequations}
\begin{align}
A_{\mu\nu\lambda}&=\frac{3a_3^{xxx}}{8\xi_x}\Re\Tr[P_\mu j_x^2P_\nu j_xP_\lambda j_x],\\
\tilde{B}_{\mu\nu\lambda}&=\frac{1}{2}\frac{a_3^{xxx}}{16\xi_xj_0^2}\sum_{\eta\neq\mu}\frac{1}{\mu-\eta}\Re\Tr[(P_\eta j_x^3P_\mu+P_\mu j_x^3P_\eta)\notag\\
&\qquad\cdot j_xP_\nu j_xP_\lambda j_x]+(\nu\leftrightarrow\lambda)\notag\\
&=\frac{a_3^{xxx}}{16\xi_xj_0^2}\sum_{\eta\neq\mu}
\frac{\Re\Tr[P_\eta j_x^3P_\mu j_xP_\nu j_xP_\lambda j_x]}{\mu-\eta}\notag\\
&\qquad\qquad+(\nu\leftrightarrow\lambda).
\end{align}
\end{subequations}
We can consider
\begin{align}
B_{\mu\nu\lambda}=\frac{a_3^{xxx}}{8\xi_xj_0^2}\sum_{\eta\neq\mu}
\frac{\Re\Tr[P_\eta j_x^3P_\mu j_xP_\nu j_xP_\lambda j_x]}{\mu-\eta},
\end{align}
instead of $\tilde{B}_{\mu\nu\lambda}$, owing to the symmetry of the summand of Eq.~\eqref{eq_SM:summand_xxx}.

We are interested in the divergent terms as $\bepsilon\to0$.
For this purpose, we only have to consider terms where at least either one of $\mu,\nu$, and $\lambda$ is the lowest Landau level, as is clear from Eq.~\eqref{eq_SM:summand_xxx}
(Note that the matrix elements $A_{\mu\nu\lambda}, B_{\mu\nu\lambda}$ do not depend on $\bepsilon$ and thus can not be singular).
We obtain after cumbersome calculations shown in the next section,
\begin{subequations}
\begin{align}
A_{\mu\nu\lambda}&\sim \frac{a_3^{xxx}j_0^4}{8\xi_x}\left(3\delta_{{001}}^{{\mu\nu\lambda}}+12\delta_{{012}}^{{\mu\nu\lambda}}+9\delta_{{011}}^{{\mu\nu\lambda}}\right),\\
B_{\mu\nu\lambda}&\sim\frac{a_3^{xxx}j_0^4}{8\xi_x}\left(-3\delta_{{001}}^{{\mu\nu\lambda}}+4\delta_{{012}}^{{\mu\nu\lambda}}-9\delta_{{011}}^{{\mu\nu\lambda}}\right),
\end{align}
\end{subequations}
with $\delta^{\mu\nu\lambda}_{ijk}=\delta_{\mu i}\delta_{\nu j}\delta_{\nu k}$.
Only the combinations of $(\mu,\nu,\lambda)$ with divergent contributions are kept here, which is represented by the symbol ``$\sim$".
To obtain these expressions, we made permutations of variables $\mu,\nu,\lambda$ by using the symmetry of the summand of Eq.~\eqref{eq_SM:summand_xxx}.
Thus, we obtain
\begin{align}
A_{\mu\nu\lambda}+B_{\mu\nu\lambda}\sim \frac{2a_3^{xxx}j_0^4}{\xi_x}\delta_{012}^{\mu\nu\lambda},
\end{align}
and
\begin{align}
\sigma_2^{xxx}&\sim {\frac{2g}{\beta V}}\frac{2a_3^{xxx}j_0^4}{\xi_x}\sum_{p_y}\frac{\alpha_0+\alpha_1+\alpha_2}{\alpha_0\alpha_1\alpha_2(\alpha_0+\alpha_1)(\alpha_1+\alpha_2)(\alpha_2+\alpha_0)}\notag\\
&\sim \frac{4ga_3^{xxx}j_0^4\bar{L_y}}{\beta V\xi_x}\int\frac{dp_y}{2\pi}\frac{1}{\alpha_0\alpha_1\alpha_2(\alpha_0+\alpha_1)(\alpha_2+\alpha_0)}\notag\\
&\sim\frac{4^5a_3^{xxx}|\mB|^3\xi_x^3}{\pi\beta \xi_x\xi_y\xi_z}\cdot\frac{1}{2048|\mB|^4\sqrt{\bepsilon}}\notag\\
&={\frac{\tau_0^2T}{2\pi|\mB|\sqrt{\bepsilon}}\mA^{xxx},}
\end{align}
where the argument $p_y$ of $\alpha_m(p_y)$ is abbreviated and $\Gamma_0/N_0=\tau_0$ is recovered on the last line.

\subsection{Details of the matrix-element calculations for $a_3^{xxx}$}
In the following, we show the calculations of the matrix elements to discuss the effect of $a_3^{xxx}$.
\subsubsection{$A_{\mu\nu\lambda}$ with $\mu=\nu=0$}
The matrix element is 
\begin{align}
&\propto\braket{0|j_x^2|0}\braket{0|j_x|1}\braket{1|j_x|0}\delta_{\mu,0}\delta_{\nu,0}\delta_{\lambda,1}\notag\\
&=\braket{0|j_x|1}\braket{1|j_x|0}\braket{0|j_x|1}\braket{1|j_x|0}\delta_{\mu,0}\delta_{\nu,0}\delta_{\lambda,1}\notag\\
&=j_0^4\delta_{\mu,0}\delta_{\nu,0}\delta_{\lambda,1}.
\end{align}
\subsubsection{$A_{\mu\nu\lambda}$ with $\mu=0$, $\nu\neq0$}
The matrix element is 
\begin{align}
&\propto\braket{0|j_x^2|2}\braket{2|j_x|1}\braket{1|j_x|0}\delta_{\mu,0}\delta_{\nu,2}\delta_{\lambda,1}\notag\\
&=\braket{0|j_x|1}\braket{1|j_x|2}\braket{2|j_x|1}\braket{1|j_x|0}\delta_{\mu,0}\delta_{\nu,2}\delta_{\lambda,1}\notag\\
&=2j_0^4\delta_{\mu,0}\delta_{\nu,2}\delta_{\lambda,1}.
\end{align}
\subsubsection{$A_{\mu\nu\lambda}$ with $\nu=0$, $\mu\neq0$}
The matrix element is 
\begin{align}
&\propto\braket{2|j_x^2|0}\braket{0|j_x|1}\braket{1|j_x|2}\delta_{\mu,2}\delta_{\nu,0}\delta_{\lambda,1}\notag\\
&=\braket{2|j_x|1}\braket{1|j_x|0}\braket{0|j_x|1}\braket{1|j_x|2}\delta_{\mu,2}\delta_{\nu,0}\delta_{\lambda,1}\notag\\
&=2j_0^4\delta_{\mu,2}\delta_{\nu,0}\delta_{\lambda,1}.
\end{align}

\subsubsection{$A_{\mu\nu\lambda}$ with $\lambda=0$}
The matrix element is 
\begin{align}
&\propto\braket{1|j_x^2|1}\braket{1|j_x|0}\braket{0|j_x|1}\delta_{\mu,1}\delta_{\nu,1}\delta_{\lambda,0}\notag\\
&=(\braket{1|j_x|0}\braket{0|j_x|1}+\braket{1|j_x|2}\braket{2|j_x|1})\notag\\
&\quad\cdot\braket{1|j_x|0}\braket{0|j_x|1}\delta_{\mu,1}\delta_{\nu,1}\delta_{\lambda,0}\notag\\
&=3j_0^4\delta_{\mu,1}\delta_{\nu,1}\delta_{\lambda,0}.
\end{align}

\subsubsection{$B_{\mu\nu\lambda}$ with $\mu=\lambda=0$}
The matrix element is 
\begin{align}
&\propto -\braket{1|j_x^3|0}\braket{0|j_x|1}\braket{1|j_x|0}\braket{0|j_x|1}\delta_{\mu,0}\delta_{\nu,1}\delta_{\lambda,0}\notag\\
&=-\braket{1|j_x^2|1}\braket{1|j_x|0}\braket{0|j_x|1}\braket{1|j_x|0}\braket{0|j_x|1}\delta_{\mu,0}\delta_{\nu,1}\delta_{\lambda,0}\notag\\
&=-(\braket{1|j_x|0}\braket{0|j_x|1}+\braket{1|j_x|2}\braket{2|j_x|1})\notag\\
&\cdot\braket{1|j_x|0}\braket{0|j_x|1}\braket{1|j_x|0}\braket{0|j_x|1}\delta_{\mu,0}\delta_{\nu,1}\delta_{\lambda,0}\notag\\
&=-3j_0^6\delta_{\mu,0}\delta_{\nu,1}\delta_{\lambda,0}.
\end{align}
\subsubsection{$B_{\mu\nu\lambda}$ with $\mu=0$, $\lambda\neq0$}
The matrix element is 
\begin{align}
&\propto \sum_{\eta\neq0}\frac{1}{-\eta}\braket{\eta|j_x^3|0}\braket{0|j_x|1}\braket{1|j_x|2}\braket{2|j_x|\eta}\delta_{\mu,0}\delta_{\nu,1}\delta_{\lambda,2}\notag\\
&=-\Bigl(\braket{1|j_x^3|0}\braket{0|j_x|1}\braket{1|j_x|2}\braket{2|j_x|1}\notag\\
&\quad+
\frac{1}{3}\braket{3|j_x^3|0}\braket{0|j_x|1}\braket{1|j_x|2}\braket{2|j_x|3}\Bigr)\delta_{\mu,0}\delta_{\nu,1}\delta_{\lambda,2}\notag\\
&=-\Bigl(\braket{1|j_x^2|1}\braket{1|j_x|0}\braket{0|j_x|1}2j_0^2\notag\\
&\quad+
\frac{1}{3}\braket{3|j_x|2}\braket{2|j_x|1}\braket{1|j_x|0}\braket{0|j_x|1}\braket{1|j_x|2}\braket{2|j_x|3}\Bigr)\notag\\
&\qquad\cdot\delta_{\mu,0}\delta_{\nu,1}\delta_{\lambda,2}\notag\\
&=-j_0^6\delta_{\mu,0}\delta_{\nu,1}\delta_{\lambda,2}(3\cdot2 +\frac{1}{3}3\cdot2)\notag\\
&=-8j_0^6\delta_{\mu,0}\delta_{\nu,1}\delta_{\lambda,2}.
\end{align}

\subsubsection{$B_{\mu\nu\lambda}$ with $\nu=0$}
The matrix element is 
\begin{align}
&\propto \sum_{\eta\neq1}\frac{1}{1-\eta}\braket{\eta|j_x^3|1}\braket{1|j_x|0}\braket{0|j_x|1}\braket{1|j_x|\eta}\delta_{\mu,1}\delta_{\nu,0}\delta_{\lambda,1}\notag\\
&=(\braket{0|j_x^3|1}\braket{1|j_x|0}\braket{0|j_x|1}\braket{1|j_x|0}\notag\\
&-\braket{2|j_x^3|1}\braket{1|j_x|0}\braket{0|j_x|1}\braket{1|j_x|2})
\delta_{\mu,1}\delta_{\nu,0}\delta_{\lambda,1}\notag\\
&=(3j_0^6-j_0^2\braket{2|j_x^3|1}\braket{1|j_x|2})\delta_{\mu,1}\delta_{\nu,0}\delta_{\lambda,1}\notag\\
&=-9j_0^6\delta_{\mu,1}\delta_{\nu,0}\delta_{\lambda,1}.
\end{align}
Here, we used
\begin{align}
\braket{2|j_x^3|1}
&=\braket{2|j_x^2|2}\braket{2|j_x|1}+\braket{2|j_x^2|0}\braket{0|j_x|1}\notag\\
&=(\braket{2|j_x|1}\braket{1|j_x|2}+\braket{2|j_x|3}\braket{3|j_x|2})\braket{2|j_x|1}\notag\\
&\qquad+\braket{2|j_x|1}\braket{1|j_x|0}\braket{0|j_x|1}\notag\\
&=6j_0^2\braket{2|j_x|1}.
\end{align}

\subsubsection{$B_{\mu\nu\lambda}$ with $\lambda=0$, $\mu\neq0$}
The matrix element is 
\begin{align}
&\propto \frac{1}{2-1}\braket{1|j_x^3|2}\braket{2|j_x|1}\braket{1|j_x|0}\braket{0|j_x|1}\delta_{\mu,2}\delta_{\nu,1}\delta_{\lambda,0}\notag\\
&=6j_0^2\cdot 2j_0^2\cdot j_0^2\delta_{\mu,2}\delta_{\nu,1}\delta_{\lambda,0}\notag\\
&=12j_0^6\delta_{\mu,2}\delta_{\nu,1}\delta_{\lambda,0}.
\end{align}

\section{Estimate of nonreciprocity under the orbital magnetic field}
Here, we estimate NCT of three-dimensional noncentrosymmetric superconductors under the orbital magnetic field.
In the following, the normal-state contribution $\sigma_{2\normal}^{xxx}$ is abbreviated. 
We also neglect the magnetoconductivity in the normal state for simplicity and thus $\sigma_{1\normal}^{xx}=\sigma_{1\normal}^{yy}=\sigma_{1\normal}$.

\subsection{Rectification}
Let us first consider the rectification in the $x$ direction, which is perpendicular to the magnetic field.
The reduced temperature $\bepsilon_*$ is given by
\begin{align}
\!\!\!\!\!\frac{1}{3}\sigma_{1\normal}^{xx}=\sigma_{1\SC}^{xx}(\bepsilon_*)&=\frac{\tau_0 T}{\pi}\frac{\xi_x}{\xi_y\xi_z}\frac{1}{\sqrt{\bepsilon_*}}
=\frac{2\tau_0 T}{\RK}\frac{\xi_x}{\xi_y\xi_z}\frac{1}{\sqrt{\bepsilon_*}},
\label{eq_SM:def_ebs}
\end{align}
with recovering $2\pi/\RK=1$,
where the second equality follows from Eq.~\eqref{eq_SM:linear_formula82} [see also Ref.~\cite{Larkin2005-lb}].
We also obtain
\begin{align}
\sigma_{2\SC}^{xxx}(\bepsilon_*)&= \frac{2\pi e}{\RK}\frac{\tau_0^2T\mA^{xxx}}{2\pi^2[B/\phi_0]\xi_x\xi_z}\frac{1}{\sqrt{\bepsilon_*}}\notag\\
&=\frac{8\tau_0T}{\pi}\,\frac{e\sigma_{1\normal}^{xx}\xi_y\xi_z}{48\xi_x}\frac{\mA^{xxx}}{T[B/\phi_0]\xi_x\xi_z},
\end{align}
with $e=1$ and $\phi_0=\pi\hbar/e=\pi$ recovered.
Let us introduce
$\mA_{d=2}^{ijk}\equiv\alpha_3^{ijk}/N_0(\xi_x\xi_y)^{3/2}$ to estimate $\mA^{xxx}$, since $\alpha_3^{ijk}$ rather than $\mA^{ijk}$ in two and three dimensions will directly correspond to each other.
We obtain
\begin{align}
\sigma_{2\SC}^{xxx}(\bepsilon_*)
&= \frac{8\tau_0T}{\pi}\,\frac{\sigma_{1\normal}^{xx}}{24\RK\xi_z\xi_y[B/\phi_0]}\notag\\
&\qquad\qquad\cdot\left(\frac{e\RK}{2\pi}\frac{\pi\sqrt{\xi_x\xi_y}}{T}\frac{\xi_y^2}{\xi_x^2}\mA^{xxx}_{d=2}\right).
\end{align}
The quantity inside the parentheses corresponds to the nonreciprocity $\eta_{\SC,d=2}^{xxx}$ of a thin film whose thickness is the unit length, and is written as $\eta_{\SC,d=2}^{xxx}|_{L_z\to1}$.
By using this, the nonreciprocity is obtained by
\begin{align}
\eta_*^{xxx}&=\left(\frac{3}{4\sigma_{1\normal}^{xx}}\right)^2\sigma_{2\SC}^{xxx}(\bepsilon_*)\\
&=\left(\frac{3}{128}\frac{8\tau_0T}{\pi}\frac{1}{\RK\sigma_{1\normal}^{xx}\xi_z\xi_y[B/\phi_0]}\right)\eta_{\SC,d=2}^{xxx}|_{L_z\to1}.\notag
\end{align}
This means that the nonreciprocity of quasi-two-dimensional superconductors coincides with the intrinsic nonreciprocity of a two-dimensional superconductor with the effective thickness
\begin{align}
l^{xxx}&\equiv \frac{3}{128}\frac{8\tau_0T}{\pi}\frac{1}{\RK\sigma_{1\normal}^{xx}\xi_z\xi_y[B/\phi_0]}.
\end{align}
Thus, we obtain
\begin{align}
l^{xxx}\sim 10^{-2}\, {\rm nm},\quad \eta_*^{xxx}\sim 10^{-5}\, {\rm \text{\textmu} m^2/\text{\textmu} A},
\end{align}
and
\begin{align}
\rho_{2*}^{xxx}=\frac{3}{4}\rho_{1\normal}^{xx}\eta_*^{xxx}\sim 10^{-5}\, {\rm \Omega \text{\textmu} m^3/\text{\textmu} A},
\end{align}
assuming
$8\tau_0T/\pi\sim1$, $[\sigma_{1\normal}^{xx}]^{-1}\sim 5\times 10^{-7}\ \mathrm{\Omega m},$ $\phi_0
\sim 2\times 10^{-16}\, {\rm me}{\rm V\cdot m^2}$, 
$\eta_{\SC,d=2}^{xxx}|_{L_z\to1}\sim 1\ {\rm \text{\textmu} m/\text{\textmu} A}$,
$\xi_y\sim 2\xi_{y}(h=0)\sim 10\ \mathrm{nm}$, $\xi_z\sim 2\ \mathrm{nm}$, and $B\sim 2\Tcn\sim 0.4\ \mathrm{me}\mathrm{V}$ focusing on the crossover region.


\subsection{NHE}
NHE can be evaluated similarly.
Note that the nonlinear Hall resistance is given by
\begin{align}
\rho_2^{xyy}&=-[\sigma_{1}^{-1}]^{xa}\sigma_2^{abc}[\sigma_{1}^{-1}]^{by}[\sigma_1^{-1}]^{cy}\notag\\
&=-[\sigma_{1}^{-1}]^{xx}\sigma_2^{xyy}[\sigma_{1}^{-1}]^{yy}[\sigma_1^{-1}]^{yy},
\end{align}
since $\sigma_{1}^{xy}=0$ according to the $y$ mirror plane.
The linear conductivity in the $y$ direction is given by
\begin{align}
\sigma_{1\SC}^{yy}(\bepsilon)&=\frac{\tau_0T|\mB|\xi_y^2}{2\pi\xi_x\xi_y\xi_z}\frac{1}{\bepsilon^{3/2}}=\frac{8\tau_0T}{\pi} \frac{2\pi }{\RK}\frac{\pi B\xi_y}{16\phi_0\bepsilon^{3/2}}.
\end{align}
We here define $\bepsilon_*$ by 
\begin{align}
\frac{1}{3}\sigma_{1\normal}=\sigma_{1\SC}^{yy}(\bepsilon_*),
\end{align}
to estimate NHE, instead of Eq.~\eqref{eq_SM:def_ebs}.
This corresponds to the $25\%$ drop of the longitudinal resistance under the applied electric current in the $y$ direction.
We obtain
\begin{align}
&\frac{1}{3}=\frac{\sigma_{1\SC}^{yy}(\bepsilon_*)}{\sigma_{1\normal}}=\frac{\pi}{16}c_1c_2\frac{1}{\bepsilon_*^{3/2}},\notag\\
&\frac{\sigma_{1\SC}^{xx}(\bepsilon_*)}{\sigma_{1\normal}}=\frac{\xi_x}{8\xi_y}c_1\frac{1}{\bepsilon_*^{1/2}},
\end{align}
with dimensionless parameters 
\begin{align}
c_1\equiv \frac{8\tau_0T}{\pi}\frac{2\pi}{\RK\sigma_{1\normal}^{xx}\xi_z},\quad c_2\equiv\frac{B\xi_y\xi_z}{\phi_0}.
\end{align}
The nonlinear conductivity is given by
\begin{align}
\frac{\sigma_{2\SC}^{xyy}(\bepsilon_*)}{\sigma_{1\normal}^2}&=\left(\frac{8\tau_0T}{\pi}\right)^2 \frac{2\pi e}{\RK\sigma_{1\normal}^2}\frac{3\pi}{256T\bepsilon_*^{3/2}}\mA^{xyy}\notag\\
&=\frac{3}{256\bepsilon_*^{3/2}}c_1^2{\xi_z}\frac{\xi_y^2}{\xi_x^2}\left(\frac{\RK e}{2\pi}\frac{\pi\sqrt{\xi_x\xi_y}}{T}\frac{\xi_x^2}{\xi_y^2}\mA^{xyy}_{d=2}\right)\notag\\
&=\frac{3}{256\bepsilon_*^{3/2}}c_1^2{\xi_z}\frac{\xi_y^2}{\xi_x^2}\eta_{\SC,d=2}^{xyy}|_{L_z\to1},
\end{align}
by using $\mA^{ijk}=(\sqrt{\xi_x\xi_y}/\xi_z)\mA^{ijk}_{d=2}$.
With the parameters used to evaluate $\rho_{2*}^{xxx}$ and by setting $\xi_x\sim\xi_y/2$ and $\eta_{\SC,d=2}^{xyy}|_{L_z\to1}\sim 0.1\, {\rm \text{\textmu} m/\text{\textmu} A}$, 
we obtain
\begin{align}
\rho_{2*}^{xyy}&=-\frac{1}{\sigma_{1\normal}+\sigma_{1\SC}^{xx}(\bepsilon_*)}\left(\frac{1}{\sigma_{1\normal}+\sigma_{1\SC}^{yy}(\bepsilon_*)}\right)^2\sigma_{2\SC}^{xyy}(\bepsilon_*)\notag\\
&=-\frac{\rho_{1\normal}}{1+\frac{\sigma_{1\SC}^{xx}(\bepsilon_*)}{\sigma_{1\normal}}}\left(\frac{3}{4}\right)^2\frac{\sigma_{2\SC}^{xyy}(\bepsilon_*)}{\sigma_{1\normal}^2}\notag\\
&\sim 10^{-5}\, \mathrm{\Omega \text{\textmu} m^3/\text{\textmu} A}.
\end{align}


\begin{thebibliography}{81}%
\makeatletter
\providecommand \@ifxundefined [1]{%
 \@ifx{#1\undefined}
}%
\providecommand \@ifnum [1]{%
 \ifnum #1\expandafter \@firstoftwo
 \else \expandafter \@secondoftwo
 \fi
}%
\providecommand \@ifx [1]{%
 \ifx #1\expandafter \@firstoftwo
 \else \expandafter \@secondoftwo
 \fi
}%
\providecommand \natexlab [1]{#1}%
\providecommand \enquote  [1]{``#1''}%
\providecommand \bibnamefont  [1]{#1}%
\providecommand \bibfnamefont [1]{#1}%
\providecommand \citenamefont [1]{#1}%
\providecommand \href@noop [0]{\@secondoftwo}%
\providecommand \href [0]{\begingroup \@sanitize@url \@href}%
\providecommand \@href[1]{\@@startlink{#1}\@@href}%
\providecommand \@@href[1]{\endgroup#1\@@endlink}%
\providecommand \@sanitize@url [0]{\catcode `\\12\catcode `\$12\catcode
  `\&12\catcode `\#12\catcode `\^12\catcode `\_12\catcode `\%12\relax}%
\providecommand \@@startlink[1]{}%
\providecommand \@@endlink[0]{}%
\providecommand \url  [0]{\begingroup\@sanitize@url \@url }%
\providecommand \@url [1]{\endgroup\@href {#1}{\urlprefix }}%
\providecommand \urlprefix  [0]{URL }%
\providecommand \Eprint [0]{\href }%
\providecommand \doibase [0]{https://doi.org/}%
\providecommand \selectlanguage [0]{\@gobble}%
\providecommand \bibinfo  [0]{\@secondoftwo}%
\providecommand \bibfield  [0]{\@secondoftwo}%
\providecommand \translation [1]{[#1]}%
\providecommand \BibitemOpen [0]{}%
\providecommand \bibitemStop [0]{}%
\providecommand \bibitemNoStop [0]{.\EOS\space}%
\providecommand \EOS [0]{\spacefactor3000\relax}%
\providecommand \BibitemShut  [1]{\csname bibitem#1\endcsname}%
\let\auto@bib@innerbib\@empty
\bibitem [{\citenamefont {Tokura}\ and\ \citenamefont
  {Nagaosa}(2018)}]{Tokura2018-nb}%
  \BibitemOpen
  \bibfield  {author} {\bibinfo {author} {\bibfnamefont {Y.}~\bibnamefont
  {Tokura}}\ and\ \bibinfo {author} {\bibfnamefont {N.}~\bibnamefont
  {Nagaosa}},\ }\bibfield  {title} {\bibinfo {title} {Nonreciprocal responses
  from non-centrosymmetric quantum materials},\ }\href
  {https://doi.org/10.1038/s41467-018-05759-4} {\bibfield  {journal} {\bibinfo
  {journal} {Nat. Commun.}\ }\textbf {\bibinfo {volume} {9}},\ \bibinfo {pages}
  {3740} (\bibinfo {year} {2018})}\BibitemShut {NoStop}%
\bibitem [{\citenamefont {Ideue}\ and\ \citenamefont
  {Iwasa}(2021)}]{Ideue2021-es}%
  \BibitemOpen
  \bibfield  {author} {\bibinfo {author} {\bibfnamefont {T.}~\bibnamefont
  {Ideue}}\ and\ \bibinfo {author} {\bibfnamefont {Y.}~\bibnamefont {Iwasa}},\
  }\bibfield  {title} {\bibinfo {title} {Symmetry breaking and nonlinear
  electric transport in van der waals nanostructures},\ }\href
  {https://doi.org/10.1146/annurev-conmatphys-060220-100347} {\bibfield
  {journal} {\bibinfo  {journal} {Annu. Rev. Condens. Matter Phys.}\ }\textbf
  {\bibinfo {volume} {12}},\ \bibinfo {pages} {201} (\bibinfo {year}
  {2021})}\BibitemShut {NoStop}%
\bibitem [{\citenamefont {Rikken}\ \emph {et~al.}(2001)\citenamefont {Rikken},
  \citenamefont {F{\"o}lling},\ and\ \citenamefont {Wyder}}]{Rikken2001-il}%
  \BibitemOpen
  \bibfield  {author} {\bibinfo {author} {\bibfnamefont {G.~L.}\ \bibnamefont
  {Rikken}}, \bibinfo {author} {\bibfnamefont {J.}~\bibnamefont
  {F{\"o}lling}},\ and\ \bibinfo {author} {\bibfnamefont {P.}~\bibnamefont
  {Wyder}},\ }\bibfield  {title} {\bibinfo {title} {Electrical magnetochiral
  anisotropy},\ }\href {https://doi.org/10.1103/PhysRevLett.87.236602}
  {\bibfield  {journal} {\bibinfo  {journal} {Phys. Rev. Lett.}\ }\textbf
  {\bibinfo {volume} {87}},\ \bibinfo {pages} {236602} (\bibinfo {year}
  {2001})}\BibitemShut {NoStop}%
\bibitem [{\citenamefont {Rikken}\ and\ \citenamefont
  {Wyder}(2005)}]{Rikken2005-ew}%
  \BibitemOpen
  \bibfield  {author} {\bibinfo {author} {\bibfnamefont {G.~L. J.~A.}\
  \bibnamefont {Rikken}}\ and\ \bibinfo {author} {\bibfnamefont
  {P.}~\bibnamefont {Wyder}},\ }\bibfield  {title} {\bibinfo {title}
  {Magnetoelectric anisotropy in diffusive transport},\ }\href
  {https://doi.org/10.1103/PhysRevLett.94.016601} {\bibfield  {journal}
  {\bibinfo  {journal} {Phys. Rev. Lett.}\ }\textbf {\bibinfo {volume} {94}},\
  \bibinfo {pages} {016601} (\bibinfo {year} {2005})}\BibitemShut {NoStop}%
\bibitem [{\citenamefont {Ideue}\ \emph {et~al.}(2017)\citenamefont {Ideue},
  \citenamefont {Hamamoto}, \citenamefont {Koshikawa}, \citenamefont {Ezawa},
  \citenamefont {Shimizu}, \citenamefont {Kaneko}, \citenamefont {Tokura},
  \citenamefont {Nagaosa},\ and\ \citenamefont {Iwasa}}]{Ideue2017-vp}%
  \BibitemOpen
  \bibfield  {author} {\bibinfo {author} {\bibfnamefont {T.}~\bibnamefont
  {Ideue}}, \bibinfo {author} {\bibfnamefont {K.}~\bibnamefont {Hamamoto}},
  \bibinfo {author} {\bibfnamefont {S.}~\bibnamefont {Koshikawa}}, \bibinfo
  {author} {\bibfnamefont {M.}~\bibnamefont {Ezawa}}, \bibinfo {author}
  {\bibfnamefont {S.}~\bibnamefont {Shimizu}}, \bibinfo {author} {\bibfnamefont
  {Y.}~\bibnamefont {Kaneko}}, \bibinfo {author} {\bibfnamefont
  {Y.}~\bibnamefont {Tokura}}, \bibinfo {author} {\bibfnamefont
  {N.}~\bibnamefont {Nagaosa}},\ and\ \bibinfo {author} {\bibfnamefont
  {Y.}~\bibnamefont {Iwasa}},\ }\bibfield  {title} {\bibinfo {title} {Bulk
  rectification effect in a polar semiconductor},\ }\href
  {https://doi.org/10.1038/nphys4056} {\bibfield  {journal} {\bibinfo
  {journal} {Nat. Phys.}\ }\textbf {\bibinfo {volume} {13}},\ \bibinfo {pages}
  {578} (\bibinfo {year} {2017})}\BibitemShut {NoStop}%
\bibitem [{\citenamefont {Wakatsuki}\ \emph {et~al.}(2017)\citenamefont
  {Wakatsuki}, \citenamefont {Saito}, \citenamefont {Hoshino}, \citenamefont
  {Itahashi}, \citenamefont {Ideue}, \citenamefont {Ezawa}, \citenamefont
  {Iwasa},\ and\ \citenamefont {Nagaosa}}]{Wakatsuki2017-dp}%
  \BibitemOpen
  \bibfield  {author} {\bibinfo {author} {\bibfnamefont {R.}~\bibnamefont
  {Wakatsuki}}, \bibinfo {author} {\bibfnamefont {Y.}~\bibnamefont {Saito}},
  \bibinfo {author} {\bibfnamefont {S.}~\bibnamefont {Hoshino}}, \bibinfo
  {author} {\bibfnamefont {Y.~M.}\ \bibnamefont {Itahashi}}, \bibinfo {author}
  {\bibfnamefont {T.}~\bibnamefont {Ideue}}, \bibinfo {author} {\bibfnamefont
  {M.}~\bibnamefont {Ezawa}}, \bibinfo {author} {\bibfnamefont
  {Y.}~\bibnamefont {Iwasa}},\ and\ \bibinfo {author} {\bibfnamefont
  {N.}~\bibnamefont {Nagaosa}},\ }\bibfield  {title} {\bibinfo {title}
  {Nonreciprocal charge transport in noncentrosymmetric superconductors},\
  }\href {https://doi.org/10.1126/sciadv.1602390} {\bibfield  {journal}
  {\bibinfo  {journal} {Science Advances}\ }\textbf {\bibinfo {volume} {3}},\
  \bibinfo {pages} {e1602390} (\bibinfo {year} {2017})}\BibitemShut {NoStop}%
\bibitem [{\citenamefont {Qin}\ \emph {et~al.}(2017)\citenamefont {Qin},
  \citenamefont {Shi}, \citenamefont {Ideue}, \citenamefont {Yoshida},
  \citenamefont {Zak}, \citenamefont {Tenne}, \citenamefont {Kikitsu},
  \citenamefont {Inoue}, \citenamefont {Hashizume},\ and\ \citenamefont
  {Iwasa}}]{Qin2017-vd}%
  \BibitemOpen
  \bibfield  {author} {\bibinfo {author} {\bibfnamefont {F.}~\bibnamefont
  {Qin}}, \bibinfo {author} {\bibfnamefont {W.}~\bibnamefont {Shi}}, \bibinfo
  {author} {\bibfnamefont {T.}~\bibnamefont {Ideue}}, \bibinfo {author}
  {\bibfnamefont {M.}~\bibnamefont {Yoshida}}, \bibinfo {author} {\bibfnamefont
  {A.}~\bibnamefont {Zak}}, \bibinfo {author} {\bibfnamefont {R.}~\bibnamefont
  {Tenne}}, \bibinfo {author} {\bibfnamefont {T.}~\bibnamefont {Kikitsu}},
  \bibinfo {author} {\bibfnamefont {D.}~\bibnamefont {Inoue}}, \bibinfo
  {author} {\bibfnamefont {D.}~\bibnamefont {Hashizume}},\ and\ \bibinfo
  {author} {\bibfnamefont {Y.}~\bibnamefont {Iwasa}},\ }\bibfield  {title}
  {\bibinfo {title} {Superconductivity in a chiral nanotube},\ }\href
  {https://doi.org/10.1038/ncomms14465} {\bibfield  {journal} {\bibinfo
  {journal} {Nat. Commun.}\ }\textbf {\bibinfo {volume} {8}},\ \bibinfo {pages}
  {14465} (\bibinfo {year} {2017})}\BibitemShut {NoStop}%
\bibitem [{\citenamefont {Yasuda}\ \emph {et~al.}(2019)\citenamefont {Yasuda},
  \citenamefont {Yasuda}, \citenamefont {Liang}, \citenamefont {Yoshimi},
  \citenamefont {Tsukazaki}, \citenamefont {Takahashi}, \citenamefont
  {Nagaosa}, \citenamefont {Kawasaki},\ and\ \citenamefont
  {Tokura}}]{Yasuda2019-jw}%
  \BibitemOpen
  \bibfield  {author} {\bibinfo {author} {\bibfnamefont {K.}~\bibnamefont
  {Yasuda}}, \bibinfo {author} {\bibfnamefont {H.}~\bibnamefont {Yasuda}},
  \bibinfo {author} {\bibfnamefont {T.}~\bibnamefont {Liang}}, \bibinfo
  {author} {\bibfnamefont {R.}~\bibnamefont {Yoshimi}}, \bibinfo {author}
  {\bibfnamefont {A.}~\bibnamefont {Tsukazaki}}, \bibinfo {author}
  {\bibfnamefont {K.~S.}\ \bibnamefont {Takahashi}}, \bibinfo {author}
  {\bibfnamefont {N.}~\bibnamefont {Nagaosa}}, \bibinfo {author} {\bibfnamefont
  {M.}~\bibnamefont {Kawasaki}},\ and\ \bibinfo {author} {\bibfnamefont
  {Y.}~\bibnamefont {Tokura}},\ }\bibfield  {title} {\bibinfo {title}
  {Nonreciprocal charge transport at topological insulator/superconductor
  interface},\ }\href {https://doi.org/10.1038/s41467-019-10658-3} {\bibfield
  {journal} {\bibinfo  {journal} {Nat. Commun.}\ }\textbf {\bibinfo {volume}
  {10}},\ \bibinfo {pages} {2734} (\bibinfo {year} {2019})}\BibitemShut
  {NoStop}%
\bibitem [{\citenamefont {Zhang}\ \emph {et~al.}(2020)\citenamefont {Zhang},
  \citenamefont {Xu}, \citenamefont {Zou}, \citenamefont {Ai}, \citenamefont
  {Dong}, \citenamefont {Huang}, \citenamefont {Leng}, \citenamefont {Liu},
  \citenamefont {Zhang}, \citenamefont {Jia}, \citenamefont {Peng},
  \citenamefont {Zhao}, \citenamefont {Yang}, \citenamefont {Li}, \citenamefont
  {Guo}, \citenamefont {Haigh}, \citenamefont {Nagaosa}, \citenamefont {Shen},\
  and\ \citenamefont {Xiu}}]{Zhang2020-al}%
  \BibitemOpen
  \bibfield  {author} {\bibinfo {author} {\bibfnamefont {E.}~\bibnamefont
  {Zhang}}, \bibinfo {author} {\bibfnamefont {X.}~\bibnamefont {Xu}}, \bibinfo
  {author} {\bibfnamefont {Y.-C.}\ \bibnamefont {Zou}}, \bibinfo {author}
  {\bibfnamefont {L.}~\bibnamefont {Ai}}, \bibinfo {author} {\bibfnamefont
  {X.}~\bibnamefont {Dong}}, \bibinfo {author} {\bibfnamefont {C.}~\bibnamefont
  {Huang}}, \bibinfo {author} {\bibfnamefont {P.}~\bibnamefont {Leng}},
  \bibinfo {author} {\bibfnamefont {S.}~\bibnamefont {Liu}}, \bibinfo {author}
  {\bibfnamefont {Y.}~\bibnamefont {Zhang}}, \bibinfo {author} {\bibfnamefont
  {Z.}~\bibnamefont {Jia}}, \bibinfo {author} {\bibfnamefont {X.}~\bibnamefont
  {Peng}}, \bibinfo {author} {\bibfnamefont {M.}~\bibnamefont {Zhao}}, \bibinfo
  {author} {\bibfnamefont {Y.}~\bibnamefont {Yang}}, \bibinfo {author}
  {\bibfnamefont {Z.}~\bibnamefont {Li}}, \bibinfo {author} {\bibfnamefont
  {H.}~\bibnamefont {Guo}}, \bibinfo {author} {\bibfnamefont {S.~J.}\
  \bibnamefont {Haigh}}, \bibinfo {author} {\bibfnamefont {N.}~\bibnamefont
  {Nagaosa}}, \bibinfo {author} {\bibfnamefont {J.}~\bibnamefont {Shen}},\ and\
  \bibinfo {author} {\bibfnamefont {F.}~\bibnamefont {Xiu}},\ }\bibfield
  {title} {\bibinfo {title} {Nonreciprocal superconducting {NbSe2} antenna},\
  }\href {https://doi.org/10.1038/s41467-020-19459-5} {\bibfield  {journal}
  {\bibinfo  {journal} {Nat. Commun.}\ }\textbf {\bibinfo {volume} {11}},\
  \bibinfo {pages} {5634} (\bibinfo {year} {2020})}\BibitemShut {NoStop}%
\bibitem [{\citenamefont {Ando}\ \emph {et~al.}(2020)\citenamefont {Ando},
  \citenamefont {Miyasaka}, \citenamefont {Li}, \citenamefont {Ishizuka},
  \citenamefont {Arakawa}, \citenamefont {Shiota}, \citenamefont {Moriyama},
  \citenamefont {Yanase},\ and\ \citenamefont {Ono}}]{Ando2020-om}%
  \BibitemOpen
  \bibfield  {author} {\bibinfo {author} {\bibfnamefont {F.}~\bibnamefont
  {Ando}}, \bibinfo {author} {\bibfnamefont {Y.}~\bibnamefont {Miyasaka}},
  \bibinfo {author} {\bibfnamefont {T.}~\bibnamefont {Li}}, \bibinfo {author}
  {\bibfnamefont {J.}~\bibnamefont {Ishizuka}}, \bibinfo {author}
  {\bibfnamefont {T.}~\bibnamefont {Arakawa}}, \bibinfo {author} {\bibfnamefont
  {Y.}~\bibnamefont {Shiota}}, \bibinfo {author} {\bibfnamefont
  {T.}~\bibnamefont {Moriyama}}, \bibinfo {author} {\bibfnamefont
  {Y.}~\bibnamefont {Yanase}},\ and\ \bibinfo {author} {\bibfnamefont
  {T.}~\bibnamefont {Ono}},\ }\bibfield  {title} {\bibinfo {title} {Observation
  of superconducting diode effect},\ }\href
  {https://doi.org/10.1038/s41586-020-2590-4} {\bibfield  {journal} {\bibinfo
  {journal} {Nature}\ }\textbf {\bibinfo {volume} {584}},\ \bibinfo {pages}
  {373} (\bibinfo {year} {2020})}\BibitemShut {NoStop}%
\bibitem [{\citenamefont {Lyu}\ \emph {et~al.}(2021)\citenamefont {Lyu},
  \citenamefont {Jiang}, \citenamefont {Wang}, \citenamefont {Xiao},
  \citenamefont {Dong}, \citenamefont {Chen}, \citenamefont {Milo{\v
  s}evi{\'c}}, \citenamefont {Wang}, \citenamefont {Divan}, \citenamefont
  {Pearson}, \citenamefont {Wu}, \citenamefont {Peeters},\ and\ \citenamefont
  {Kwok}}]{Lyu2021-sm}%
  \BibitemOpen
  \bibfield  {author} {\bibinfo {author} {\bibfnamefont {Y.-Y.}\ \bibnamefont
  {Lyu}}, \bibinfo {author} {\bibfnamefont {J.}~\bibnamefont {Jiang}}, \bibinfo
  {author} {\bibfnamefont {Y.-L.}\ \bibnamefont {Wang}}, \bibinfo {author}
  {\bibfnamefont {Z.-L.}\ \bibnamefont {Xiao}}, \bibinfo {author}
  {\bibfnamefont {S.}~\bibnamefont {Dong}}, \bibinfo {author} {\bibfnamefont
  {Q.-H.}\ \bibnamefont {Chen}}, \bibinfo {author} {\bibfnamefont {M.~V.}\
  \bibnamefont {Milo{\v s}evi{\'c}}}, \bibinfo {author} {\bibfnamefont
  {H.}~\bibnamefont {Wang}}, \bibinfo {author} {\bibfnamefont {R.}~\bibnamefont
  {Divan}}, \bibinfo {author} {\bibfnamefont {J.~E.}\ \bibnamefont {Pearson}},
  \bibinfo {author} {\bibfnamefont {P.}~\bibnamefont {Wu}}, \bibinfo {author}
  {\bibfnamefont {F.~M.}\ \bibnamefont {Peeters}},\ and\ \bibinfo {author}
  {\bibfnamefont {W.-K.}\ \bibnamefont {Kwok}},\ }\bibfield  {title} {\bibinfo
  {title} {Superconducting diode effect via conformal-mapped nanoholes},\
  }\href {https://doi.org/10.1038/s41467-021-23077-0} {\bibfield  {journal}
  {\bibinfo  {journal} {Nat. Commun.}\ }\textbf {\bibinfo {volume} {12}},\
  \bibinfo {pages} {2703} (\bibinfo {year} {2021})}\BibitemShut {NoStop}%
\bibitem [{\citenamefont {Wu}\ \emph {et~al.}(2022{\natexlab{a}})\citenamefont
  {Wu}, \citenamefont {Wang}, \citenamefont {Xu}, \citenamefont {Sivakumar},
  \citenamefont {Pasco}, \citenamefont {Filippozzi}, \citenamefont {Parkin},
  \citenamefont {Zeng}, \citenamefont {McQueen},\ and\ \citenamefont
  {Ali}}]{Wu2022-ey}%
  \BibitemOpen
  \bibfield  {author} {\bibinfo {author} {\bibfnamefont {H.}~\bibnamefont
  {Wu}}, \bibinfo {author} {\bibfnamefont {Y.}~\bibnamefont {Wang}}, \bibinfo
  {author} {\bibfnamefont {Y.}~\bibnamefont {Xu}}, \bibinfo {author}
  {\bibfnamefont {P.~K.}\ \bibnamefont {Sivakumar}}, \bibinfo {author}
  {\bibfnamefont {C.}~\bibnamefont {Pasco}}, \bibinfo {author} {\bibfnamefont
  {U.}~\bibnamefont {Filippozzi}}, \bibinfo {author} {\bibfnamefont {S.~S.~P.}\
  \bibnamefont {Parkin}}, \bibinfo {author} {\bibfnamefont {Y.-J.}\
  \bibnamefont {Zeng}}, \bibinfo {author} {\bibfnamefont {T.}~\bibnamefont
  {McQueen}},\ and\ \bibinfo {author} {\bibfnamefont {M.~N.}\ \bibnamefont
  {Ali}},\ }\bibfield  {title} {\bibinfo {title} {The field-free josephson
  diode in a van der waals heterostructure},\ }\href
  {https://doi.org/10.1038/s41586-022-04504-8} {\bibfield  {journal} {\bibinfo
  {journal} {Nature}\ }\textbf {\bibinfo {volume} {604}},\ \bibinfo {pages}
  {653} (\bibinfo {year} {2022}{\natexlab{a}})}\BibitemShut {NoStop}%
\bibitem [{\citenamefont {Baumgartner}\ \emph {et~al.}(2022)\citenamefont
  {Baumgartner}, \citenamefont {Fuchs}, \citenamefont {Costa}, \citenamefont
  {Reinhardt}, \citenamefont {Gronin}, \citenamefont {Gardner}, \citenamefont
  {Lindemann}, \citenamefont {Manfra}, \citenamefont {Faria~Junior},
  \citenamefont {Kochan}, \citenamefont {Fabian}, \citenamefont {Paradiso},\
  and\ \citenamefont {Strunk}}]{Baumgartner2022-lg}%
  \BibitemOpen
  \bibfield  {author} {\bibinfo {author} {\bibfnamefont {C.}~\bibnamefont
  {Baumgartner}}, \bibinfo {author} {\bibfnamefont {L.}~\bibnamefont {Fuchs}},
  \bibinfo {author} {\bibfnamefont {A.}~\bibnamefont {Costa}}, \bibinfo
  {author} {\bibfnamefont {S.}~\bibnamefont {Reinhardt}}, \bibinfo {author}
  {\bibfnamefont {S.}~\bibnamefont {Gronin}}, \bibinfo {author} {\bibfnamefont
  {G.~C.}\ \bibnamefont {Gardner}}, \bibinfo {author} {\bibfnamefont
  {T.}~\bibnamefont {Lindemann}}, \bibinfo {author} {\bibfnamefont {M.~J.}\
  \bibnamefont {Manfra}}, \bibinfo {author} {\bibfnamefont {P.~E.}\
  \bibnamefont {Faria~Junior}}, \bibinfo {author} {\bibfnamefont
  {D.}~\bibnamefont {Kochan}}, \bibinfo {author} {\bibfnamefont
  {J.}~\bibnamefont {Fabian}}, \bibinfo {author} {\bibfnamefont
  {N.}~\bibnamefont {Paradiso}},\ and\ \bibinfo {author} {\bibfnamefont
  {C.}~\bibnamefont {Strunk}},\ }\bibfield  {title} {\bibinfo {title}
  {Supercurrent rectification and magnetochiral effects in symmetric josephson
  junctions},\ }\href {https://doi.org/10.1038/s41565-021-01009-9} {\bibfield
  {journal} {\bibinfo  {journal} {Nat. Nanotechnol.}\ }\textbf {\bibinfo
  {volume} {17}},\ \bibinfo {pages} {39} (\bibinfo {year} {2022})}\BibitemShut
  {NoStop}%
\bibitem [{\citenamefont {Bauriedl}\ \emph {et~al.}(2022)\citenamefont
  {Bauriedl}, \citenamefont {B{\"a}uml}, \citenamefont {Fuchs}, \citenamefont
  {Baumgartner}, \citenamefont {Paulik}, \citenamefont {Bauer}, \citenamefont
  {Lin}, \citenamefont {Lupton}, \citenamefont {Taniguchi}, \citenamefont
  {Watanabe}, \citenamefont {Strunk},\ and\ \citenamefont
  {Paradiso}}]{Bauriedl2022-nq}%
  \BibitemOpen
  \bibfield  {author} {\bibinfo {author} {\bibfnamefont {L.}~\bibnamefont
  {Bauriedl}}, \bibinfo {author} {\bibfnamefont {C.}~\bibnamefont {B{\"a}uml}},
  \bibinfo {author} {\bibfnamefont {L.}~\bibnamefont {Fuchs}}, \bibinfo
  {author} {\bibfnamefont {C.}~\bibnamefont {Baumgartner}}, \bibinfo {author}
  {\bibfnamefont {N.}~\bibnamefont {Paulik}}, \bibinfo {author} {\bibfnamefont
  {J.~M.}\ \bibnamefont {Bauer}}, \bibinfo {author} {\bibfnamefont {K.-Q.}\
  \bibnamefont {Lin}}, \bibinfo {author} {\bibfnamefont {J.~M.}\ \bibnamefont
  {Lupton}}, \bibinfo {author} {\bibfnamefont {T.}~\bibnamefont {Taniguchi}},
  \bibinfo {author} {\bibfnamefont {K.}~\bibnamefont {Watanabe}}, \bibinfo
  {author} {\bibfnamefont {C.}~\bibnamefont {Strunk}},\ and\ \bibinfo {author}
  {\bibfnamefont {N.}~\bibnamefont {Paradiso}},\ }\bibfield  {title} {\bibinfo
  {title} {Supercurrent diode effect and magnetochiral anisotropy in few-layer
  {NbSe2}},\ }\href {https://doi.org/10.1038/s41467-022-31954-5} {\bibfield
  {journal} {\bibinfo  {journal} {Nat. Commun.}\ }\textbf {\bibinfo {volume}
  {13}},\ \bibinfo {pages} {4266} (\bibinfo {year} {2022})}\BibitemShut
  {NoStop}%
\bibitem [{\citenamefont {Lin}\ \emph {et~al.}(2022)\citenamefont {Lin},
  \citenamefont {Siriviboon}, \citenamefont {Scammell}, \citenamefont {Liu},
  \citenamefont {Rhodes}, \citenamefont {Watanabe}, \citenamefont {Taniguchi},
  \citenamefont {Hone}, \citenamefont {Scheurer},\ and\ \citenamefont
  {Li}}]{Lin2022-cz}%
  \BibitemOpen
  \bibfield  {author} {\bibinfo {author} {\bibfnamefont {J.-X.}\ \bibnamefont
  {Lin}}, \bibinfo {author} {\bibfnamefont {P.}~\bibnamefont {Siriviboon}},
  \bibinfo {author} {\bibfnamefont {H.~D.}\ \bibnamefont {Scammell}}, \bibinfo
  {author} {\bibfnamefont {S.}~\bibnamefont {Liu}}, \bibinfo {author}
  {\bibfnamefont {D.}~\bibnamefont {Rhodes}}, \bibinfo {author} {\bibfnamefont
  {K.}~\bibnamefont {Watanabe}}, \bibinfo {author} {\bibfnamefont
  {T.}~\bibnamefont {Taniguchi}}, \bibinfo {author} {\bibfnamefont
  {J.}~\bibnamefont {Hone}}, \bibinfo {author} {\bibfnamefont {M.~S.}\
  \bibnamefont {Scheurer}},\ and\ \bibinfo {author} {\bibfnamefont {J.~I.~A.}\
  \bibnamefont {Li}},\ }\bibfield  {title} {\bibinfo {title} {Zero-field
  superconducting diode effect in small-twist-angle trilayer graphene},\ }\href
  {https://doi.org/10.1038/s41567-022-01700-1} {\bibfield  {journal} {\bibinfo
  {journal} {Nat. Phys.}\ }\textbf {\bibinfo {volume} {18}},\ \bibinfo {pages}
  {1221} (\bibinfo {year} {2022})}\BibitemShut {NoStop}%
\bibitem [{\citenamefont {Narita}\ \emph {et~al.}(2022)\citenamefont {Narita},
  \citenamefont {Ishizuka}, \citenamefont {Kawarazaki}, \citenamefont {Kan},
  \citenamefont {Shiota}, \citenamefont {Moriyama}, \citenamefont {Shimakawa},
  \citenamefont {Ognev}, \citenamefont {Samardak}, \citenamefont {Yanase},\
  and\ \citenamefont {Ono}}]{Narita2022-od}%
  \BibitemOpen
  \bibfield  {author} {\bibinfo {author} {\bibfnamefont {H.}~\bibnamefont
  {Narita}}, \bibinfo {author} {\bibfnamefont {J.}~\bibnamefont {Ishizuka}},
  \bibinfo {author} {\bibfnamefont {R.}~\bibnamefont {Kawarazaki}}, \bibinfo
  {author} {\bibfnamefont {D.}~\bibnamefont {Kan}}, \bibinfo {author}
  {\bibfnamefont {Y.}~\bibnamefont {Shiota}}, \bibinfo {author} {\bibfnamefont
  {T.}~\bibnamefont {Moriyama}}, \bibinfo {author} {\bibfnamefont
  {Y.}~\bibnamefont {Shimakawa}}, \bibinfo {author} {\bibfnamefont {A.~V.}\
  \bibnamefont {Ognev}}, \bibinfo {author} {\bibfnamefont {A.~S.}\ \bibnamefont
  {Samardak}}, \bibinfo {author} {\bibfnamefont {Y.}~\bibnamefont {Yanase}},\
  and\ \bibinfo {author} {\bibfnamefont {T.}~\bibnamefont {Ono}},\ }\bibfield
  {title} {\bibinfo {title} {Field-free superconducting diode effect in
  noncentrosymmetric superconductor/ferromagnet multilayers},\ }\href
  {https://doi.org/10.1038/s41565-022-01159-4} {\bibfield  {journal} {\bibinfo
  {journal} {Nat. Nanotechnol.}\ }\textbf {\bibinfo {volume} {17}},\ \bibinfo
  {pages} {823} (\bibinfo {year} {2022})}\BibitemShut {NoStop}%
\bibitem [{\citenamefont {Mizuno}\ \emph {et~al.}(2022)\citenamefont {Mizuno},
  \citenamefont {Tsuchiya}, \citenamefont {Awaji},\ and\ \citenamefont
  {Yoshida}}]{Mizuno2022-bd}%
  \BibitemOpen
  \bibfield  {author} {\bibinfo {author} {\bibfnamefont {A.}~\bibnamefont
  {Mizuno}}, \bibinfo {author} {\bibfnamefont {Y.}~\bibnamefont {Tsuchiya}},
  \bibinfo {author} {\bibfnamefont {S.}~\bibnamefont {Awaji}},\ and\ \bibinfo
  {author} {\bibfnamefont {Y.}~\bibnamefont {Yoshida}},\ }\bibfield  {title}
  {\bibinfo {title} {Rectification at various temperatures in {YBa2Cu3Oy}
  coated conductors with {PrBa2Cu3Oy} buffer layers},\ }\href
  {https://doi.org/10.1109/TASC.2022.3154325} {\bibfield  {journal} {\bibinfo
  {journal} {IEEE Trans. Appl. Supercond.}\ }\textbf {\bibinfo {volume} {32}},\
  \bibinfo {pages} {1} (\bibinfo {year} {2022})}\BibitemShut {NoStop}%
\bibitem [{\citenamefont {Ideue}\ and\ \citenamefont
  {Iwasa}(2020)}]{Ideue2020-kg}%
  \BibitemOpen
  \bibfield  {author} {\bibinfo {author} {\bibfnamefont {T.}~\bibnamefont
  {Ideue}}\ and\ \bibinfo {author} {\bibfnamefont {Y.}~\bibnamefont {Iwasa}},\
  }\bibfield  {title} {\bibinfo {title} {One-way supercurrent achieved in an
  electrically polar film},\ }\href
  {https://doi.org/10.1038/d41586-020-02380-8} {\bibfield  {journal} {\bibinfo
  {journal} {Nature}\ }\textbf {\bibinfo {volume} {584}},\ \bibinfo {pages}
  {349} (\bibinfo {year} {2020})}\BibitemShut {NoStop}%
\bibitem [{\citenamefont {Yuan}\ and\ \citenamefont {Fu}(2022)}]{Yuan2022-pz}%
  \BibitemOpen
  \bibfield  {author} {\bibinfo {author} {\bibfnamefont {N.~F.~Q.}\
  \bibnamefont {Yuan}}\ and\ \bibinfo {author} {\bibfnamefont {L.}~\bibnamefont
  {Fu}},\ }\bibfield  {title} {\bibinfo {title} {Supercurrent diode effect and
  finite momentum superconductors},\ }\href
  {https://doi.org/10.1073/pnas.2119548119} {\bibfield  {journal} {\bibinfo
  {journal} {Proceedings of the National Academy of Sciences}\ }\textbf
  {\bibinfo {volume} {119}},\ \bibinfo {pages} {e2119548119} (\bibinfo {year}
  {2022})},\ \Eprint {https://arxiv.org/abs/2106.01909} {arXiv:2106.01909
  [cond-mat.supr-con]} \BibitemShut {NoStop}%
\bibitem [{\citenamefont {Daido}\ \emph {et~al.}(2022)\citenamefont {Daido},
  \citenamefont {Ikeda},\ and\ \citenamefont {Yanase}}]{Daido2022-ox}%
  \BibitemOpen
  \bibfield  {author} {\bibinfo {author} {\bibfnamefont {A.}~\bibnamefont
  {Daido}}, \bibinfo {author} {\bibfnamefont {Y.}~\bibnamefont {Ikeda}},\ and\
  \bibinfo {author} {\bibfnamefont {Y.}~\bibnamefont {Yanase}},\ }\bibfield
  {title} {\bibinfo {title} {Intrinsic superconducting diode effect},\ }\href
  {https://doi.org/10.1103/PhysRevLett.128.037001} {\bibfield  {journal}
  {\bibinfo  {journal} {Phys. Rev. Lett.}\ }\textbf {\bibinfo {volume} {128}},\
  \bibinfo {pages} {037001} (\bibinfo {year} {2022})}\BibitemShut {NoStop}%
\bibitem [{\citenamefont {He}\ \emph {et~al.}(2022)\citenamefont {He},
  \citenamefont {Tanaka},\ and\ \citenamefont {Nagaosa}}]{He2022-px}%
  \BibitemOpen
  \bibfield  {author} {\bibinfo {author} {\bibfnamefont {J.~J.}\ \bibnamefont
  {He}}, \bibinfo {author} {\bibfnamefont {Y.}~\bibnamefont {Tanaka}},\ and\
  \bibinfo {author} {\bibfnamefont {N.}~\bibnamefont {Nagaosa}},\ }\bibfield
  {title} {\bibinfo {title} {A phenomenological theory of superconductor
  diodes},\ }\href {https://doi.org/10.1088/1367-2630/ac6766} {\bibfield
  {journal} {\bibinfo  {journal} {New J. Phys.}\ }\textbf {\bibinfo {volume}
  {24}},\ \bibinfo {pages} {053014} (\bibinfo {year} {2022})}\BibitemShut
  {NoStop}%
\bibitem [{\citenamefont {Ili{\'c}}\ and\ \citenamefont
  {Bergeret}(2022)}]{Ilic2022-kh}%
  \BibitemOpen
  \bibfield  {author} {\bibinfo {author} {\bibfnamefont {S.}~\bibnamefont
  {Ili{\'c}}}\ and\ \bibinfo {author} {\bibfnamefont {F.~S.}\ \bibnamefont
  {Bergeret}},\ }\bibfield  {title} {\bibinfo {title} {Theory of the
  supercurrent diode effect in rashba superconductors with arbitrary
  disorder},\ }\href {https://doi.org/10.1103/PhysRevLett.128.177001}
  {\bibfield  {journal} {\bibinfo  {journal} {Phys. Rev. Lett.}\ }\textbf
  {\bibinfo {volume} {128}},\ \bibinfo {pages} {177001} (\bibinfo {year}
  {2022})}\BibitemShut {NoStop}%
\bibitem [{\citenamefont {He}\ \emph {et~al.}(2023)\citenamefont {He},
  \citenamefont {Tanaka},\ and\ \citenamefont {Nagaosa}}]{He2023-qo}%
  \BibitemOpen
  \bibfield  {author} {\bibinfo {author} {\bibfnamefont {J.~J.}\ \bibnamefont
  {He}}, \bibinfo {author} {\bibfnamefont {Y.}~\bibnamefont {Tanaka}},\ and\
  \bibinfo {author} {\bibfnamefont {N.}~\bibnamefont {Nagaosa}},\ }\bibfield
  {title} {\bibinfo {title} {The supercurrent diode effect and nonreciprocal
  paraconductivity due to the chiral structure of nanotubes},\ }\href
  {https://doi.org/10.1038/s41467-023-39083-3} {\bibfield  {journal} {\bibinfo
  {journal} {Nat. Commun.}\ }\textbf {\bibinfo {volume} {14}},\ \bibinfo
  {pages} {3330} (\bibinfo {year} {2023})},\ \Eprint
  {https://arxiv.org/abs/2212.02183} {arXiv:2212.02183 [cond-mat.supr-con]}
  \BibitemShut {NoStop}%
\bibitem [{\citenamefont {Du}\ \emph {et~al.}(2021)\citenamefont {Du},
  \citenamefont {Lu},\ and\ \citenamefont {Xie}}]{Du2021-fg}%
  \BibitemOpen
  \bibfield  {author} {\bibinfo {author} {\bibfnamefont {Z.~Z.}\ \bibnamefont
  {Du}}, \bibinfo {author} {\bibfnamefont {H.-Z.}\ \bibnamefont {Lu}},\ and\
  \bibinfo {author} {\bibfnamefont {X.~C.}\ \bibnamefont {Xie}},\ }\bibfield
  {title} {\bibinfo {title} {Nonlinear hall effects},\ }\href
  {https://doi.org/10.1038/s42254-021-00359-6} {\bibfield  {journal} {\bibinfo
  {journal} {Nature Reviews Physics}\ }\textbf {\bibinfo {volume} {3}},\
  \bibinfo {pages} {744} (\bibinfo {year} {2021})}\BibitemShut {NoStop}%
\bibitem [{\citenamefont {Sodemann}\ and\ \citenamefont
  {Fu}(2015)}]{Sodemann2015-vj}%
  \BibitemOpen
  \bibfield  {author} {\bibinfo {author} {\bibfnamefont {I.}~\bibnamefont
  {Sodemann}}\ and\ \bibinfo {author} {\bibfnamefont {L.}~\bibnamefont {Fu}},\
  }\bibfield  {title} {\bibinfo {title} {Quantum nonlinear hall effect induced
  by berry curvature dipole in {Time-Reversal} invariant materials},\ }\href
  {https://doi.org/10.1103/PhysRevLett.115.216806} {\bibfield  {journal}
  {\bibinfo  {journal} {Phys. Rev. Lett.}\ }\textbf {\bibinfo {volume} {115}},\
  \bibinfo {pages} {216806} (\bibinfo {year} {2015})}\BibitemShut {NoStop}%
\bibitem [{\citenamefont {Ma}\ \emph {et~al.}(2019)\citenamefont {Ma},
  \citenamefont {Xu}, \citenamefont {Shen}, \citenamefont {MacNeill},
  \citenamefont {Fatemi}, \citenamefont {Chang}, \citenamefont {Mier~Valdivia},
  \citenamefont {Wu}, \citenamefont {Du}, \citenamefont {Hsu}, \citenamefont
  {Fang}, \citenamefont {Gibson}, \citenamefont {Watanabe}, \citenamefont
  {Taniguchi}, \citenamefont {Cava}, \citenamefont {Kaxiras}, \citenamefont
  {Lu}, \citenamefont {Lin}, \citenamefont {Fu}, \citenamefont {Gedik},\ and\
  \citenamefont {Jarillo-Herrero}}]{Ma2019-om}%
  \BibitemOpen
  \bibfield  {author} {\bibinfo {author} {\bibfnamefont {Q.}~\bibnamefont
  {Ma}}, \bibinfo {author} {\bibfnamefont {S.-Y.}\ \bibnamefont {Xu}}, \bibinfo
  {author} {\bibfnamefont {H.}~\bibnamefont {Shen}}, \bibinfo {author}
  {\bibfnamefont {D.}~\bibnamefont {MacNeill}}, \bibinfo {author}
  {\bibfnamefont {V.}~\bibnamefont {Fatemi}}, \bibinfo {author} {\bibfnamefont
  {T.-R.}\ \bibnamefont {Chang}}, \bibinfo {author} {\bibfnamefont {A.~M.}\
  \bibnamefont {Mier~Valdivia}}, \bibinfo {author} {\bibfnamefont
  {S.}~\bibnamefont {Wu}}, \bibinfo {author} {\bibfnamefont {Z.}~\bibnamefont
  {Du}}, \bibinfo {author} {\bibfnamefont {C.-H.}\ \bibnamefont {Hsu}},
  \bibinfo {author} {\bibfnamefont {S.}~\bibnamefont {Fang}}, \bibinfo {author}
  {\bibfnamefont {Q.~D.}\ \bibnamefont {Gibson}}, \bibinfo {author}
  {\bibfnamefont {K.}~\bibnamefont {Watanabe}}, \bibinfo {author}
  {\bibfnamefont {T.}~\bibnamefont {Taniguchi}}, \bibinfo {author}
  {\bibfnamefont {R.~J.}\ \bibnamefont {Cava}}, \bibinfo {author}
  {\bibfnamefont {E.}~\bibnamefont {Kaxiras}}, \bibinfo {author} {\bibfnamefont
  {H.-Z.}\ \bibnamefont {Lu}}, \bibinfo {author} {\bibfnamefont
  {H.}~\bibnamefont {Lin}}, \bibinfo {author} {\bibfnamefont {L.}~\bibnamefont
  {Fu}}, \bibinfo {author} {\bibfnamefont {N.}~\bibnamefont {Gedik}},\ and\
  \bibinfo {author} {\bibfnamefont {P.}~\bibnamefont {Jarillo-Herrero}},\
  }\bibfield  {title} {\bibinfo {title} {Observation of the nonlinear hall
  effect under time-reversal-symmetric conditions},\ }\href
  {https://doi.org/10.1038/s41586-018-0807-6} {\bibfield  {journal} {\bibinfo
  {journal} {Nature}\ }\textbf {\bibinfo {volume} {565}},\ \bibinfo {pages}
  {337} (\bibinfo {year} {2019})}\BibitemShut {NoStop}%
\bibitem [{\citenamefont {Kang}\ \emph {et~al.}(2019)\citenamefont {Kang},
  \citenamefont {Li}, \citenamefont {Sohn}, \citenamefont {Shan},\ and\
  \citenamefont {Mak}}]{Kang2019-dr}%
  \BibitemOpen
  \bibfield  {author} {\bibinfo {author} {\bibfnamefont {K.}~\bibnamefont
  {Kang}}, \bibinfo {author} {\bibfnamefont {T.}~\bibnamefont {Li}}, \bibinfo
  {author} {\bibfnamefont {E.}~\bibnamefont {Sohn}}, \bibinfo {author}
  {\bibfnamefont {J.}~\bibnamefont {Shan}},\ and\ \bibinfo {author}
  {\bibfnamefont {K.~F.}\ \bibnamefont {Mak}},\ }\bibfield  {title} {\bibinfo
  {title} {Nonlinear anomalous hall effect in few-layer {WTe2}},\ }\href
  {https://doi.org/10.1038/s41563-019-0294-7} {\bibfield  {journal} {\bibinfo
  {journal} {Nat. Mater.}\ }\textbf {\bibinfo {volume} {18}},\ \bibinfo {pages}
  {324} (\bibinfo {year} {2019})}\BibitemShut {NoStop}%
\bibitem [{\citenamefont {Kumar}\ \emph {et~al.}(2021)\citenamefont {Kumar},
  \citenamefont {Hsu}, \citenamefont {Sharma}, \citenamefont {Chang},
  \citenamefont {Yu}, \citenamefont {Wang}, \citenamefont {Eda}, \citenamefont
  {Liang},\ and\ \citenamefont {Yang}}]{Kumar2021-wf}%
  \BibitemOpen
  \bibfield  {author} {\bibinfo {author} {\bibfnamefont {D.}~\bibnamefont
  {Kumar}}, \bibinfo {author} {\bibfnamefont {C.-H.}\ \bibnamefont {Hsu}},
  \bibinfo {author} {\bibfnamefont {R.}~\bibnamefont {Sharma}}, \bibinfo
  {author} {\bibfnamefont {T.-R.}\ \bibnamefont {Chang}}, \bibinfo {author}
  {\bibfnamefont {P.}~\bibnamefont {Yu}}, \bibinfo {author} {\bibfnamefont
  {J.}~\bibnamefont {Wang}}, \bibinfo {author} {\bibfnamefont {G.}~\bibnamefont
  {Eda}}, \bibinfo {author} {\bibfnamefont {G.}~\bibnamefont {Liang}},\ and\
  \bibinfo {author} {\bibfnamefont {H.}~\bibnamefont {Yang}},\ }\bibfield
  {title} {\bibinfo {title} {Room-temperature nonlinear hall effect and
  wireless radiofrequency rectification in weyl semimetal {TaIrTe4}},\ }\href
  {https://doi.org/10.1038/s41565-020-00839-3} {\bibfield  {journal} {\bibinfo
  {journal} {Nat. Nanotechnol.}\ }\textbf {\bibinfo {volume} {16}},\ \bibinfo
  {pages} {421} (\bibinfo {year} {2021})}\BibitemShut {NoStop}%
\bibitem [{\citenamefont {Itahashi}\ \emph {et~al.}(2022)\citenamefont
  {Itahashi}, \citenamefont {Ideue}, \citenamefont {Hoshino}, \citenamefont
  {Goto}, \citenamefont {Namiki}, \citenamefont {Sasagawa},\ and\ \citenamefont
  {Iwasa}}]{Itahashi2022-ja}%
  \BibitemOpen
  \bibfield  {author} {\bibinfo {author} {\bibfnamefont {Y.~M.}\ \bibnamefont
  {Itahashi}}, \bibinfo {author} {\bibfnamefont {T.}~\bibnamefont {Ideue}},
  \bibinfo {author} {\bibfnamefont {S.}~\bibnamefont {Hoshino}}, \bibinfo
  {author} {\bibfnamefont {C.}~\bibnamefont {Goto}}, \bibinfo {author}
  {\bibfnamefont {H.}~\bibnamefont {Namiki}}, \bibinfo {author} {\bibfnamefont
  {T.}~\bibnamefont {Sasagawa}},\ and\ \bibinfo {author} {\bibfnamefont
  {Y.}~\bibnamefont {Iwasa}},\ }\bibfield  {title} {\bibinfo {title} {Giant
  second harmonic transport under time-reversal symmetry in a trigonal
  superconductor},\ }\href {https://doi.org/10.1038/s41467-022-29314-4}
  {\bibfield  {journal} {\bibinfo  {journal} {Nat. Commun.}\ }\textbf {\bibinfo
  {volume} {13}},\ \bibinfo {pages} {1659} (\bibinfo {year}
  {2022})}\BibitemShut {NoStop}%
\bibitem [{\citenamefont {Zhang}\ and\ \citenamefont
  {Fu}(2021)}]{Zhang2021-hz}%
  \BibitemOpen
  \bibfield  {author} {\bibinfo {author} {\bibfnamefont {Y.}~\bibnamefont
  {Zhang}}\ and\ \bibinfo {author} {\bibfnamefont {L.}~\bibnamefont {Fu}},\
  }\bibfield  {title} {\bibinfo {title} {Terahertz detection based on nonlinear
  hall effect without magnetic field},\ }\bibfield  {journal} {\bibinfo
  {journal} {Proc. Natl. Acad. Sci. U. S. A.}\ }\textbf {\bibinfo {volume}
  {118}},\ \href {https://doi.org/10.1073/pnas.2100736118}
  {10.1073/pnas.2100736118} (\bibinfo {year} {2021})\BibitemShut {NoStop}%
\bibitem [{\citenamefont {Toshio}\ and\ \citenamefont
  {Kawakami}(2022)}]{Toshio2022-de}%
  \BibitemOpen
  \bibfield  {author} {\bibinfo {author} {\bibfnamefont {R.}~\bibnamefont
  {Toshio}}\ and\ \bibinfo {author} {\bibfnamefont {N.}~\bibnamefont
  {Kawakami}},\ }\bibfield  {title} {\bibinfo {title} {Plasmonic quantum
  nonlinear hall effect in noncentrosymmetric two-dimensional materials},\
  }\href {https://doi.org/10.1103/PhysRevB.106.L201301} {\bibfield  {journal}
  {\bibinfo  {journal} {Phys. Rev. B Condens. Matter}\ }\textbf {\bibinfo
  {volume} {106}},\ \bibinfo {pages} {L201301} (\bibinfo {year}
  {2022})}\BibitemShut {NoStop}%
\bibitem [{\citenamefont {Wakatsuki}\ and\ \citenamefont
  {Nagaosa}(2018)}]{Wakatsuki2018-ll}%
  \BibitemOpen
  \bibfield  {author} {\bibinfo {author} {\bibfnamefont {R.}~\bibnamefont
  {Wakatsuki}}\ and\ \bibinfo {author} {\bibfnamefont {N.}~\bibnamefont
  {Nagaosa}},\ }\bibfield  {title} {\bibinfo {title} {Nonreciprocal current in
  noncentrosymmetric rashba superconductors},\ }\href
  {https://doi.org/10.1103/PhysRevLett.121.026601} {\bibfield  {journal}
  {\bibinfo  {journal} {Phys. Rev. Lett.}\ }\textbf {\bibinfo {volume} {121}},\
  \bibinfo {pages} {026601} (\bibinfo {year} {2018})}\BibitemShut {NoStop}%
\bibitem [{\citenamefont {Hoshino}\ \emph {et~al.}(2018)\citenamefont
  {Hoshino}, \citenamefont {Wakatsuki}, \citenamefont {Hamamoto},\ and\
  \citenamefont {Nagaosa}}]{Hoshino2018-sa}%
  \BibitemOpen
  \bibfield  {author} {\bibinfo {author} {\bibfnamefont {S.}~\bibnamefont
  {Hoshino}}, \bibinfo {author} {\bibfnamefont {R.}~\bibnamefont {Wakatsuki}},
  \bibinfo {author} {\bibfnamefont {K.}~\bibnamefont {Hamamoto}},\ and\
  \bibinfo {author} {\bibfnamefont {N.}~\bibnamefont {Nagaosa}},\ }\bibfield
  {title} {\bibinfo {title} {Nonreciprocal charge transport in two-dimensional
  noncentrosymmetric superconductors},\ }\href
  {https://doi.org/10.1103/PhysRevB.98.054510} {\bibfield  {journal} {\bibinfo
  {journal} {Phys. Rev. B Condens. Matter}\ }\textbf {\bibinfo {volume} {98}},\
  \bibinfo {pages} {054510} (\bibinfo {year} {2018})}\BibitemShut {NoStop}%
\bibitem [{\citenamefont {Wu}\ \emph {et~al.}(2022{\natexlab{b}})\citenamefont
  {Wu}, \citenamefont {Wang}, \citenamefont {Zhou}, \citenamefont {Wang},
  \citenamefont {Dong}, \citenamefont {He}, \citenamefont {Ding}, \citenamefont
  {Teng}, \citenamefont {Zhang}, \citenamefont {Li}, \citenamefont {Zhao},
  \citenamefont {Zhang}, \citenamefont {Liu}, \citenamefont {Qi}, \citenamefont
  {Watanabe}, \citenamefont {Taniguchi},\ and\ \citenamefont {Li}}]{Wu2022-kw}%
  \BibitemOpen
  \bibfield  {author} {\bibinfo {author} {\bibfnamefont {Y.}~\bibnamefont
  {Wu}}, \bibinfo {author} {\bibfnamefont {Q.}~\bibnamefont {Wang}}, \bibinfo
  {author} {\bibfnamefont {X.}~\bibnamefont {Zhou}}, \bibinfo {author}
  {\bibfnamefont {J.}~\bibnamefont {Wang}}, \bibinfo {author} {\bibfnamefont
  {P.}~\bibnamefont {Dong}}, \bibinfo {author} {\bibfnamefont {J.}~\bibnamefont
  {He}}, \bibinfo {author} {\bibfnamefont {Y.}~\bibnamefont {Ding}}, \bibinfo
  {author} {\bibfnamefont {B.}~\bibnamefont {Teng}}, \bibinfo {author}
  {\bibfnamefont {Y.}~\bibnamefont {Zhang}}, \bibinfo {author} {\bibfnamefont
  {Y.}~\bibnamefont {Li}}, \bibinfo {author} {\bibfnamefont {C.}~\bibnamefont
  {Zhao}}, \bibinfo {author} {\bibfnamefont {H.}~\bibnamefont {Zhang}},
  \bibinfo {author} {\bibfnamefont {J.}~\bibnamefont {Liu}}, \bibinfo {author}
  {\bibfnamefont {Y.}~\bibnamefont {Qi}}, \bibinfo {author} {\bibfnamefont
  {K.}~\bibnamefont {Watanabe}}, \bibinfo {author} {\bibfnamefont
  {T.}~\bibnamefont {Taniguchi}},\ and\ \bibinfo {author} {\bibfnamefont
  {J.}~\bibnamefont {Li}},\ }\bibfield  {title} {\bibinfo {title}
  {Nonreciprocal charge transport in topological kagome superconductor
  {CsV3Sb5}},\ }\href {https://doi.org/10.1038/s41535-022-00516-9} {\bibfield
  {journal} {\bibinfo  {journal} {npj Quantum Materials}\ }\textbf {\bibinfo
  {volume} {7}},\ \bibinfo {pages} {1} (\bibinfo {year}
  {2022}{\natexlab{b}})}\BibitemShut {NoStop}%
\bibitem [{\citenamefont {Guo}\ \emph {et~al.}(2022)\citenamefont {Guo},
  \citenamefont {Putzke}, \citenamefont {Konyzheva}, \citenamefont {Huang},
  \citenamefont {Gutierrez-Amigo}, \citenamefont {Errea}, \citenamefont {Chen},
  \citenamefont {Vergniory}, \citenamefont {Felser}, \citenamefont {Fischer},
  \citenamefont {Neupert},\ and\ \citenamefont {Moll}}]{Guo2022-ey}%
  \BibitemOpen
  \bibfield  {author} {\bibinfo {author} {\bibfnamefont {C.}~\bibnamefont
  {Guo}}, \bibinfo {author} {\bibfnamefont {C.}~\bibnamefont {Putzke}},
  \bibinfo {author} {\bibfnamefont {S.}~\bibnamefont {Konyzheva}}, \bibinfo
  {author} {\bibfnamefont {X.}~\bibnamefont {Huang}}, \bibinfo {author}
  {\bibfnamefont {M.}~\bibnamefont {Gutierrez-Amigo}}, \bibinfo {author}
  {\bibfnamefont {I.}~\bibnamefont {Errea}}, \bibinfo {author} {\bibfnamefont
  {D.}~\bibnamefont {Chen}}, \bibinfo {author} {\bibfnamefont {M.~G.}\
  \bibnamefont {Vergniory}}, \bibinfo {author} {\bibfnamefont {C.}~\bibnamefont
  {Felser}}, \bibinfo {author} {\bibfnamefont {M.~H.}\ \bibnamefont {Fischer}},
  \bibinfo {author} {\bibfnamefont {T.}~\bibnamefont {Neupert}},\ and\ \bibinfo
  {author} {\bibfnamefont {P.~J.~W.}\ \bibnamefont {Moll}},\ }\bibfield
  {title} {\bibinfo {title} {Switchable chiral transport in charge-ordered
  kagome metal {CsV3Sb5}},\ }\href {https://doi.org/10.1038/s41586-022-05127-9}
  {\bibfield  {journal} {\bibinfo  {journal} {Nature}\ }\textbf {\bibinfo
  {volume} {611}},\ \bibinfo {pages} {461} (\bibinfo {year}
  {2022})}\BibitemShut {NoStop}%
\bibitem [{\citenamefont {Bauer}\ and\ \citenamefont
  {Sigrist}(2012)}]{Bauer2012-xi}%
  \BibitemOpen
  \bibfield  {author} {\bibinfo {author} {\bibfnamefont {E.}~\bibnamefont
  {Bauer}}\ and\ \bibinfo {author} {\bibfnamefont {M.}~\bibnamefont
  {Sigrist}},\ }\href@noop {} {\emph {\bibinfo {title} {{Non-Centrosymmetric}
  Superconductors: Introduction and Overview}}}\ (\bibinfo  {publisher}
  {Springer Science \& Business Media},\ \bibinfo {year} {2012})\BibitemShut
  {NoStop}%
\bibitem [{\citenamefont {Smidman}\ \emph {et~al.}(2017)\citenamefont
  {Smidman}, \citenamefont {Salamon}, \citenamefont {Yuan},\ and\ \citenamefont
  {Agterberg}}]{Smidman2017-hb}%
  \BibitemOpen
  \bibfield  {author} {\bibinfo {author} {\bibfnamefont {M.}~\bibnamefont
  {Smidman}}, \bibinfo {author} {\bibfnamefont {M.~B.}\ \bibnamefont
  {Salamon}}, \bibinfo {author} {\bibfnamefont {H.~Q.}\ \bibnamefont {Yuan}},\
  and\ \bibinfo {author} {\bibfnamefont {D.~F.}\ \bibnamefont {Agterberg}},\
  }\bibfield  {title} {\bibinfo {title} {Superconductivity and spin--orbit
  coupling in non-centrosymmetric materials: a review},\ }\href
  {https://doi.org/10.1088/1361-6633/80/3/036501} {\bibfield  {journal}
  {\bibinfo  {journal} {Rep. Prog. Phys.}\ }\textbf {\bibinfo {volume} {80}},\
  \bibinfo {pages} {036501} (\bibinfo {year} {2017})}\BibitemShut {NoStop}%
\bibitem [{\citenamefont {Agterberg}(2003)}]{Agterberg2003-jn}%
  \BibitemOpen
  \bibfield  {author} {\bibinfo {author} {\bibfnamefont {D.~F.}\ \bibnamefont
  {Agterberg}},\ }\bibfield  {title} {\bibinfo {title} {Novel magnetic field
  effects in unconventional superconductors},\ }\href
  {https://doi.org/10.1016/s0921-4534(03)00634-8} {\bibfield  {journal}
  {\bibinfo  {journal} {Physica C Supercond.}\ }\textbf {\bibinfo {volume}
  {387}},\ \bibinfo {pages} {13} (\bibinfo {year} {2003})}\BibitemShut
  {NoStop}%
\bibitem [{\citenamefont {Barzykin}\ and\ \citenamefont
  {Gor'kov}(2002)}]{Barzykin2002-eh}%
  \BibitemOpen
  \bibfield  {author} {\bibinfo {author} {\bibfnamefont {V.}~\bibnamefont
  {Barzykin}}\ and\ \bibinfo {author} {\bibfnamefont {L.~P.}\ \bibnamefont
  {Gor'kov}},\ }\bibfield  {title} {\bibinfo {title} {Inhomogeneous stripe
  phase revisited for surface superconductivity},\ }\href
  {https://doi.org/10.1103/PhysRevLett.89.227002} {\bibfield  {journal}
  {\bibinfo  {journal} {Phys. Rev. Lett.}\ }\textbf {\bibinfo {volume} {89}},\
  \bibinfo {pages} {227002} (\bibinfo {year} {2002})}\BibitemShut {NoStop}%
\bibitem [{\citenamefont {Dimitrova}\ and\ \citenamefont
  {Feigel'man}(2003)}]{Dimitrova2003-mo}%
  \BibitemOpen
  \bibfield  {author} {\bibinfo {author} {\bibfnamefont {O.~V.}\ \bibnamefont
  {Dimitrova}}\ and\ \bibinfo {author} {\bibfnamefont {M.~V.}\ \bibnamefont
  {Feigel'man}},\ }\bibfield  {title} {\bibinfo {title} {Phase diagram of a
  surface superconductor in parallel magnetic field},\ }\href
  {https://doi.org/10.1134/1.1644308} {\bibfield  {journal} {\bibinfo
  {journal} {JETP Lett.}\ }\textbf {\bibinfo {volume} {78}},\ \bibinfo {pages}
  {637} (\bibinfo {year} {2003})}\BibitemShut {NoStop}%
\bibitem [{\citenamefont {Kaur}\ \emph {et~al.}(2005)\citenamefont {Kaur},
  \citenamefont {Agterberg},\ and\ \citenamefont {Sigrist}}]{Kaur2005-jf}%
  \BibitemOpen
  \bibfield  {author} {\bibinfo {author} {\bibfnamefont {R.~P.}\ \bibnamefont
  {Kaur}}, \bibinfo {author} {\bibfnamefont {D.~F.}\ \bibnamefont
  {Agterberg}},\ and\ \bibinfo {author} {\bibfnamefont {M.}~\bibnamefont
  {Sigrist}},\ }\bibfield  {title} {\bibinfo {title} {Helical vortex phase in
  the noncentrosymmetric {CePt3Si}},\ }\href
  {https://doi.org/10.1103/PhysRevLett.94.137002} {\bibfield  {journal}
  {\bibinfo  {journal} {Phys. Rev. Lett.}\ }\textbf {\bibinfo {volume} {94}},\
  \bibinfo {pages} {137002} (\bibinfo {year} {2005})}\BibitemShut {NoStop}%
\bibitem [{\citenamefont {Agterberg}\ and\ \citenamefont
  {Kaur}(2007)}]{Agterberg2007-vl}%
  \BibitemOpen
  \bibfield  {author} {\bibinfo {author} {\bibfnamefont {D.~F.}\ \bibnamefont
  {Agterberg}}\ and\ \bibinfo {author} {\bibfnamefont {R.~P.}\ \bibnamefont
  {Kaur}},\ }\bibfield  {title} {\bibinfo {title} {Magnetic-field-induced
  helical and stripe phases in rashba superconductors},\ }\href
  {https://doi.org/10.1103/PhysRevB.75.064511} {\bibfield  {journal} {\bibinfo
  {journal} {Phys. Rev. B Condens. Matter}\ }\textbf {\bibinfo {volume} {75}},\
  \bibinfo {pages} {064511} (\bibinfo {year} {2007})}\BibitemShut {NoStop}%
\bibitem [{\citenamefont {Dimitrova}\ and\ \citenamefont
  {Feigel'man}(2007)}]{Dimitrova2007-hp}%
  \BibitemOpen
  \bibfield  {author} {\bibinfo {author} {\bibfnamefont {O.}~\bibnamefont
  {Dimitrova}}\ and\ \bibinfo {author} {\bibfnamefont {M.~V.}\ \bibnamefont
  {Feigel'man}},\ }\bibfield  {title} {\bibinfo {title} {Theory of a
  two-dimensional superconductor with broken inversion symmetry},\ }\href
  {https://doi.org/10.1103/PhysRevB.76.014522} {\bibfield  {journal} {\bibinfo
  {journal} {Phys. Rev. B Condens. Matter}\ }\textbf {\bibinfo {volume} {76}},\
  \bibinfo {pages} {014522} (\bibinfo {year} {2007})}\BibitemShut {NoStop}%
\bibitem [{\citenamefont {Samokhin}(2008)}]{Samokhin2008-nv}%
  \BibitemOpen
  \bibfield  {author} {\bibinfo {author} {\bibfnamefont {K.~V.}\ \bibnamefont
  {Samokhin}},\ }\bibfield  {title} {\bibinfo {title} {Upper critical field in
  noncentrosymmetric superconductors},\ }\href
  {https://doi.org/10.1103/PhysRevB.78.224520} {\bibfield  {journal} {\bibinfo
  {journal} {Phys. Rev. B Condens. Matter}\ }\textbf {\bibinfo {volume} {78}},\
  \bibinfo {pages} {224520} (\bibinfo {year} {2008})}\BibitemShut {NoStop}%
\bibitem [{\citenamefont {Yanase}\ and\ \citenamefont
  {Sigrist}(2008)}]{Yanase2008-yb}%
  \BibitemOpen
  \bibfield  {author} {\bibinfo {author} {\bibfnamefont {Y.}~\bibnamefont
  {Yanase}}\ and\ \bibinfo {author} {\bibfnamefont {M.}~\bibnamefont
  {Sigrist}},\ }\bibfield  {title} {\bibinfo {title} {Helical superconductivity
  in non-centrosymmetric superconductors with dominantly spin triplet
  pairing},\ }\href {https://doi.org/10.1143/JPSJS.77SA.342} {\bibfield
  {journal} {\bibinfo  {journal} {J. Phys. Soc. Jpn.}\ }\textbf {\bibinfo
  {volume} {77}},\ \bibinfo {pages} {342} (\bibinfo {year} {2008})}\BibitemShut
  {NoStop}%
\bibitem [{\citenamefont {Michaeli}\ \emph {et~al.}(2012)\citenamefont
  {Michaeli}, \citenamefont {Potter},\ and\ \citenamefont
  {Lee}}]{Michaeli2012-gl}%
  \BibitemOpen
  \bibfield  {author} {\bibinfo {author} {\bibfnamefont {K.}~\bibnamefont
  {Michaeli}}, \bibinfo {author} {\bibfnamefont {A.~C.}\ \bibnamefont
  {Potter}},\ and\ \bibinfo {author} {\bibfnamefont {P.~A.}\ \bibnamefont
  {Lee}},\ }\bibfield  {title} {\bibinfo {title} {Superconducting and
  ferromagnetic phases in {SrTiO3/LaAlO3} oxide interface structures:
  possibility of finite momentum pairing},\ }\href
  {https://doi.org/10.1103/PhysRevLett.108.117003} {\bibfield  {journal}
  {\bibinfo  {journal} {Phys. Rev. Lett.}\ }\textbf {\bibinfo {volume} {108}},\
  \bibinfo {pages} {117003} (\bibinfo {year} {2012})}\BibitemShut {NoStop}%
\bibitem [{\citenamefont {Houzet}\ and\ \citenamefont
  {Meyer}(2015)}]{Houzet2015-iy}%
  \BibitemOpen
  \bibfield  {author} {\bibinfo {author} {\bibfnamefont {M.}~\bibnamefont
  {Houzet}}\ and\ \bibinfo {author} {\bibfnamefont {J.~S.}\ \bibnamefont
  {Meyer}},\ }\bibfield  {title} {\bibinfo {title} {Quasiclassical theory of
  disordered rashba superconductors},\ }\href
  {https://doi.org/10.1103/PhysRevB.92.014509} {\bibfield  {journal} {\bibinfo
  {journal} {Phys. Rev. B Condens. Matter}\ }\textbf {\bibinfo {volume} {92}},\
  \bibinfo {pages} {014509} (\bibinfo {year} {2015})}\BibitemShut {NoStop}%
\bibitem [{\citenamefont {Fulde}\ and\ \citenamefont
  {Ferrell}(1964)}]{Fulde1964-qq}%
  \BibitemOpen
  \bibfield  {author} {\bibinfo {author} {\bibfnamefont {P.}~\bibnamefont
  {Fulde}}\ and\ \bibinfo {author} {\bibfnamefont {R.~A.}\ \bibnamefont
  {Ferrell}},\ }\bibfield  {title} {\bibinfo {title} {Superconductivity in a
  strong {Spin-Exchange} field},\ }\href
  {https://doi.org/10.1103/PhysRev.135.A550} {\bibfield  {journal} {\bibinfo
  {journal} {Phys. Rev.}\ }\textbf {\bibinfo {volume} {135}},\ \bibinfo {pages}
  {A550} (\bibinfo {year} {1964})}\BibitemShut {NoStop}%
\bibitem [{\citenamefont {Larkin}\ and\ \citenamefont
  {Ovchinnikov}(1964)}]{Larkin1964-en}%
  \BibitemOpen
  \bibfield  {author} {\bibinfo {author} {\bibfnamefont {A.~I.}\ \bibnamefont
  {Larkin}}\ and\ \bibinfo {author} {\bibfnamefont {Y.~N.}\ \bibnamefont
  {Ovchinnikov}},\ }\bibfield  {title} {\bibinfo {title} {Nonuniform state of
  superconductors},\ }\href@noop {} {\bibfield  {journal} {\bibinfo  {journal}
  {Zh. Eksperim. i Teor. Fiz.}\ }\textbf {\bibinfo {volume} {47}},\ \bibinfo
  {pages} {1136} (\bibinfo {year} {1964})}\BibitemShut {NoStop}%
\bibitem [{\citenamefont {Matsuda}\ and\ \citenamefont
  {Shimahara}(2007)}]{Matsuda2007-em}%
  \BibitemOpen
  \bibfield  {author} {\bibinfo {author} {\bibfnamefont {Y.}~\bibnamefont
  {Matsuda}}\ and\ \bibinfo {author} {\bibfnamefont {H.}~\bibnamefont
  {Shimahara}},\ }\bibfield  {title} {\bibinfo {title}
  {{Fulde--Ferrell--Larkin--Ovchinnikov} state in heavy fermion
  superconductors},\ }\href {https://doi.org/10.1143/JPSJ.76.051005} {\bibfield
   {journal} {\bibinfo  {journal} {J. Phys. Soc. Jpn.}\ }\textbf {\bibinfo
  {volume} {76}},\ \bibinfo {pages} {051005} (\bibinfo {year}
  {2007})}\BibitemShut {NoStop}%
\bibitem [{\citenamefont {Wosnitza}(2018)}]{Wosnitza2018-qh}%
  \BibitemOpen
  \bibfield  {author} {\bibinfo {author} {\bibfnamefont {J.}~\bibnamefont
  {Wosnitza}},\ }\bibfield  {title} {\bibinfo {title} {{FFLO} states in layered
  organic superconductors},\ }\href {https://doi.org/10.1002/andp.201700282}
  {\bibfield  {journal} {\bibinfo  {journal} {Ann. Phys.}\ }\textbf {\bibinfo
  {volume} {530}},\ \bibinfo {pages} {1700282} (\bibinfo {year}
  {2018})}\BibitemShut {NoStop}%
\bibitem [{\citenamefont {Agterberg}\ \emph {et~al.}(2020)\citenamefont
  {Agterberg}, \citenamefont {Davis}, \citenamefont {Edkins}, \citenamefont
  {Fradkin}, \citenamefont {Van~Harlingen}, \citenamefont {Kivelson},
  \citenamefont {Lee}, \citenamefont {Radzihovsky}, \citenamefont {Tranquada},\
  and\ \citenamefont {Wang}}]{Agterberg2020-gs}%
  \BibitemOpen
  \bibfield  {author} {\bibinfo {author} {\bibfnamefont {D.~F.}\ \bibnamefont
  {Agterberg}}, \bibinfo {author} {\bibfnamefont {J.~C.~S.}\ \bibnamefont
  {Davis}}, \bibinfo {author} {\bibfnamefont {S.~D.}\ \bibnamefont {Edkins}},
  \bibinfo {author} {\bibfnamefont {E.}~\bibnamefont {Fradkin}}, \bibinfo
  {author} {\bibfnamefont {D.~J.}\ \bibnamefont {Van~Harlingen}}, \bibinfo
  {author} {\bibfnamefont {S.~A.}\ \bibnamefont {Kivelson}}, \bibinfo {author}
  {\bibfnamefont {P.~A.}\ \bibnamefont {Lee}}, \bibinfo {author} {\bibfnamefont
  {L.}~\bibnamefont {Radzihovsky}}, \bibinfo {author} {\bibfnamefont {J.~M.}\
  \bibnamefont {Tranquada}},\ and\ \bibinfo {author} {\bibfnamefont
  {Y.}~\bibnamefont {Wang}},\ }\bibfield  {title} {\bibinfo {title} {The
  physics of {Pair-Density} waves: Cuprate superconductors and beyond},\ }\href
  {https://doi.org/10.1146/annurev-conmatphys-031119-050711} {\bibfield
  {journal} {\bibinfo  {journal} {Annu. Rev. Condens. Matter Phys.}\ }\textbf
  {\bibinfo {volume} {11}},\ \bibinfo {pages} {231} (\bibinfo {year}
  {2020})}\BibitemShut {NoStop}%
\bibitem [{\citenamefont {Kumagai}\ \emph {et~al.}(2011)\citenamefont
  {Kumagai}, \citenamefont {Shishido}, \citenamefont {Shibauchi},\ and\
  \citenamefont {Matsuda}}]{Kumagai2011-ao}%
  \BibitemOpen
  \bibfield  {author} {\bibinfo {author} {\bibfnamefont {K.}~\bibnamefont
  {Kumagai}}, \bibinfo {author} {\bibfnamefont {H.}~\bibnamefont {Shishido}},
  \bibinfo {author} {\bibfnamefont {T.}~\bibnamefont {Shibauchi}},\ and\
  \bibinfo {author} {\bibfnamefont {Y.}~\bibnamefont {Matsuda}},\ }\bibfield
  {title} {\bibinfo {title} {Evolution of paramagnetic quasiparticle
  excitations emerged in the high-field superconducting phase of {CeCoIn5}},\
  }\href {https://doi.org/10.1103/PhysRevLett.106.137004} {\bibfield  {journal}
  {\bibinfo  {journal} {Phys. Rev. Lett.}\ }\textbf {\bibinfo {volume} {106}},\
  \bibinfo {pages} {137004} (\bibinfo {year} {2011})}\BibitemShut {NoStop}%
\bibitem [{\citenamefont {Mayaffre}\ \emph {et~al.}(2014)\citenamefont
  {Mayaffre}, \citenamefont {Kr{\"a}mer}, \citenamefont {Horvati{\'c}},
  \citenamefont {Berthier}, \citenamefont {Miyagawa}, \citenamefont {Kanoda},\
  and\ \citenamefont {Mitrovi{\'c}}}]{Mayaffre2014-kt}%
  \BibitemOpen
  \bibfield  {author} {\bibinfo {author} {\bibfnamefont {H.}~\bibnamefont
  {Mayaffre}}, \bibinfo {author} {\bibfnamefont {S.}~\bibnamefont
  {Kr{\"a}mer}}, \bibinfo {author} {\bibfnamefont {M.}~\bibnamefont
  {Horvati{\'c}}}, \bibinfo {author} {\bibfnamefont {C.}~\bibnamefont
  {Berthier}}, \bibinfo {author} {\bibfnamefont {K.}~\bibnamefont {Miyagawa}},
  \bibinfo {author} {\bibfnamefont {K.}~\bibnamefont {Kanoda}},\ and\ \bibinfo
  {author} {\bibfnamefont {V.~F.}\ \bibnamefont {Mitrovi{\'c}}},\ }\bibfield
  {title} {\bibinfo {title} {Evidence of andreev bound states as a hallmark of
  the {FFLO} phase in {$\kappa$-(BEDT-TTF)2Cu(NCS)2}},\ }\href
  {https://doi.org/10.1038/nphys3121} {\bibfield  {journal} {\bibinfo
  {journal} {Nat. Phys.}\ }\textbf {\bibinfo {volume} {10}},\ \bibinfo {pages}
  {928} (\bibinfo {year} {2014})}\BibitemShut {NoStop}%
\bibitem [{\citenamefont {Koutroulakis}\ \emph {et~al.}(2016)\citenamefont
  {Koutroulakis}, \citenamefont {K{\"u}hne}, \citenamefont {Schlueter},
  \citenamefont {Wosnitza},\ and\ \citenamefont {Brown}}]{Koutroulakis2016-wg}%
  \BibitemOpen
  \bibfield  {author} {\bibinfo {author} {\bibfnamefont {G.}~\bibnamefont
  {Koutroulakis}}, \bibinfo {author} {\bibfnamefont {H.}~\bibnamefont
  {K{\"u}hne}}, \bibinfo {author} {\bibfnamefont {J.~A.}\ \bibnamefont
  {Schlueter}}, \bibinfo {author} {\bibfnamefont {J.}~\bibnamefont
  {Wosnitza}},\ and\ \bibinfo {author} {\bibfnamefont {S.~E.}\ \bibnamefont
  {Brown}},\ }\bibfield  {title} {\bibinfo {title} {Microscopic study of the
  {Fulde-Ferrell-Larkin-Ovchinnikov} state in an {All-Organic}
  superconductor},\ }\href {https://doi.org/10.1103/PhysRevLett.116.067003}
  {\bibfield  {journal} {\bibinfo  {journal} {Phys. Rev. Lett.}\ }\textbf
  {\bibinfo {volume} {116}},\ \bibinfo {pages} {067003} (\bibinfo {year}
  {2016})}\BibitemShut {NoStop}%
\bibitem [{\citenamefont {Kitagawa}\ \emph {et~al.}(2018)\citenamefont
  {Kitagawa}, \citenamefont {Nakamine}, \citenamefont {Ishida}, \citenamefont
  {Jeevan}, \citenamefont {Geibel},\ and\ \citenamefont
  {Steglich}}]{Kitagawa2018-qh}%
  \BibitemOpen
  \bibfield  {author} {\bibinfo {author} {\bibfnamefont {S.}~\bibnamefont
  {Kitagawa}}, \bibinfo {author} {\bibfnamefont {G.}~\bibnamefont {Nakamine}},
  \bibinfo {author} {\bibfnamefont {K.}~\bibnamefont {Ishida}}, \bibinfo
  {author} {\bibfnamefont {H.~S.}\ \bibnamefont {Jeevan}}, \bibinfo {author}
  {\bibfnamefont {C.}~\bibnamefont {Geibel}},\ and\ \bibinfo {author}
  {\bibfnamefont {F.}~\bibnamefont {Steglich}},\ }\bibfield  {title} {\bibinfo
  {title} {Evidence for the presence of the {Fulde-Ferrell-Larkin-Ovchinnikov}
  state in {CeCu$_{2}$Si$_{2}$} revealed using $^{63}$cu {NMR}},\ }\href
  {https://doi.org/10.1103/PhysRevLett.121.157004} {\bibfield  {journal}
  {\bibinfo  {journal} {Phys. Rev. Lett.}\ }\textbf {\bibinfo {volume} {121}},\
  \bibinfo {pages} {157004} (\bibinfo {year} {2018})}\BibitemShut {NoStop}%
\bibitem [{\citenamefont {Kasahara}\ \emph {et~al.}(2021)\citenamefont
  {Kasahara}, \citenamefont {Suzuki}, \citenamefont {Machida}, \citenamefont
  {Sato}, \citenamefont {Ukai}, \citenamefont {Murayama}, \citenamefont
  {Suetsugu}, \citenamefont {Kasahara}, \citenamefont {Shibauchi},
  \citenamefont {Hanaguri},\ and\ \citenamefont {Matsuda}}]{Kasahara2021-qa}%
  \BibitemOpen
  \bibfield  {author} {\bibinfo {author} {\bibfnamefont {S.}~\bibnamefont
  {Kasahara}}, \bibinfo {author} {\bibfnamefont {H.}~\bibnamefont {Suzuki}},
  \bibinfo {author} {\bibfnamefont {T.}~\bibnamefont {Machida}}, \bibinfo
  {author} {\bibfnamefont {Y.}~\bibnamefont {Sato}}, \bibinfo {author}
  {\bibfnamefont {Y.}~\bibnamefont {Ukai}}, \bibinfo {author} {\bibfnamefont
  {H.}~\bibnamefont {Murayama}}, \bibinfo {author} {\bibfnamefont
  {S.}~\bibnamefont {Suetsugu}}, \bibinfo {author} {\bibfnamefont
  {Y.}~\bibnamefont {Kasahara}}, \bibinfo {author} {\bibfnamefont
  {T.}~\bibnamefont {Shibauchi}}, \bibinfo {author} {\bibfnamefont
  {T.}~\bibnamefont {Hanaguri}},\ and\ \bibinfo {author} {\bibfnamefont
  {Y.}~\bibnamefont {Matsuda}},\ }\bibfield  {title} {\bibinfo {title}
  {Quasiparticle nodal plane in the {Fulde-Ferrell-Larkin-Ovchinnikov} state of
  {FeSe}},\ }\href {https://doi.org/10.1103/PhysRevLett.127.257001} {\bibfield
  {journal} {\bibinfo  {journal} {Phys. Rev. Lett.}\ }\textbf {\bibinfo
  {volume} {127}},\ \bibinfo {pages} {257001} (\bibinfo {year}
  {2021})}\BibitemShut {NoStop}%
\bibitem [{\citenamefont {Kinjo}\ \emph {et~al.}(2022)\citenamefont {Kinjo},
  \citenamefont {Manago}, \citenamefont {Kitagawa}, \citenamefont {Mao},
  \citenamefont {Yonezawa}, \citenamefont {Maeno},\ and\ \citenamefont
  {Ishida}}]{Kinjo2022-bg}%
  \BibitemOpen
  \bibfield  {author} {\bibinfo {author} {\bibfnamefont {K.}~\bibnamefont
  {Kinjo}}, \bibinfo {author} {\bibfnamefont {M.}~\bibnamefont {Manago}},
  \bibinfo {author} {\bibfnamefont {S.}~\bibnamefont {Kitagawa}}, \bibinfo
  {author} {\bibfnamefont {Z.~Q.}\ \bibnamefont {Mao}}, \bibinfo {author}
  {\bibfnamefont {S.}~\bibnamefont {Yonezawa}}, \bibinfo {author}
  {\bibfnamefont {Y.}~\bibnamefont {Maeno}},\ and\ \bibinfo {author}
  {\bibfnamefont {K.}~\bibnamefont {Ishida}},\ }\bibfield  {title} {\bibinfo
  {title} {Superconducting spin smecticity evidencing the
  {Fulde-Ferrell-Larkin-Ovchinnikov} state in {Sr$_{2}$RuO$_{4}$}},\ }\href
  {https://doi.org/10.1126/science.abb0332} {\bibfield  {journal} {\bibinfo
  {journal} {Science}\ }\textbf {\bibinfo {volume} {376}},\ \bibinfo {pages}
  {397} (\bibinfo {year} {2022})},\ \Eprint
  {https://arxiv.org/abs/https://www.science.org/doi/pdf/10.1126/science.abb0332}
  {https://www.science.org/doi/pdf/10.1126/science.abb0332} \BibitemShut
  {NoStop}%
\bibitem [{\citenamefont {Hamidian}\ \emph {et~al.}(2016)\citenamefont
  {Hamidian}, \citenamefont {Edkins}, \citenamefont {Joo}, \citenamefont
  {Kostin}, \citenamefont {Eisaki}, \citenamefont {Uchida}, \citenamefont
  {Lawler}, \citenamefont {Kim}, \citenamefont {Mackenzie}, \citenamefont
  {Fujita}, \citenamefont {Lee},\ and\ \citenamefont
  {Davis}}]{Hamidian2016-xz}%
  \BibitemOpen
  \bibfield  {author} {\bibinfo {author} {\bibfnamefont {M.~H.}\ \bibnamefont
  {Hamidian}}, \bibinfo {author} {\bibfnamefont {S.~D.}\ \bibnamefont
  {Edkins}}, \bibinfo {author} {\bibfnamefont {S.~H.}\ \bibnamefont {Joo}},
  \bibinfo {author} {\bibfnamefont {A.}~\bibnamefont {Kostin}}, \bibinfo
  {author} {\bibfnamefont {H.}~\bibnamefont {Eisaki}}, \bibinfo {author}
  {\bibfnamefont {S.}~\bibnamefont {Uchida}}, \bibinfo {author} {\bibfnamefont
  {M.~J.}\ \bibnamefont {Lawler}}, \bibinfo {author} {\bibfnamefont {E.-A.}\
  \bibnamefont {Kim}}, \bibinfo {author} {\bibfnamefont {A.~P.}\ \bibnamefont
  {Mackenzie}}, \bibinfo {author} {\bibfnamefont {K.}~\bibnamefont {Fujita}},
  \bibinfo {author} {\bibfnamefont {J.}~\bibnamefont {Lee}},\ and\ \bibinfo
  {author} {\bibfnamefont {J.~C.~S.}\ \bibnamefont {Davis}},\ }\bibfield
  {title} {\bibinfo {title} {Detection of a cooper-pair density wave in
  {Bi2Sr2CaCu2O8+x}},\ }\href {https://doi.org/10.1038/nature17411} {\bibfield
  {journal} {\bibinfo  {journal} {Nature}\ }\textbf {\bibinfo {volume} {532}},\
  \bibinfo {pages} {343} (\bibinfo {year} {2016})}\BibitemShut {NoStop}%
\bibitem [{\citenamefont {Ruan}\ \emph {et~al.}(2018)\citenamefont {Ruan},
  \citenamefont {Li}, \citenamefont {Hu}, \citenamefont {Hao}, \citenamefont
  {Li}, \citenamefont {Cai}, \citenamefont {Zhou}, \citenamefont {Lee},\ and\
  \citenamefont {Wang}}]{Ruan2018-fv}%
  \BibitemOpen
  \bibfield  {author} {\bibinfo {author} {\bibfnamefont {W.}~\bibnamefont
  {Ruan}}, \bibinfo {author} {\bibfnamefont {X.}~\bibnamefont {Li}}, \bibinfo
  {author} {\bibfnamefont {C.}~\bibnamefont {Hu}}, \bibinfo {author}
  {\bibfnamefont {Z.}~\bibnamefont {Hao}}, \bibinfo {author} {\bibfnamefont
  {H.}~\bibnamefont {Li}}, \bibinfo {author} {\bibfnamefont {P.}~\bibnamefont
  {Cai}}, \bibinfo {author} {\bibfnamefont {X.}~\bibnamefont {Zhou}}, \bibinfo
  {author} {\bibfnamefont {D.-H.}\ \bibnamefont {Lee}},\ and\ \bibinfo {author}
  {\bibfnamefont {Y.}~\bibnamefont {Wang}},\ }\bibfield  {title} {\bibinfo
  {title} {Visualization of the periodic modulation of cooper pairing in a
  cuprate superconductor},\ }\href {https://doi.org/10.1038/s41567-018-0276-8}
  {\bibfield  {journal} {\bibinfo  {journal} {Nat. Phys.}\ }\textbf {\bibinfo
  {volume} {14}},\ \bibinfo {pages} {1178} (\bibinfo {year}
  {2018})}\BibitemShut {NoStop}%
\bibitem [{\citenamefont {Chen}\ \emph {et~al.}(2021)\citenamefont {Chen},
  \citenamefont {Yang}, \citenamefont {Hu}, \citenamefont {Zhao}, \citenamefont
  {Yuan}, \citenamefont {Xing}, \citenamefont {Qian}, \citenamefont {Huang},
  \citenamefont {Li}, \citenamefont {Ye}, \citenamefont {Ma}, \citenamefont
  {Ni}, \citenamefont {Zhang}, \citenamefont {Yin}, \citenamefont {Gong},
  \citenamefont {Tu}, \citenamefont {Lei}, \citenamefont {Tan}, \citenamefont
  {Zhou}, \citenamefont {Shen}, \citenamefont {Dong}, \citenamefont {Yan},
  \citenamefont {Wang},\ and\ \citenamefont {Gao}}]{Chen2021-qp}%
  \BibitemOpen
  \bibfield  {author} {\bibinfo {author} {\bibfnamefont {H.}~\bibnamefont
  {Chen}}, \bibinfo {author} {\bibfnamefont {H.}~\bibnamefont {Yang}}, \bibinfo
  {author} {\bibfnamefont {B.}~\bibnamefont {Hu}}, \bibinfo {author}
  {\bibfnamefont {Z.}~\bibnamefont {Zhao}}, \bibinfo {author} {\bibfnamefont
  {J.}~\bibnamefont {Yuan}}, \bibinfo {author} {\bibfnamefont {Y.}~\bibnamefont
  {Xing}}, \bibinfo {author} {\bibfnamefont {G.}~\bibnamefont {Qian}}, \bibinfo
  {author} {\bibfnamefont {Z.}~\bibnamefont {Huang}}, \bibinfo {author}
  {\bibfnamefont {G.}~\bibnamefont {Li}}, \bibinfo {author} {\bibfnamefont
  {Y.}~\bibnamefont {Ye}}, \bibinfo {author} {\bibfnamefont {S.}~\bibnamefont
  {Ma}}, \bibinfo {author} {\bibfnamefont {S.}~\bibnamefont {Ni}}, \bibinfo
  {author} {\bibfnamefont {H.}~\bibnamefont {Zhang}}, \bibinfo {author}
  {\bibfnamefont {Q.}~\bibnamefont {Yin}}, \bibinfo {author} {\bibfnamefont
  {C.}~\bibnamefont {Gong}}, \bibinfo {author} {\bibfnamefont {Z.}~\bibnamefont
  {Tu}}, \bibinfo {author} {\bibfnamefont {H.}~\bibnamefont {Lei}}, \bibinfo
  {author} {\bibfnamefont {H.}~\bibnamefont {Tan}}, \bibinfo {author}
  {\bibfnamefont {S.}~\bibnamefont {Zhou}}, \bibinfo {author} {\bibfnamefont
  {C.}~\bibnamefont {Shen}}, \bibinfo {author} {\bibfnamefont {X.}~\bibnamefont
  {Dong}}, \bibinfo {author} {\bibfnamefont {B.}~\bibnamefont {Yan}}, \bibinfo
  {author} {\bibfnamefont {Z.}~\bibnamefont {Wang}},\ and\ \bibinfo {author}
  {\bibfnamefont {H.-J.}\ \bibnamefont {Gao}},\ }\bibfield  {title} {\bibinfo
  {title} {Roton pair density wave in a strong-coupling kagome
  superconductor},\ }\href {https://doi.org/10.1038/s41586-021-03983-5}
  {\bibfield  {journal} {\bibinfo  {journal} {Nature}\ }\textbf {\bibinfo
  {volume} {599}},\ \bibinfo {pages} {222} (\bibinfo {year}
  {2021})}\BibitemShut {NoStop}%
\bibitem [{\citenamefont {Gu}\ \emph {et~al.}(2023)\citenamefont {Gu},
  \citenamefont {Carroll}, \citenamefont {Wang}, \citenamefont {Ran},
  \citenamefont {Broyles}, \citenamefont {Siddiquee}, \citenamefont {Butch},
  \citenamefont {Saha}, \citenamefont {Paglione}, \citenamefont {Davis},\ and\
  \citenamefont {Liu}}]{Gu2023-ps}%
  \BibitemOpen
  \bibfield  {author} {\bibinfo {author} {\bibfnamefont {Q.}~\bibnamefont
  {Gu}}, \bibinfo {author} {\bibfnamefont {J.~P.}\ \bibnamefont {Carroll}},
  \bibinfo {author} {\bibfnamefont {S.}~\bibnamefont {Wang}}, \bibinfo {author}
  {\bibfnamefont {S.}~\bibnamefont {Ran}}, \bibinfo {author} {\bibfnamefont
  {C.}~\bibnamefont {Broyles}}, \bibinfo {author} {\bibfnamefont
  {H.}~\bibnamefont {Siddiquee}}, \bibinfo {author} {\bibfnamefont {N.~P.}\
  \bibnamefont {Butch}}, \bibinfo {author} {\bibfnamefont {S.~R.}\ \bibnamefont
  {Saha}}, \bibinfo {author} {\bibfnamefont {J.}~\bibnamefont {Paglione}},
  \bibinfo {author} {\bibfnamefont {J.~C.~S.}\ \bibnamefont {Davis}},\ and\
  \bibinfo {author} {\bibfnamefont {X.}~\bibnamefont {Liu}},\ }\bibfield
  {title} {\bibinfo {title} {Detection of a pair density wave state in
  {UTe2}},\ }\href {https://doi.org/10.1038/s41586-023-05919-7} {\bibfield
  {journal} {\bibinfo  {journal} {Nature}\ }\textbf {\bibinfo {volume} {618}},\
  \bibinfo {pages} {921} (\bibinfo {year} {2023})}\BibitemShut {NoStop}%
\bibitem [{\citenamefont {Daido}\ and\ \citenamefont
  {Yanase}(2022)}]{Daido2022-gj}%
  \BibitemOpen
  \bibfield  {author} {\bibinfo {author} {\bibfnamefont {A.}~\bibnamefont
  {Daido}}\ and\ \bibinfo {author} {\bibfnamefont {Y.}~\bibnamefont {Yanase}},\
  }\bibfield  {title} {\bibinfo {title} {Superconducting diode effect and
  nonreciprocal transition lines},\ }\href
  {https://doi.org/10.1103/PhysRevB.106.205206} {\bibfield  {journal} {\bibinfo
   {journal} {Phys. Rev. B Condens. Matter}\ }\textbf {\bibinfo {volume}
  {106}},\ \bibinfo {pages} {205206} (\bibinfo {year} {2022})}\BibitemShut
  {NoStop}%
\bibitem [{\citenamefont {Kim}\ \emph {et~al.}(2016)\citenamefont {Kim},
  \citenamefont {Park},\ and\ \citenamefont {Gilbert}}]{Kim2016-id}%
  \BibitemOpen
  \bibfield  {author} {\bibinfo {author} {\bibfnamefont {Y.}~\bibnamefont
  {Kim}}, \bibinfo {author} {\bibfnamefont {M.~J.}\ \bibnamefont {Park}},\ and\
  \bibinfo {author} {\bibfnamefont {M.~J.}\ \bibnamefont {Gilbert}},\
  }\bibfield  {title} {\bibinfo {title} {Probing unconventional
  superconductivity in inversion-symmetric doped weyl semimetal},\ }\href
  {https://doi.org/10.1103/PhysRevB.93.214511} {\bibfield  {journal} {\bibinfo
  {journal} {Phys. Rev. B Condens. Matter}\ }\textbf {\bibinfo {volume} {93}},\
  \bibinfo {pages} {214511} (\bibinfo {year} {2016})}\BibitemShut {NoStop}%
\bibitem [{\citenamefont {Tinkham}\ \emph {et~al.}(2003)\citenamefont
  {Tinkham}, \citenamefont {Free}, \citenamefont {Lau},\ and\ \citenamefont
  {Markovic}}]{Tinkham2003-tu}%
  \BibitemOpen
  \bibfield  {author} {\bibinfo {author} {\bibfnamefont {M.}~\bibnamefont
  {Tinkham}}, \bibinfo {author} {\bibfnamefont {J.~U.}\ \bibnamefont {Free}},
  \bibinfo {author} {\bibfnamefont {C.~N.}\ \bibnamefont {Lau}},\ and\ \bibinfo
  {author} {\bibfnamefont {N.}~\bibnamefont {Markovic}},\ }\bibfield  {title}
  {\bibinfo {title} {Hysteretic {$I\ensuremath\{-\}V$} curves of
  superconducting nanowires},\ }\href
  {https://doi.org/10.1103/PhysRevB.68.134515} {\bibfield  {journal} {\bibinfo
  {journal} {Phys. Rev. B Condens. Matter}\ }\textbf {\bibinfo {volume} {68}},\
  \bibinfo {pages} {134515} (\bibinfo {year} {2003})}\BibitemShut {NoStop}%
\bibitem [{\citenamefont {Hou}\ \emph {et~al.}(2023)\citenamefont {Hou},
  \citenamefont {Nichele}, \citenamefont {Chi}, \citenamefont {Lodesani},
  \citenamefont {Wu}, \citenamefont {Ritter}, \citenamefont {Haxell},
  \citenamefont {Davydova}, \citenamefont {Ili{\'c}}, \citenamefont
  {Glezakou-Elbert}, \citenamefont {Varambally}, \citenamefont {Bergeret},
  \citenamefont {Kamra}, \citenamefont {Fu}, \citenamefont {Lee},\ and\
  \citenamefont {Moodera}}]{Hou2023-hu}%
  \BibitemOpen
  \bibfield  {author} {\bibinfo {author} {\bibfnamefont {Y.}~\bibnamefont
  {Hou}}, \bibinfo {author} {\bibfnamefont {F.}~\bibnamefont {Nichele}},
  \bibinfo {author} {\bibfnamefont {H.}~\bibnamefont {Chi}}, \bibinfo {author}
  {\bibfnamefont {A.}~\bibnamefont {Lodesani}}, \bibinfo {author}
  {\bibfnamefont {Y.}~\bibnamefont {Wu}}, \bibinfo {author} {\bibfnamefont
  {M.~F.}\ \bibnamefont {Ritter}}, \bibinfo {author} {\bibfnamefont {D.~Z.}\
  \bibnamefont {Haxell}}, \bibinfo {author} {\bibfnamefont {M.}~\bibnamefont
  {Davydova}}, \bibinfo {author} {\bibfnamefont {S.}~\bibnamefont {Ili{\'c}}},
  \bibinfo {author} {\bibfnamefont {O.}~\bibnamefont {Glezakou-Elbert}},
  \bibinfo {author} {\bibfnamefont {A.}~\bibnamefont {Varambally}}, \bibinfo
  {author} {\bibfnamefont {F.~S.}\ \bibnamefont {Bergeret}}, \bibinfo {author}
  {\bibfnamefont {A.}~\bibnamefont {Kamra}}, \bibinfo {author} {\bibfnamefont
  {L.}~\bibnamefont {Fu}}, \bibinfo {author} {\bibfnamefont {P.~A.}\
  \bibnamefont {Lee}},\ and\ \bibinfo {author} {\bibfnamefont {J.~S.}\
  \bibnamefont {Moodera}},\ }\bibfield  {title} {\bibinfo {title} {Ubiquitous
  superconducting diode effect in superconductor thin films},\ }\href
  {https://doi.org/10.1103/PhysRevLett.131.027001} {\bibfield  {journal}
  {\bibinfo  {journal} {Phys. Rev. Lett.}\ }\textbf {\bibinfo {volume} {131}},\
  \bibinfo {pages} {027001} (\bibinfo {year} {2023})}\BibitemShut {NoStop}%
\bibitem [{\citenamefont {Vodolazov}\ and\ \citenamefont
  {Peeters}(2005)}]{Vodolazov2005-cg}%
  \BibitemOpen
  \bibfield  {author} {\bibinfo {author} {\bibfnamefont {D.~Y.}\ \bibnamefont
  {Vodolazov}}\ and\ \bibinfo {author} {\bibfnamefont {F.~M.}\ \bibnamefont
  {Peeters}},\ }\bibfield  {title} {\bibinfo {title} {Superconducting rectifier
  based on the asymmetric surface barrier effect},\ }\href
  {https://doi.org/10.1103/PhysRevB.72.172508} {\bibfield  {journal} {\bibinfo
  {journal} {Phys. Rev. B Condens. Matter}\ }\textbf {\bibinfo {volume} {72}},\
  \bibinfo {pages} {172508} (\bibinfo {year} {2005})}\BibitemShut {NoStop}%
\bibitem [{\citenamefont {Naritsuka}\ \emph {et~al.}(2017)\citenamefont
  {Naritsuka}, \citenamefont {Ishii}, \citenamefont {Miyake}, \citenamefont
  {Tokiwa}, \citenamefont {Toda}, \citenamefont {Shimozawa}, \citenamefont
  {Terashima}, \citenamefont {Shibauchi}, \citenamefont {Matsuda},\ and\
  \citenamefont {Kasahara}}]{Naritsuka2017-or}%
  \BibitemOpen
  \bibfield  {author} {\bibinfo {author} {\bibfnamefont {M.}~\bibnamefont
  {Naritsuka}}, \bibinfo {author} {\bibfnamefont {T.}~\bibnamefont {Ishii}},
  \bibinfo {author} {\bibfnamefont {S.}~\bibnamefont {Miyake}}, \bibinfo
  {author} {\bibfnamefont {Y.}~\bibnamefont {Tokiwa}}, \bibinfo {author}
  {\bibfnamefont {R.}~\bibnamefont {Toda}}, \bibinfo {author} {\bibfnamefont
  {M.}~\bibnamefont {Shimozawa}}, \bibinfo {author} {\bibfnamefont
  {T.}~\bibnamefont {Terashima}}, \bibinfo {author} {\bibfnamefont
  {T.}~\bibnamefont {Shibauchi}}, \bibinfo {author} {\bibfnamefont
  {Y.}~\bibnamefont {Matsuda}},\ and\ \bibinfo {author} {\bibfnamefont
  {Y.}~\bibnamefont {Kasahara}},\ }\bibfield  {title} {\bibinfo {title}
  {Emergent exotic superconductivity in artificially engineered tricolor kondo
  superlattices},\ }\href {https://doi.org/10.1103/PhysRevB.96.174512}
  {\bibfield  {journal} {\bibinfo  {journal} {Phys. Rev. B Condens. Matter}\
  }\textbf {\bibinfo {volume} {96}},\ \bibinfo {pages} {174512} (\bibinfo
  {year} {2017})}\BibitemShut {NoStop}%
\bibitem [{\citenamefont {Naritsuka}\ \emph {et~al.}(2021)\citenamefont
  {Naritsuka}, \citenamefont {Terashima},\ and\ \citenamefont
  {Matsuda}}]{Naritsuka2021-ym}%
  \BibitemOpen
  \bibfield  {author} {\bibinfo {author} {\bibfnamefont {M.}~\bibnamefont
  {Naritsuka}}, \bibinfo {author} {\bibfnamefont {T.}~\bibnamefont
  {Terashima}},\ and\ \bibinfo {author} {\bibfnamefont {Y.}~\bibnamefont
  {Matsuda}},\ }\bibfield  {title} {\bibinfo {title} {Controlling
  unconventional superconductivity in artificially engineered f-electron kondo
  superlattices},\ }\href {https://doi.org/10.1088/1361-648X/abfdf2} {\bibfield
   {journal} {\bibinfo  {journal} {J. Phys. Condens. Matter}\ }\textbf
  {\bibinfo {volume} {33}},\ \bibinfo {pages} {273001} (\bibinfo {year}
  {2021})}\BibitemShut {NoStop}%
\bibitem [{\citenamefont {Sekihara}\ \emph {et~al.}(2013)\citenamefont
  {Sekihara}, \citenamefont {Masutomi},\ and\ \citenamefont
  {Okamoto}}]{Sekihara2013-dm}%
  \BibitemOpen
  \bibfield  {author} {\bibinfo {author} {\bibfnamefont {T.}~\bibnamefont
  {Sekihara}}, \bibinfo {author} {\bibfnamefont {R.}~\bibnamefont {Masutomi}},\
  and\ \bibinfo {author} {\bibfnamefont {T.}~\bibnamefont {Okamoto}},\
  }\bibfield  {title} {\bibinfo {title} {Two-dimensional superconducting state
  of monolayer pb films grown on {GaAs(110}) in a strong parallel magnetic
  field},\ }\href {https://doi.org/10.1103/PhysRevLett.111.057005} {\bibfield
  {journal} {\bibinfo  {journal} {Phys. Rev. Lett.}\ }\textbf {\bibinfo
  {volume} {111}},\ \bibinfo {pages} {057005} (\bibinfo {year}
  {2013})}\BibitemShut {NoStop}%
\bibitem [{\citenamefont {Schumann}\ \emph {et~al.}(2020)\citenamefont
  {Schumann}, \citenamefont {Galletti}, \citenamefont {Jeong}, \citenamefont
  {Ahadi}, \citenamefont {Strickland}, \citenamefont {Salmani-Rezaie},\ and\
  \citenamefont {Stemmer}}]{Schumann2020-mw}%
  \BibitemOpen
  \bibfield  {author} {\bibinfo {author} {\bibfnamefont {T.}~\bibnamefont
  {Schumann}}, \bibinfo {author} {\bibfnamefont {L.}~\bibnamefont {Galletti}},
  \bibinfo {author} {\bibfnamefont {H.}~\bibnamefont {Jeong}}, \bibinfo
  {author} {\bibfnamefont {K.}~\bibnamefont {Ahadi}}, \bibinfo {author}
  {\bibfnamefont {W.~M.}\ \bibnamefont {Strickland}}, \bibinfo {author}
  {\bibfnamefont {S.}~\bibnamefont {Salmani-Rezaie}},\ and\ \bibinfo {author}
  {\bibfnamefont {S.}~\bibnamefont {Stemmer}},\ }\bibfield  {title} {\bibinfo
  {title} {Possible signatures of mixed-parity superconductivity in doped polar
  {SrTiO$_{3}$} films},\ }\href {https://doi.org/10.1103/PhysRevB.101.100503}
  {\bibfield  {journal} {\bibinfo  {journal} {Phys. Rev. B Condens. Matter}\
  }\textbf {\bibinfo {volume} {101}},\ \bibinfo {pages} {100503} (\bibinfo
  {year} {2020})}\BibitemShut {NoStop}%
\bibitem [{\citenamefont {Larkin}\ and\ \citenamefont
  {Varlamov}(2005)}]{Larkin2005-lb}%
  \BibitemOpen
  \bibfield  {author} {\bibinfo {author} {\bibfnamefont {A.}~\bibnamefont
  {Larkin}}\ and\ \bibinfo {author} {\bibfnamefont {A.}~\bibnamefont
  {Varlamov}},\ }\href@noop {} {\emph {\bibinfo {title} {Theory of Fluctuations
  in Superconductors}}}\ (\bibinfo  {publisher} {OUP Oxford},\ \bibinfo {year}
  {2005})\BibitemShut {NoStop}%
\bibitem [{Sup()}]{Supplemental}%
  \BibitemOpen
  \href@noop {} {}\bibinfo {note} {See Supplemental Material for more
  details.}\BibitemShut {Stop}%
\bibitem [{\citenamefont {Schmid}(1969)}]{Schmid1969-fw}%
  \BibitemOpen
  \bibfield  {author} {\bibinfo {author} {\bibfnamefont {A.}~\bibnamefont
  {Schmid}},\ }\bibfield  {title} {\bibinfo {title} {Diamagnetic susceptibility
  at the transition to the superconducting state},\ }\href
  {https://doi.org/10.1103/PhysRev.180.527} {\bibfield  {journal} {\bibinfo
  {journal} {Phys. Rev.}\ }\textbf {\bibinfo {volume} {180}},\ \bibinfo {pages}
  {527} (\bibinfo {year} {1969})}\BibitemShut {NoStop}%
\bibitem [{\citenamefont {Konschelle}\ \emph {et~al.}(2007)\citenamefont
  {Konschelle}, \citenamefont {Cayssol},\ and\ \citenamefont
  {Buzdin}}]{Konschelle2007-da}%
  \BibitemOpen
  \bibfield  {author} {\bibinfo {author} {\bibfnamefont {F.}~\bibnamefont
  {Konschelle}}, \bibinfo {author} {\bibfnamefont {J.}~\bibnamefont
  {Cayssol}},\ and\ \bibinfo {author} {\bibfnamefont {A.~I.}\ \bibnamefont
  {Buzdin}},\ }\bibfield  {title} {\bibinfo {title} {Anomalous fluctuation
  regimes at {FFLO} transition},\ }\href
  {https://doi.org/10.1209/0295-5075/79/67001} {\bibfield  {journal} {\bibinfo
  {journal} {EPL}\ }\textbf {\bibinfo {volume} {79}},\ \bibinfo {pages} {67001}
  (\bibinfo {year} {2007})}\BibitemShut {NoStop}%
\bibitem [{\citenamefont {Bollinger}\ \emph {et~al.}(2011)\citenamefont
  {Bollinger}, \citenamefont {Dubuis}, \citenamefont {Yoon}, \citenamefont
  {Pavuna}, \citenamefont {Misewich},\ and\ \citenamefont {Bo{\v
  z}ovi{\'c}}}]{Bollinger2011-ef}%
  \BibitemOpen
  \bibfield  {author} {\bibinfo {author} {\bibfnamefont {A.~T.}\ \bibnamefont
  {Bollinger}}, \bibinfo {author} {\bibfnamefont {G.}~\bibnamefont {Dubuis}},
  \bibinfo {author} {\bibfnamefont {J.}~\bibnamefont {Yoon}}, \bibinfo {author}
  {\bibfnamefont {D.}~\bibnamefont {Pavuna}}, \bibinfo {author} {\bibfnamefont
  {J.}~\bibnamefont {Misewich}},\ and\ \bibinfo {author} {\bibfnamefont
  {I.}~\bibnamefont {Bo{\v z}ovi{\'c}}},\ }\bibfield  {title} {\bibinfo {title}
  {Superconductor-insulator transition in la2 - {xSrxCuO4} at the pair quantum
  resistance},\ }\href {https://doi.org/10.1038/nature09998} {\bibfield
  {journal} {\bibinfo  {journal} {Nature}\ }\textbf {\bibinfo {volume} {472}},\
  \bibinfo {pages} {458} (\bibinfo {year} {2011})}\BibitemShut {NoStop}%
\bibitem [{\citenamefont {Leng}\ \emph {et~al.}(2011)\citenamefont {Leng},
  \citenamefont {Garcia-Barriocanal}, \citenamefont {Bose}, \citenamefont
  {Lee},\ and\ \citenamefont {Goldman}}]{Leng2011-bi}%
  \BibitemOpen
  \bibfield  {author} {\bibinfo {author} {\bibfnamefont {X.}~\bibnamefont
  {Leng}}, \bibinfo {author} {\bibfnamefont {J.}~\bibnamefont
  {Garcia-Barriocanal}}, \bibinfo {author} {\bibfnamefont {S.}~\bibnamefont
  {Bose}}, \bibinfo {author} {\bibfnamefont {Y.}~\bibnamefont {Lee}},\ and\
  \bibinfo {author} {\bibfnamefont {A.~M.}\ \bibnamefont {Goldman}},\
  }\bibfield  {title} {\bibinfo {title} {Electrostatic control of the evolution
  from a superconducting phase to an insulating phase in ultrathin
  {YBa2Cu3O(7-x}) films},\ }\href
  {https://doi.org/10.1103/PhysRevLett.107.027001} {\bibfield  {journal}
  {\bibinfo  {journal} {Phys. Rev. Lett.}\ }\textbf {\bibinfo {volume} {107}},\
  \bibinfo {pages} {027001} (\bibinfo {year} {2011})}\BibitemShut {NoStop}%
\bibitem [{\citenamefont {Nojima}\ \emph {et~al.}(2011)\citenamefont {Nojima},
  \citenamefont {Tada}, \citenamefont {Nakamura}, \citenamefont {Kobayashi},
  \citenamefont {Shimotani},\ and\ \citenamefont {Iwasa}}]{Nojima2011-il}%
  \BibitemOpen
  \bibfield  {author} {\bibinfo {author} {\bibfnamefont {T.}~\bibnamefont
  {Nojima}}, \bibinfo {author} {\bibfnamefont {H.}~\bibnamefont {Tada}},
  \bibinfo {author} {\bibfnamefont {S.}~\bibnamefont {Nakamura}}, \bibinfo
  {author} {\bibfnamefont {N.}~\bibnamefont {Kobayashi}}, \bibinfo {author}
  {\bibfnamefont {H.}~\bibnamefont {Shimotani}},\ and\ \bibinfo {author}
  {\bibfnamefont {Y.}~\bibnamefont {Iwasa}},\ }\bibfield  {title} {\bibinfo
  {title} {Hole reduction and electron accumulation in
  {YBa$_{2}$Cu$_{3}$O$_{y}$} thin films using an electrochemical technique:
  Evidence for an n-type metallic state},\ }\href
  {https://doi.org/10.1103/PhysRevB.84.020502} {\bibfield  {journal} {\bibinfo
  {journal} {Phys. Rev. B Condens. Matter}\ }\textbf {\bibinfo {volume} {84}},\
  \bibinfo {pages} {020502} (\bibinfo {year} {2011})}\BibitemShut {NoStop}%
\bibitem [{\citenamefont {Liao}\ \emph {et~al.}(2018)\citenamefont {Liao},
  \citenamefont {Zhu}, \citenamefont {Zhang}, \citenamefont {Zhong},
  \citenamefont {Schneeloch}, \citenamefont {Gu}, \citenamefont {Jiang},
  \citenamefont {Zhang}, \citenamefont {Ma},\ and\ \citenamefont
  {Xue}}]{Liao2018-ty}%
  \BibitemOpen
  \bibfield  {author} {\bibinfo {author} {\bibfnamefont {M.}~\bibnamefont
  {Liao}}, \bibinfo {author} {\bibfnamefont {Y.}~\bibnamefont {Zhu}}, \bibinfo
  {author} {\bibfnamefont {J.}~\bibnamefont {Zhang}}, \bibinfo {author}
  {\bibfnamefont {R.}~\bibnamefont {Zhong}}, \bibinfo {author} {\bibfnamefont
  {J.}~\bibnamefont {Schneeloch}}, \bibinfo {author} {\bibfnamefont
  {G.}~\bibnamefont {Gu}}, \bibinfo {author} {\bibfnamefont {K.}~\bibnamefont
  {Jiang}}, \bibinfo {author} {\bibfnamefont {D.}~\bibnamefont {Zhang}},
  \bibinfo {author} {\bibfnamefont {X.}~\bibnamefont {Ma}},\ and\ \bibinfo
  {author} {\bibfnamefont {Q.-K.}\ \bibnamefont {Xue}},\ }\bibfield  {title}
  {\bibinfo {title} {{Superconductor-Insulator} transitions in exfoliated
  {Bi2Sr2CaCu2O8+$\delta$} flakes},\ }\href
  {https://doi.org/10.1021/acs.nanolett.8b02183} {\bibfield  {journal}
  {\bibinfo  {journal} {Nano Lett.}\ }\textbf {\bibinfo {volume} {18}},\
  \bibinfo {pages} {5660} (\bibinfo {year} {2018})}\BibitemShut {NoStop}%
\bibitem [{\citenamefont {Abrahams}\ and\ \citenamefont
  {Tsuneto}(1966)}]{Abrahams1966-xl}%
  \BibitemOpen
  \bibfield  {author} {\bibinfo {author} {\bibfnamefont {E.}~\bibnamefont
  {Abrahams}}\ and\ \bibinfo {author} {\bibfnamefont {T.}~\bibnamefont
  {Tsuneto}},\ }\bibfield  {title} {\bibinfo {title} {Time variation of the
  {Ginzburg-Landau} order parameter},\ }\href
  {https://doi.org/10.1103/PhysRev.152.416} {\bibfield  {journal} {\bibinfo
  {journal} {Phys. Rev.}\ }\textbf {\bibinfo {volume} {152}},\ \bibinfo {pages}
  {416} (\bibinfo {year} {1966})}\BibitemShut {NoStop}%
\bibitem [{\citenamefont {Asaba}\ and\ \citenamefont
  {Matsuda}()}]{Asaba_private}%
  \BibitemOpen
  \bibfield  {author} {\bibinfo {author} {\bibfnamefont {T.}~\bibnamefont
  {Asaba}}\ and\ \bibinfo {author} {\bibfnamefont {Y.}~\bibnamefont
  {Matsuda}},\ }\href@noop {} {}\bibinfo {note} {Private
  communication.}\BibitemShut {Stop}%
\end{thebibliography}
\end{document}